\documentclass[twocolumn]{aastex63}

\usepackage{amsmath}
\usepackage{natbib}
\usepackage{multirow} 
\usepackage{anyfontsize}
\usepackage{indentfirst}
\usepackage{comment}
\usepackage{booktabs}
\usepackage{bigints}
\usepackage{mathrsfs,amsmath} 
\usepackage{makecell}
\usepackage{graphicx}

\usepackage[normalem]{ulem}
\useunder{\uline}{\ul}{}
\usepackage{hyperref}
\usepackage{lineno}

\hypersetup{linkcolor=magenta,citecolor=cyan,filecolor=yellow,urlcolor=blue}

\received{ddmmyyyy}
\revised{\today}
\accepted{ddmmyyyy}
\published{ddmmyyyy}

\submitjournal{ApJS}

\shorttitle{Main Bursts versus Extended Emission}
\shortauthors{Li et al.}

\begin{document}

\title{Pulse-resolved Classification and Characteristics of Long-duration GRBs with \emph{Swift}-BAT Data.II. \\ Main Burst versus Extended Emission}

\author[0000-0002-1343-3089]{Liang Li}
\affiliation{Institute of Fundamental Physics and Quantum Technology, Ningbo University, Ningbo, Zhejiang 315211, People's Republic of China}
\affiliation{School of Physical Science and Technology, Ningbo University, Ningbo, Zhejiang 315211, People's Republic of China}

\author{Xiao~Wang}
\affiliation{Institute of Fundamental Physics and Quantum Technology, Ningbo University, Ningbo, Zhejiang 315211, People's Republic of China}
\affiliation{School of Physical Science and Technology, Ningbo University, Ningbo, Zhejiang 315211, People's Republic of China}

\author{Zhi-Li~Cui}
\affiliation{Institute of Fundamental Physics and Quantum Technology, Ningbo University, Ningbo, Zhejiang 315211, People's Republic of China}
\affiliation{School of Physical Science and Technology, Ningbo University, Ningbo, Zhejiang 315211, People's Republic of China}

\author{Cheng-Long~Xiao}
\affiliation{Institute of Fundamental Physics and Quantum Technology, Ningbo University, Ningbo, Zhejiang 315211, People's Republic of China}
\affiliation{School of Physical Science and Technology, Ningbo University, Ningbo, Zhejiang 315211, People's Republic of China}

\author{Wen~Li}
\affiliation{Institute of Fundamental Physics and Quantum Technology, Ningbo University, Ningbo, Zhejiang 315211, People's Republic of China}
\affiliation{School of Physical Science and Technology, Ningbo University, Ningbo, Zhejiang 315211, People's Republic of China}

\author{Yu Wang}
\affiliation{ICRANet, Piazza della Repubblica 10, I-65122 Pescara, Italy}
\affiliation{ICRA and Dipartimento di Fisica, Universit\`a  di Roma ``La Sapienza'', Piazzale Aldo Moro 5, I-00185 Roma, Italy}
\affiliation{INAF -- Osservatorio Astronomico d'Abruzzo, Via M. Maggini snc, I-64100, Teramo, Italy}

\author{Zi-Gao Dai}
\affiliation{Department of Astronomy, University of Science and Technology of China, Hefei 230026, China\label{USTC}.}

\author{Rong-Gen~Cai}
\affiliation{Institute of Fundamental Physics and Quantum Technology, Ningbo University, Ningbo, Zhejiang 315211, People's Republic of China}

\correspondingauthor{Liang-Li, and Rong-Gen~Cai}
\email{liliang@nbu.edu.cn; cairg@itp.ac.cn}

\begin{abstract}

Long gamma-ray bursts (GRBs) frequently exhibit complex prompt emission structures with multiple temporally distinct episodes, such as a main emission (ME) phase followed by a weak extended emission (EE) tail. Whether these subcomponents from a common physical origin with similar classification properties, or instead represent fundamentally different emission mechanisms within a single event, remains an open question. Here, we present a systematic, pulse-resolved analysis of 22 \emph{Swift}/BAT long-duration GRBs, each exhibiting a well-separated, bright ME ($G_1$) followed by a fainter EE ($G_2$) after a background-consistent quiescent gap. For each component, we independently measure standard classification diagnostics, including duration ($T_{90}$), spectral hardness ratio (HR), minimum variability timescale (MVT), and spectral lag. We then compare these properties between the ME and EE within individual bursts. We find that the EE is systematically softer (lower HR in 19 of 22 events), smoother (longer MVT in 17 of 22 events), and more diverse in spectral lag than the ME. However, both components still occupy the long-GRB track in the traditional duration-hardness and duration-MVT planes, indicating a common Type~II (collapsar) origin. These results suggest that the EE in long GRBs represents a physically distinct regime of the central engine, characterized by a lower luminosity, longer emission timescales, and evolved spectral properties, rather than a simple continuation of the main burst. This picture is consistent with late-time fallback accretion onto a black hole or proto-magnetar spin-down.

\end{abstract}


\section{Introduction} \label{sec:intro}

Gamma-ray bursts (GRBs) are among the most energetic transients in the Universe, powered by compact relativistic central engines that launch collimated jets from cosmological distances \citep{Rees1994,Kumar2015}. Historically, GRBs have been classified into two populations based on their prompt-emission duration ($T_{90}$): long GRBs ($T_{90}\gtrsim 2$~s), associated with the core collapse of massive stars (Type~II), and short GRBs ($T_{90} < 2$~s), associated with compact binary mergers (Type~I) \citep{Kouveliotou1993,Woosley2006,Berger2014}. Multi-messenger and multi-wavelength observations strongly support this progenitor dichotomy, including supernova associations in long GRBs  \citep[e.g., GRB~030329/SN 2003dh,][]{Hjorth2003} and kilonova/gravitational-wave counterparts in short GRBs \citep[e.g., GRB~170817A/GW170817,][]{Abbott2017}.

However, the canonical $T_{90}\simeq 2$~s boundary is only a rough guideline and is increasingly recognized as ambiguous. The observed $T_{90}$ is instrument-dependent, varying with detector sensitivity, energy band, and trigger thresholds \citep{Koshut1996,QinY2013,Lien2016}, and does not always reflect a single physical timescale of the central engine. Significant overlap in the $\sim$1-3~s regime, where events of intermediate duration may belong to either class \citep{Bromberg2013,Sakamoto2011BATcat,Steinhardt2023}, with roughly 10–20\% of bursts having ambiguous classification based on $T_{90}$ alone \citep{Steinhardt2023}. Several well-studied outliers, the so-called ``hybrid" events, demonstrate that duration alone can be misleading. For example, the $\sim 1$ s burst GRB~200826A exhibits collapsar-like behavior (Type~II) \citep{Ahumada2021NatAs,ZhangBB2021NatAs,Rossi2022}, whereas the long-lived events GRB~060614 and GRB~211211A (both $T_{90}\gtrsim 20$\,s) carry merger signatures (Type~I), including a kilonova in the latter case \citep{Gehrels2006Nature,Fynbo2006Nature,GalYam2006Nature,Troja2022Nature,Rastinejad2022Nature,Yang2022Nature}. 

A principal source of this classification ambiguity is that many GRBs consist of multiple distinct emission episodes separated by quiescent intervals, rather than single continuous emission pulses \citep[e.g.,][]{Li2019a,Li2021b}. When sub-bursts are aggregated, the resulting global $T_{90}$ can be substantially inflated and obscure the duration-to-progenitor mapping \citep{ZhangBB2012,2025arXiv251223660L}. Two multi-episode scenarios are especially noteworthy. First, short GRBs frequently display a sub-second hard spike followed by a softer, long-lived extended emission (EE) tail persisting for tens to hundreds of seconds \citep{Norris2006, Gehrels2006,Perley2009, Kaneko2015, Hu2014}. Second, many long GRBs exhibit an early, low-fluence precursor preceding the main burst by several to hundreds of seconds \citep{Koshut1995, Lazzati2005, Burlon2009, Hu2014, Charisi2015, Coppin2020,2026arXiv260121693L}. In both scenarios, the total $T_{90}$ reflects the sum of disjoint episodes and intervening quiescent gaps rather than a single physical timescale. Consequently, robust inference about the central engine requires an episode-level approach: each sub-burst must be characterized independently in terms of duration, hardness, lag, and variability before global conclusions are drawn.

Apart from duration, long and short GRBs diverge in several prompt-emission properties that probe their emission physics. Short GRBs typically exhibit harder spectra, negligible spectral lags, and shorter minimum variability timescales (MVT) compared to long GRBs \citep[e.g.,][]{Norris2000, Norris2005, Yi2006, Horvath2010, Golkhou2014, Golkhou2015}. Long GRBs often show smoother, multi-peaked temporal structures, measurable positive lags (hard photons arriving earlier than softer ones), and longer MVT. These differences in hardness, lag, variability, and light curve morphology collectively reflect underlying differences in the GRB central engine duration and jet dynamics \citep{Zhang2009,Kumar2015}.

While precursors in long GRBs and extended emission in short GRBs have been individually catalogued, comparatively few systematic studies have compared ME and EE within the same long GRB using a uniform set of prompt emission diagnostics. This gap matters for at least two reasons. First, the spectral and temporal properties of EE, which typically exhibit softer spectra, longer variability timescales, and sometimes large or even sign-reversed lags relative to the main pulse, could indicate a distinct dissipation regime or a transition in jet composition, such as from an initially highly relativistic, shock-dominated outflow to a slower, magnetically modulated, or late-accretion phase \citep{Norris2006, Kaneko2015}. Second, the apparent kinship between long-GRB ME+EE and short-GRB ME+EE phenomenology raises the question of whether EE is a partially unified engine behavior across progenitors: an impulsive, high-power episode followed by a lower-power, longer-lived one \citep{Zhang2011Review,Metzger2011}.

Traditionally, GRBs are categorized into Type I (compact object merger) or Type II (collapsar) events based on $T_{90}$, spectral hardness, lag, and afterglow properties \citep{Zhang2009}. Our analysis (Paper I, \citealt{2026arXiv260121693L}) shows that both $G_1$ (precursor) and $G_2$ (main emission), which constitute \emph{Group I} as defined in Paper I, generally fall within the Type II category across these parameters. However, MVT and spectral lag highlight subtle but systematic differences: precursors typically have MVTs $\sim$3-10 times longer and diversely spectral lags, whereas main pulses exhibit shorter MVTs and significant positive lags ($\tau \sim$ 100-600~ms). Such divergence suggests that while both emissions arise from a collapsar, they trace different stages of jet evolution or emission mechanisms.

Motivated by these considerations, this paper (Paper II) presents a comparative, episode (pulse)-resolved study of long GRBs that exhibit an initial ME (denoted as $G_1$) followed, after a quiescent interval, by a distinct and weaker EE (denoted as $G_2$) tail (\emph{Group II} defined in Paper I, \citealt{2026arXiv260121693L}). We assemble a clean subsample of two-episode long GRBs observed by \emph{Swift}/BAT, requiring a conservative flux-ratio criterion $F_{\rm p}(\mathrm{G_2})/F_{\rm p}(\mathrm{G_1})<0.5$ to ensure that the second episode is genuinely weaker than the first. For each episode we measure, in a uniform bandpass, the episode-level duration ($T_{90}$), hardness ratio (50–100 keV over 25–50 keV), spectral lag (50–100 vs. 25–50 keV), and MVT. Our goals are: (i) to quantify systematic differences between ME and EE in long GRBs; (ii) to test whether EE constitutes a physically distinct prompt emission phase with different variability and spectral properties; and (iii) to discuss implications for central-engine physics and for the phenomenological bridges between long-GRB ME+EE and short-GRB ME+EE events. A companion paper (Paper I) analyzed long GRBs with early-time precursor emission; here we focus on the ME+EE morphology and its episode-level characteristics and classification properties.

The paper is organized as follows. Section~\ref{sec:methodology} describes sample selection, data reduction, and analysis methodology. Section~\ref{sec:results} presents results of the comparative ME versus EE analysis. Section ~\ref{sec:discussion} discusses implications for GRB classification and central-engine physics. Section~\ref{sec:conclusion} summarizes our conclusions. Throughout the paper, the standard $\Lambda$-CDM cosmology with the parameters $H_{0}= 67.4$ ${\rm km s^{-1}}$ ${\rm Mpc^{-1}}$, $\Omega_{M}=0.315$, and $\Omega_{\Lambda}=0.685$ are adopted \citep{PlanckCollaboration2018}.

\section{Methodology}\label{sec:methodology}

\subsection{Sample selection}\label{sec:sample}

To investigate the Type I/II classification and the characteristics of distinct prompt-emission episodes in long-duration gamma-ray bursts, we searched the \textit{Swift}/Burst Alert Telescope (BAT) archive for events detected between 2004 November and 2024 December. Specifically, we selected long GRBs exhibiting two distinct emission episodes separated by a background-consistent quiescent interval. Candidate events were identified by visual inspection and by requiring that the count rate return to background level between $G_1$ and $G_2$. In practice, our selection criteria are: 

\begin{itemize}
\item The light curve must clearly show a bright ME followed by a fainter EE, with a distinct quiescent interval ($\Delta T_{\rm gap}$) between them where the flux drop to background levels. We exclude cases with overlapping pulses or more than two episodes to avoid ambiguity.

\item To ensure $G_2$ is an extended emission rather than a second main-burst peak, we impose a conservative flux criterion:
\begin{equation}
F_{\rm p}(G_2) / F_{\rm p}(G_1) < 0.5,
\end{equation}
where $F_{\rm p}$ denotes the 1-second peak photon flux in the 15--150 keV band. This criterion effectively filters out cases of two comparably bright pulses.

\item Both episodes must have sufficient signal-to-noise in mask-weighted light curves to allow reliable measurement of spectral hardness, variability timescale, and lag. We required that both episodes exceed 5$\sigma$ significance \citep{Lien2016} and that the partial coding fraction was sufficient for reliable background modeling and energy-resolved products to ensure that bursts have stable \emph{Swift} pointing throughout the prompt emission \citep{Barthelmy2005}.

\item We limit our sample to GRBs with known redshifts, as the derivation of several key observables in our classification scheme require rest-frame corrections.

\end{itemize}

Applying these criteria, we identified a final sample of 22 long GRBs with clean well-separated two-episode morphology (Table~\ref{tab:pulse}). Each burst shows an initial bright main burst followed by weaker emission after a noticeable quiescent gap. The episode properties for each event are summarized in Table \ref{tab:pulse}.
Further details regarding the sample selection methods can be found in \cite{2026arXiv260121693L}.

\subsection{Standard Classification Diagnostics}\label{sec:classification}

For each burst, we performed uniform analysis on BAT data. We extracted mask-weighted light curves in the full BAT band (15-350~keV) and four standard \emph{Swift}-BAT energy bands: 15-25~keV, 25-50~keV, 50-100~keV, and 100-350~keV (for hardness ratio and spectral lag analysis). 

For each episode ($G_1$ and $G_2$), we measured the following prompt-emission properties independently:

\begin{itemize}
\item Episode (pulse)-wise duration ($T_{90}$): We computed $T_{90}$ for each episode independently using the 15-150 keV cumulative fluence within the episode boundaries, excluding the quiescent gap, considering only the time interval $[t_{1}, t_{2}]$ of that episode and its internal background,
\begin{equation}
T^{G_1}_{90} = t^{G_1}_2-t^{G_1}_1,
\qquad
T^{G_2}_{90} = t^{G_2}_2-t^{G_2}_1,
\end{equation}
where the start $t_1$ of an episode was defined at the time when the cumulative count reaches 5\% of that episode’s total counts, and the end $t_2$ at 95\% of the total counts (\citealt{Kouveliotou1993}, the standard $T_{90}$ definition, but applied locally to each episode). 
These times were refined by requiring that the count rate falls to the background level at the boundaries. By construction, the interval between $t_2(G_1)$ and $t_1(G_2)$ showed no significant excess counts, confirming a true quiescent gap. The quiescent interval $\Delta T_{\rm gap} \equiv t_1(G_2) - t_2(G_1)$ therefore can be defined as the time from the end of $G_1$ to the start of $G_2$, during which no significant emission was detected.
In other words, $T_{90}^{\rm Episode}$ measures the duration of the emission in $G_1$ or $G_2$ alone, excluding any quiescent time before or after. For each burst, this gives $T_{90}^{\rm ME}$ for the main emission and $T_{90}^{\rm EE}$ for the extended emission\footnote{We stress that by isolating episodes we intentionally discard the quiescent interval from calculations like $T_{90}$. For instance, if a burst has $T_{90}^{\rm global} = 100$~s but consists of two 30~s episodes separated by 40~s of silence, our analysis treats it as two episodes of 30~s each (plus an 40~s gap), rather than as one 100~s event. This approach removes the bias introduced by long gaps, which can artificially inflate $T_{90}$ \citep{MacLachlan2013}. It allows a like-for-like comparison of the ME vs. EE under the same analysis conditions.}. The uncertainties in $T_{90}^{\rm Episode}$ were estimated through standard error propagation, accounting for both photon-counting statistics and background fluctuations. 

\item Episode (pulse)-wise peak flux ($F_{\rm p}$): The peak 1-s (or shorter, if relevant, see Table \ref{tab:pulse}) count rate in each episode’s 15–150 keV light curve, used to the $F_{\rm p}(G_2)/F_{\rm p}(G_1) < 0.5$ selection criterion.

\item Episode (pulse)-wise spectral hardness ratio: We define a hardness ratio for each episode as the ratio of background-subtracted fluences in the 50-100~keV band to that in the 25-50~keV band,
\begin{equation}
{\rm HR} \equiv S_{50-100~{\rm keV}}/S_{25-50~{\rm keV}},
\end{equation}
with Poisson uncertainties propagated accordingly. This HR is a dimensionless measure of the spectrum’s hardness within BAT. 
A larger HR means a harder spectrum (relatively more high-energy photons). We chose these bands to straddle the $\sim50$ keV typical peak energy of BAT-detected long GRBs, so that HR $>$ 1 indicates more weight in higher energies. 

\item Episode (pulse)-wise MVT: We estimated MVT $\Delta t_{\min}$ in each episode following the procedure described in \cite{MacLachlan2013,Golkhou2014,Golkhou2015} using a Haar wavelet transform and an autocorrelation function (ACF) analysis on high time-resolution light curves (down to 100~$\mu$s if signal allows) to find the shortest timescale at which significant signal variation exceeds statistical noise \citep{Ahumada2021NatAs,MacLachlan2013}. Uncertainties were estimated by adding Poisson noise to the light curves via bootstrap resampling. 

\item Episode (pulse)-wise spectral lag: Using the 50–100~keV and 25–50~keV light curves, we use the cross-correlation function (CCF) to determined the lag as the time offset that maximizes the CCF. We required a well-defined CCF peak with significance $>5\sigma$ for a lag measurement. Uncertainties on lag were derived by fitting the CCF peak with a smooth function (e.g. Gaussian) and via Monte Carlo simulations \citep{Band1997,Norris2000}. A positive lag indicates the hard band leads the soft band \citep{Band1997}. Lags were measured in the observer frame. Observationally, a “typical” long GRB has a positive lag of order 0.1–0.5~s, while short GRBs usually have negligible or zero lag \citep{Band1997,MacLachlan2013}. 

\end{itemize}

All episode-level products for $G_1$ and $G_2$ are derived in the identical bandpass and with identical algorithms to ensure unbiased pairwise comparisons. All measurements are reported in the observer frame to maintain homogeneity across the sample.

\section{Results}\label{sec:results}

\begin{figure*}
\includegraphics[angle=0,scale=0.32]{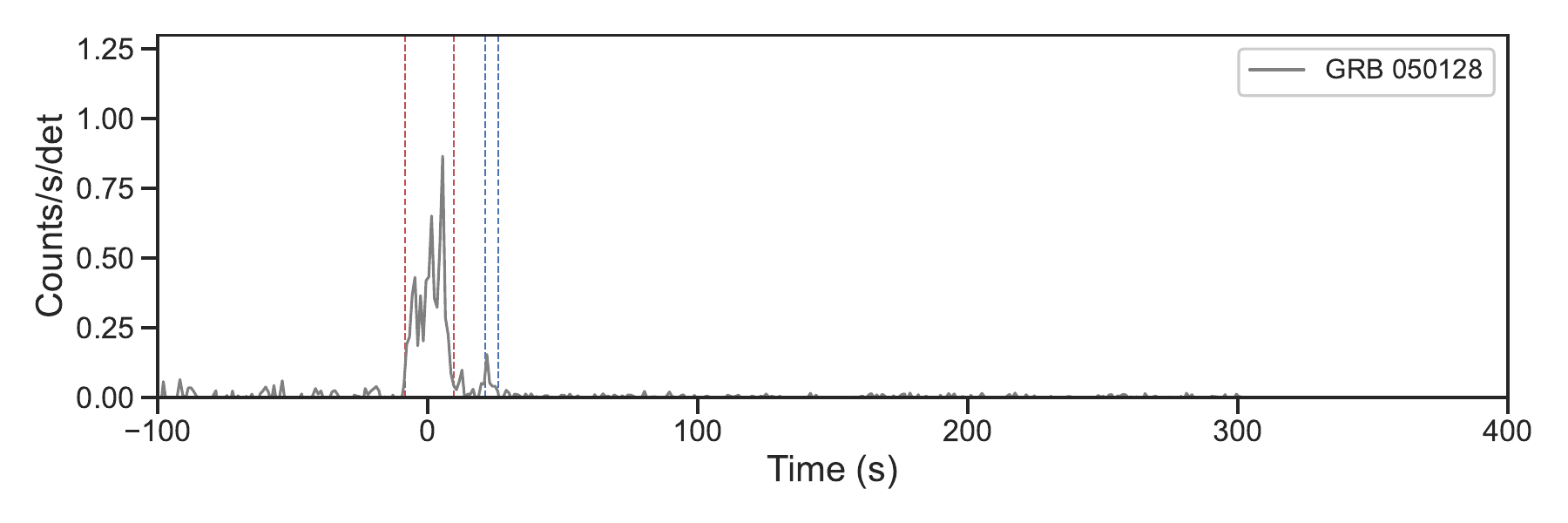}
\includegraphics[angle=0,scale=0.32]{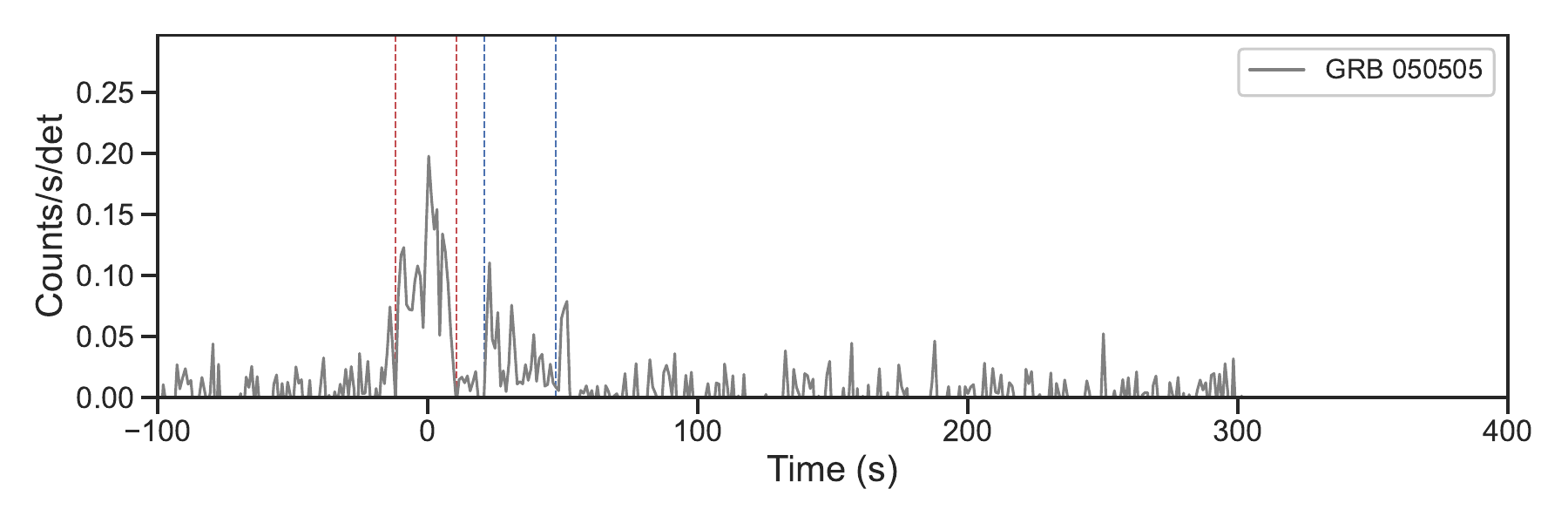}
\includegraphics[angle=0,scale=0.32]{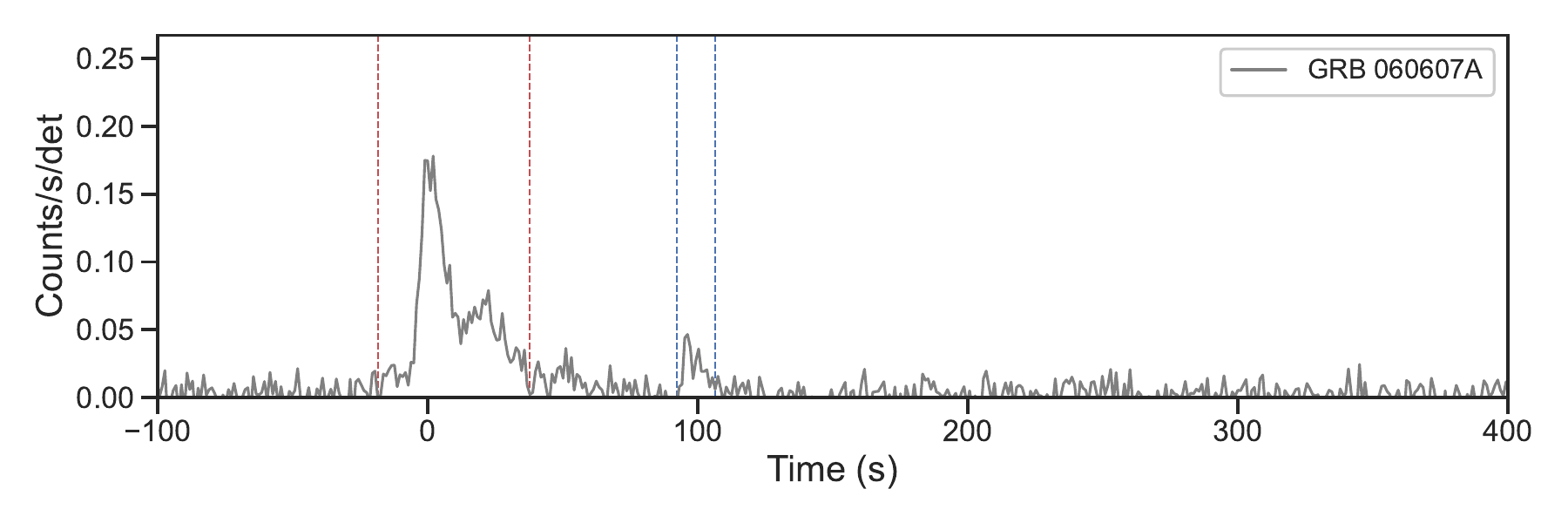}
\includegraphics[angle=0,scale=0.32]{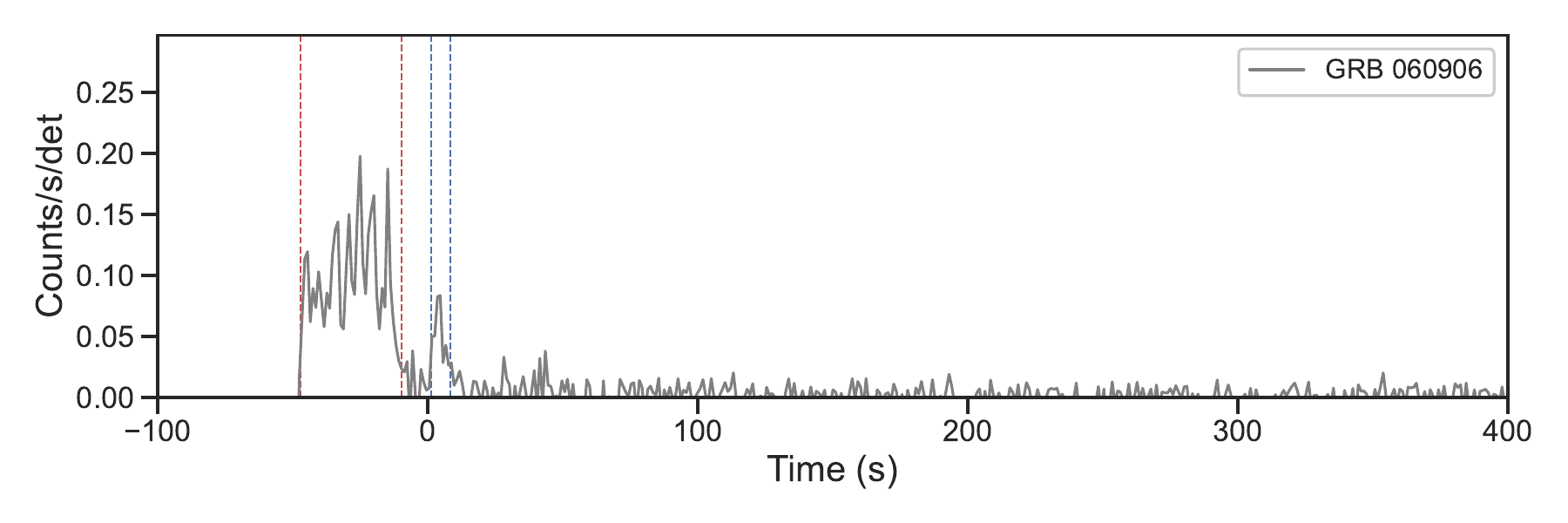}
\includegraphics[angle=0,scale=0.32]{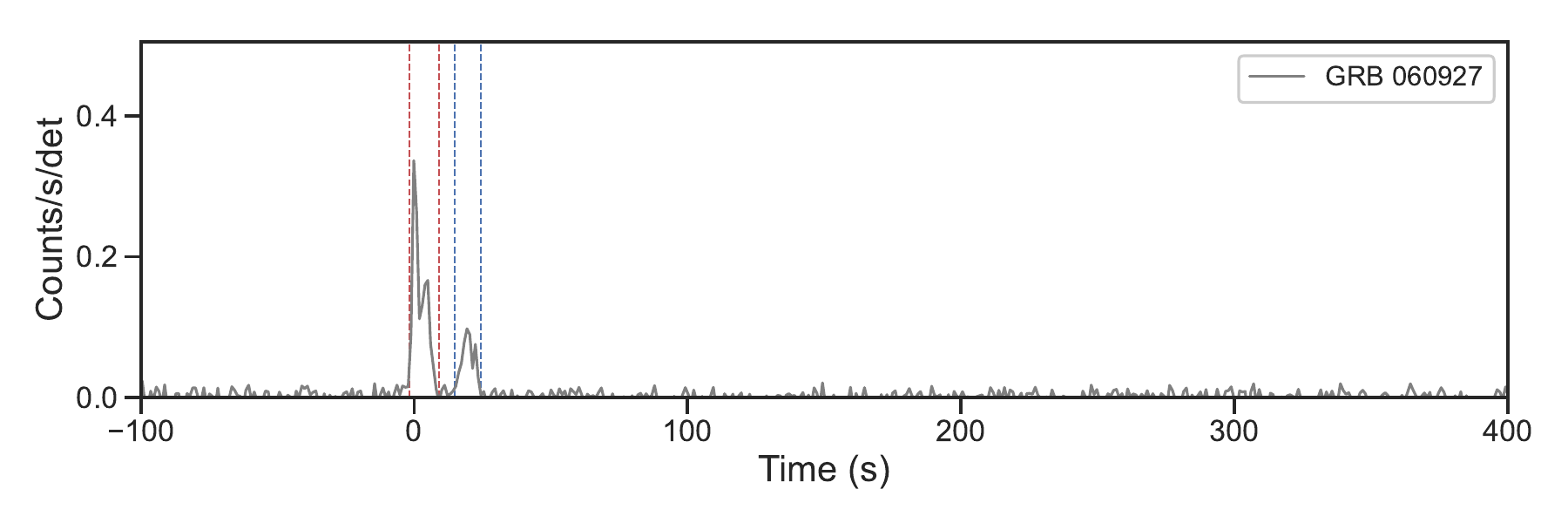}
\includegraphics[angle=0,scale=0.32]{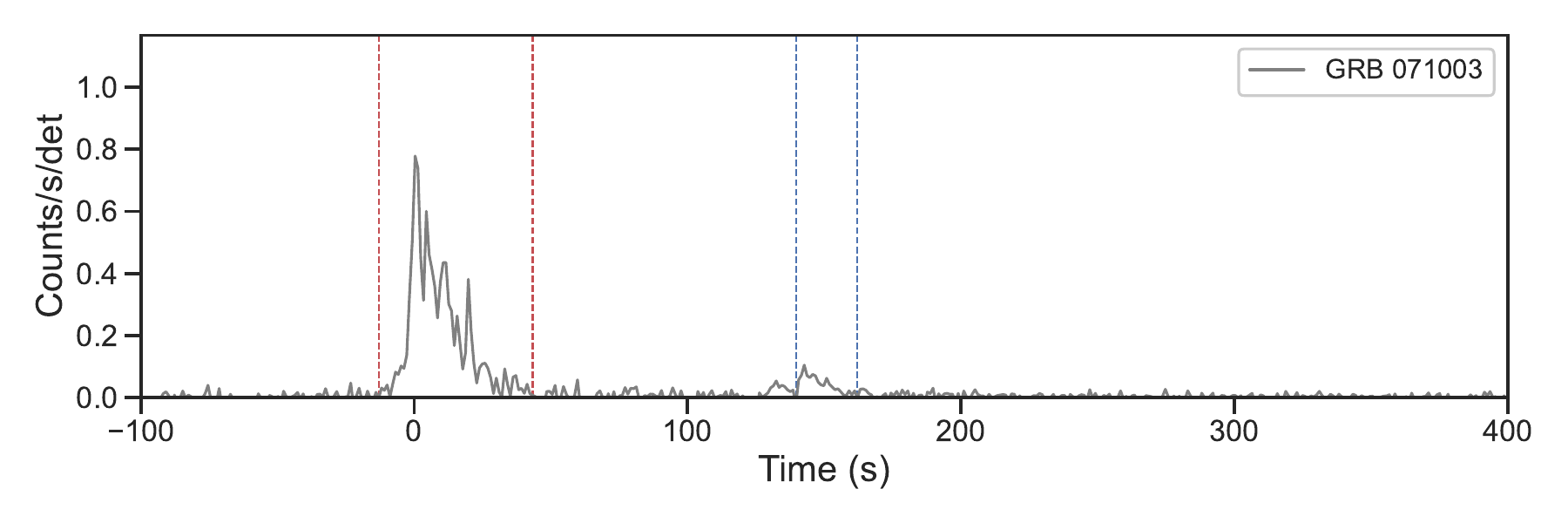}
\includegraphics[angle=0,scale=0.32]{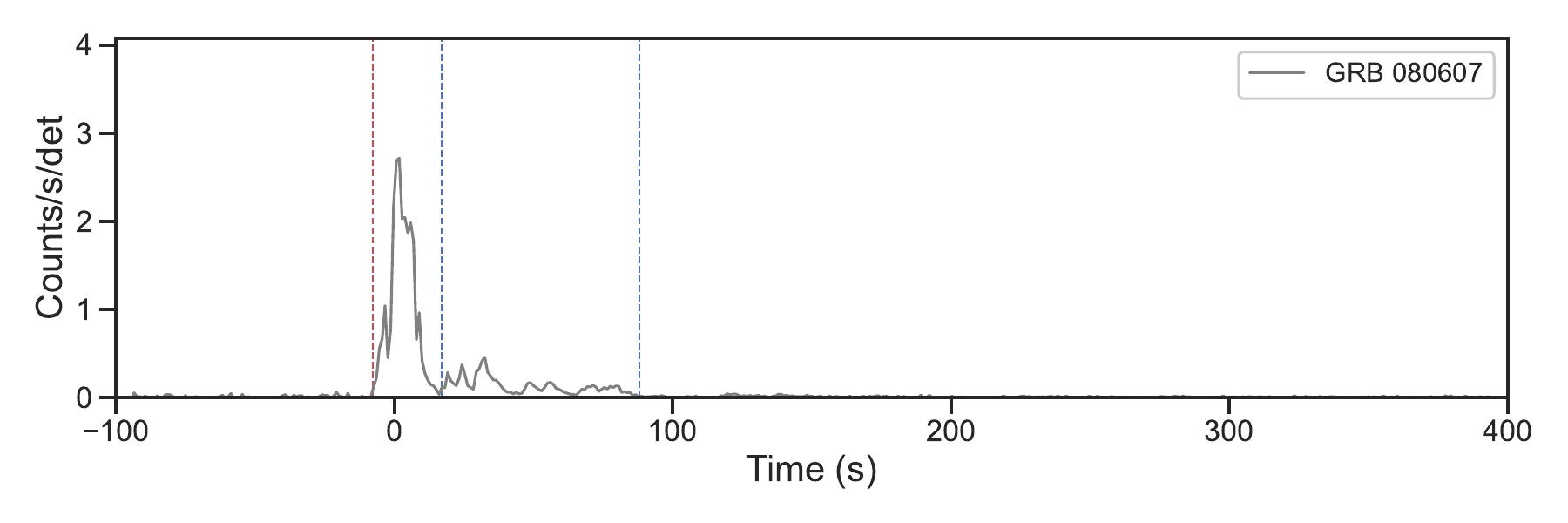}
\includegraphics[angle=0,scale=0.32]{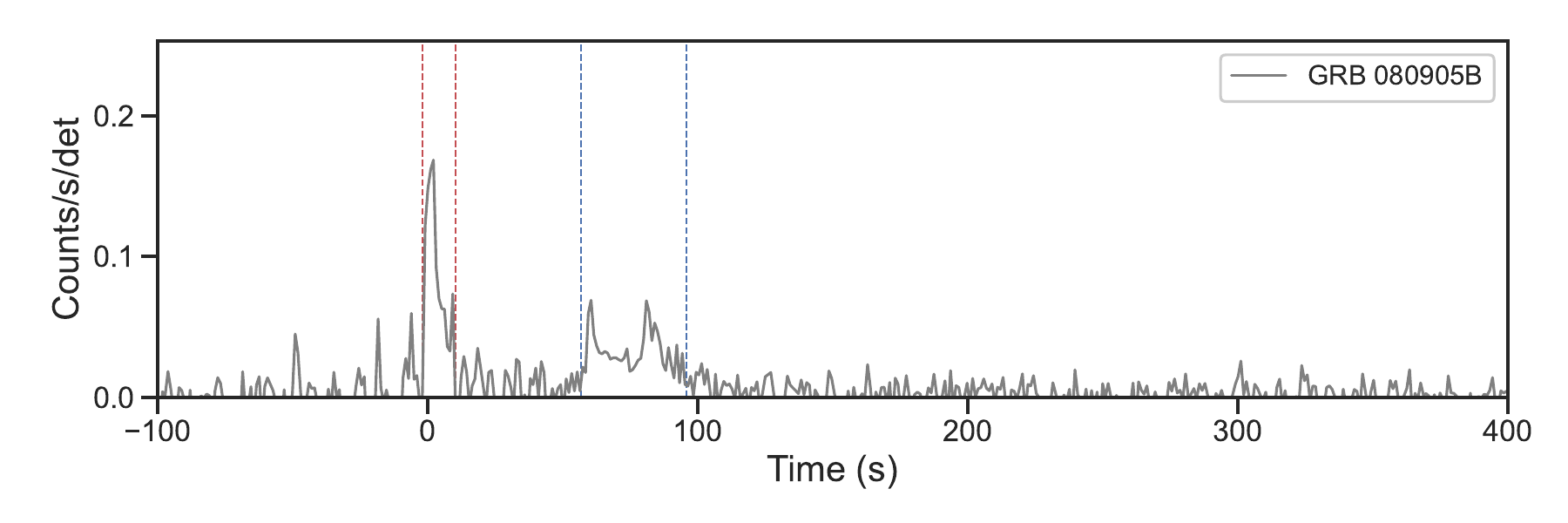}
\caption{The mask-weighted net \emph{Swift}/BAT light curves for the GRBs in our sample. Each panel shows a prompt emission consisting of a main burst sipke ($G_1$, indicated by red dashed lines) followed by a extended emission ($G_2$, blue dashed lines). The quiescent interval between the two components is clearly visible and corresponds to a return to background-level count rates.}\label{fig:lc_sample}
\end{figure*}

Figure~\ref{fig:lc_sample} presents the mask–weighted net \emph{Swift}/BAT light curves for the twenty-two ME+EE events of our target sample analyzed in this study. In each panel, two well–separated prompt emission episodes are bracketed by vertical dashed lines (red for $G_1$, blue for $G_2$). In all cases, the count rate during the gap drops to background levels, affirming that we are dealing with two distinct emission episodes. 
The ME is a bright, typically spiky or irregular pulse, whereas the EE is fainter and sometimes appears as a broader, smoother hump of emission. 
These episode windows define the intervals over which we measure \emph{episode–wise} $T_{90}$, hardness ratio, minimum variability timescale, and spectral lag. 
Visually, $G_2$ is fainter than $G_1$ in all cases, consistent with our selection $F_{\rm p}(G_{2})/F_{\rm p}(G_{1})<0.5$, and often appears longer and smoother, motivating the ME–EE comparisons in the following analysis. 

We next examine whether the standard prompt-emission diagnostics show systematic ME-EE contrasts. Because $G_1$ (ME) and $G_2$ (EE) are paired within each GRB, we employ paired tests throughout. For each burst we adopt (i) the hardness ratio evaluated from background-subtracted fluences in the two standard BAT energy bands, (ii) the minimum variability timescale measured with the Haar-wavelet method, and (iii) the spectral lag between 25-50~keV and 15-25~keV derived from the cross-correlation function analysis. For each quantity we use a single representative value per episode following our uniform-selection rule (finest available time bin for HR and MVT; finest available resolution for lag).

Episode-level $T_{90}$, $F_{\rm p}$, HR, MVT, and lag measurements for $G_1$ and $G_2$ are reported in Tables \ref{tab:t90}-\ref{tab:lag}. We also report rest-frame timing versions ($T_{90}/(1+z)$, ${\rm MVT}/(1+z)$, ${\rm lag}/(1+z)$) for completeness\footnote{Note that the rest-frame scaling would shorten $T_{90}$ and lag by a factor of $(1+z)^{-1}$ and leave HR unchanged since it is a ratio.}. We compare the main emission ($G_1$) and extended emission ($G_2$) episodes across our sample of twenty-two long GRBs in terms of four key prompt-emission diagnostics: duration, spectral hardness, minimum variability timescale, and spectral lag. Below we detail each of these comparisons, highlighting systematic trends. For ease of discussion, we sometimes refer to $G_1$ as the “ME” and $G_2$ as the “EE” component.

\subsection{Episode (pulse)-wise Parameter Distribution}\label{sec:dis}

\begin{figure*}[ht!]
\includegraphics[width=1.\hsize,clip]{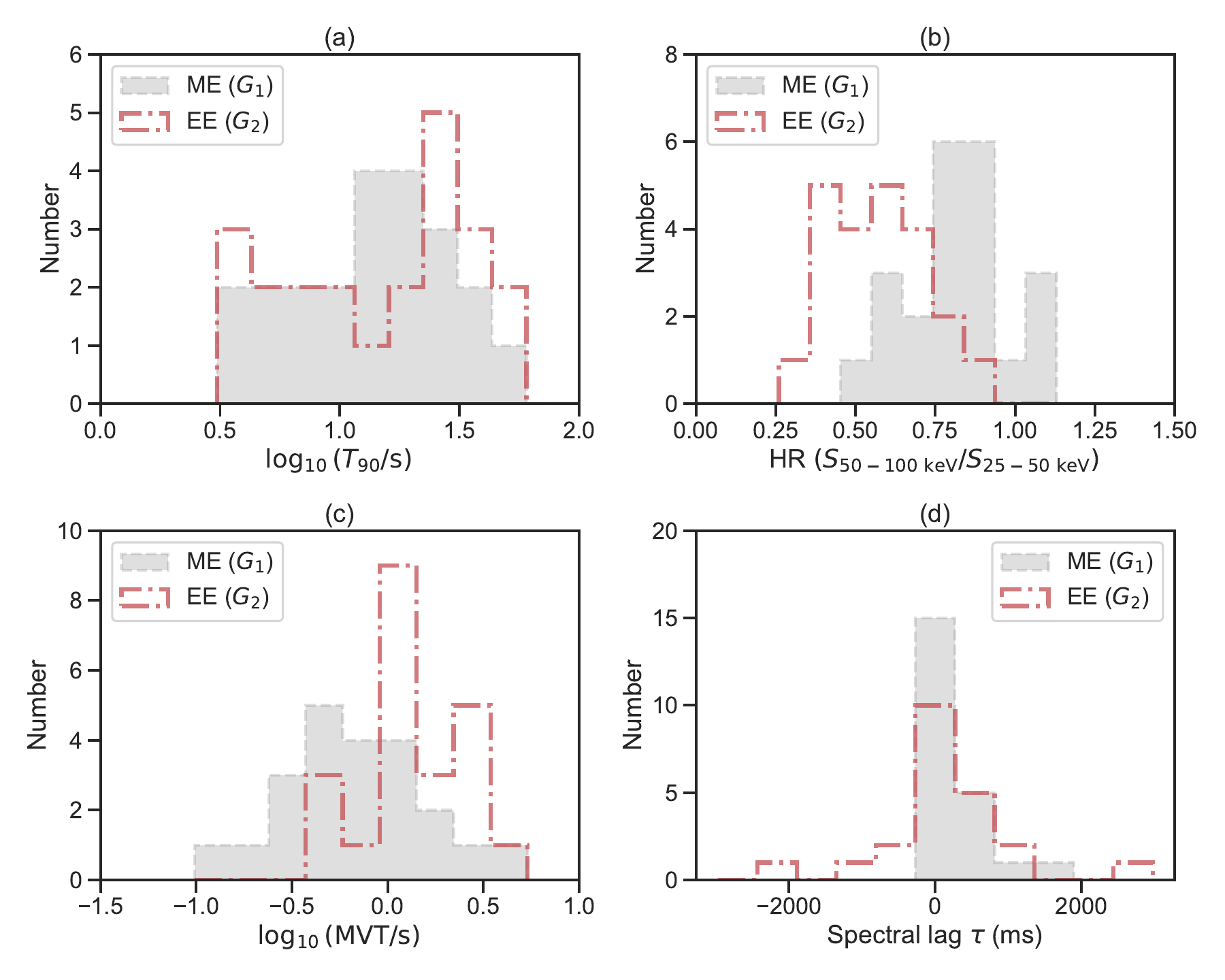}
\caption{Observer-frame distribution comparisons between the main-emission (ME; $G_1$) and extended-emission (EE; $G_2$) components for the 22 long GRBs in our ME+EE sample. Each panel shows histogram counts (not normalized) for paired episode-level measurements derived in a uniform \emph{Swift}/BAT bandpass. (a): $\log_{10}(T_{90}/{\rm s})$. (b): hardness ratio ${\rm HR}\equiv S_{50\text{--}100\,{\rm keV}}/S_{25\text{--}50\,{\rm keV}}$. (c): $\log_{10}({\rm MVT}/{\rm s})$, where MVT is the minimum variability timescale measured with the Haar-wavelet method. (d): spectral lag $\tau$ (ms) between the 25--50~keV and 15--25~keV bands (positive values indicate that the higher-energy band leads). The ME distributions are shown as filled histograms, and the EE distributions are shown as step histograms.}
\label{fig:para_dis}
\end{figure*}

Figure~\ref{fig:para_dis} summarizes, in a uniform and episode-resolved manner, how the EE systematically differs from the ME within the same long GRB. The $T_{90}$ distributions (Fig.~\ref{fig:para_dis}a) overlap substantially, indicating that episode-wise duration alone is not a robust discriminator between ME and EE across the sample. In contrast, Figure~\ref{fig:para_dis}b shows that EE is systematically shifted toward lower hardness ratios, consistent with a softer spectrum in the later component. Figure~\ref{fig:para_dis}c further shows that EE tends to occupy larger MVT values, implying smoother temporal structure and reduced short-timescale variability relative to the ME. Finally, the lag distributions (Fig.~\ref{fig:para_dis}d) suggest that EE exhibits broader and more heterogeneous lag behavior, including large magnitudes and sign changes, whereas ME lags cluster more narrowly. Consequently, the distributions support a picture in which EE represents a spectrally softer and temporally smoother prompt-emission phase.

The quiescent gaps $\Delta T_{\rm gap}$ between episodes are wide \citep{Ramirez-Ruiz2001,Nakar2002}, typically seconds to tens of seconds, a few can reach to hundreds of seconds for all twenty-two events, ranging from $\sim 0.3$~s (GRB 080607) up to $\sim 139.0$~s (GRB 140430A) with a median-sample value of 31.8~s. We find no obvious correlation between the length of the gap $\Delta T_{\rm gap}$ and the durations of either episode. For example, GRB 080607 has one of the shorter gaps ($\sim$0.3~s) but a relatively long main burst (14.1~s), whereas GRB 151027A has a longer gap ($\sim$ 79.6~s) following a very short main pulse ($\sim$ 5.6~s). 

\begin{figure*}[ht!]
\includegraphics[width=1.\hsize,clip]{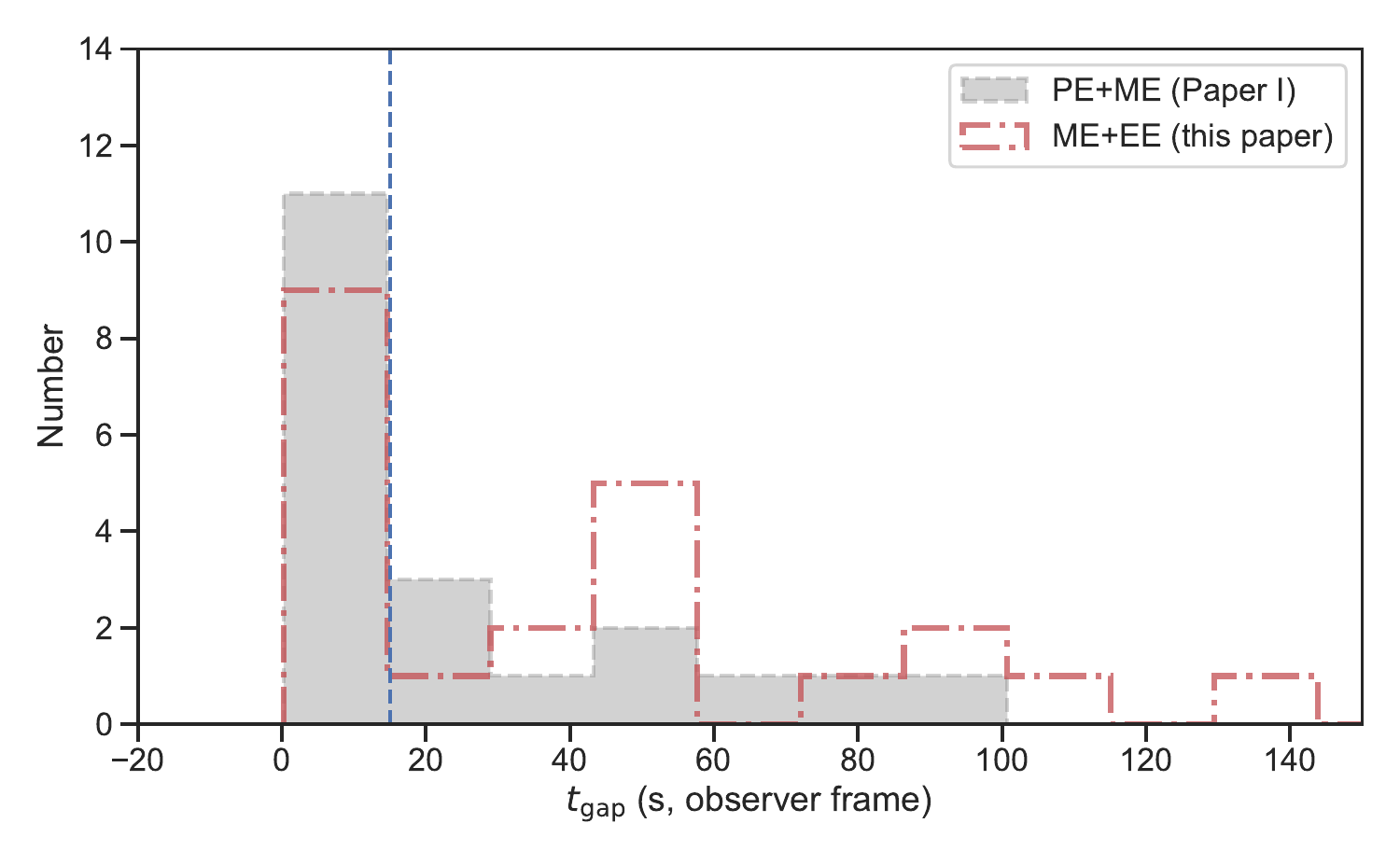}
\caption{Distribution of observer-frame quiescent-gap durations, $t_{\rm gap}\equiv t_{1}(G_{2})-t_{2}(G_{1})$, for the two pulse-resolved samples analyzed in our series. The filled histogram shows the 22 PE+ME GRBs from Paper~I \citep{2026arXiv260121693L}, while the step histogram shows the 22 ME+EE GRBs from Paper~II (this work). The vertical dashed line marks $t_{\rm gap}=15$~s, illustrating that most events in both samples cluster at short gaps ($t_{\rm gap}\lesssim 15$~s), with a smaller fraction extending to longer quiescent intervals. Histogram counts are shown without normalization.}
\label{fig:tgap}
\end{figure*}

Figure~\ref{fig:tgap} compares the observer-frame quiescent-gap distributions for both the PE+ME sample (Paper I) and the ME+EE sample (this work). Both distributions concentrate at short gaps, with the majority of bursts having $\Delta T_{\rm gap} \lesssim 15$~s (vertical dashed line in Figure~\ref{fig:tgap}), indicating that $G_2$ typically follows $G_1$ after only a brief background-consistent interval. A smaller fraction of events exhibits substantially longer gaps, extending to $\gtrsim 10^2$~s, forming a long-gap tail in both samples. This comparison suggests that short quiescent intervals dominate the configuration for two-episode prompt emission in \emph{Swift}/BAT long GRBs, while rare long-gap events represent extreme cases of engine intermittency rather than the norm.

\subsection{Episode (pulse)-wise Duration}\label{sec:t90}

We first preform a case-by-case test the episode-wise durations of the ME and the subsequent EE differ systematically in our ME+EE sample. For each burst, we adopt the episode-wise $T_{90}$ measured within the corresponding $[t_1,t_2]$ interval (Table~\ref{tab:pulse}), thereby excluding the quiescent gap from the duration estimate. 
The $T_{90}$ values in ME vary from $\sim$3.3-48.4~s with a median of 15.0~s, while the $T_{90}$ values in EE lie between $\sim$3.1-60.2~s with a median of 18.6~s. Both the ME and EE of every burst have $T_{90}$ well above 2~s, placing each episode individually in the long-GRB regime by the standard duration criterion. 
Notably, across the twenty-two events, there is no universal ordering between $T_{90}^{\rm ME}$ and $T_{90}^{\rm EE}$. In some bursts (e.g., GRB 100704A, GRB 140430A), EE lasts 2-3 times longer than ME, while in others (e.g., GRB 071003), EE is shorter. The episode-wise $T_{90}$ values for ME and EE  with asymmetric $1\sigma$ uncertainties
are reported in Table \ref{tab:t90}. 

To quantify the effect size, we compute $R_{T_{90}} \equiv \log_{10}(T_{90}^{\rm EE}/T_{90}^{\rm ME})$. The median $R_{T_{90}} =-0.016$, corresponding to a typical ratio $T_{90}^{\rm EE}/T_{90}^{\rm ME} \approx 0.96$. Bootstrap resampling yields a 95\% confidence interval $R_{T_{90}} \in [-0.17, 0.24]$, consistent with no systematic shift. Episode-wise duration therefore does not provide a robust discriminator between ME and EE, motivating the use of additional prompt emission diagnostics to characterize the physical contrast between the two phases. 

In general, the quiescent interval seems independent of the episode durations. This suggests that the gap is truly an “engine off” period rather than, say, the tail end of the ME pulse or the beginning of the EE pulse disguised at low signal level. If the gap were just a low-level continuation of $G_1$, we might expect longer $G_1$ pulses to correlate with shorter gaps or vice versa, but we do not see such trends. 

\subsection{Episode (pulse)-wise hardness ratio}\label{sec:HR}

Across all twenty-two GRBs, $R_{\rm HR}$ ranges from about 0.25 up to 1.25 with a median-sample value of $0.70$, indicating that EE is significantly \emph{softer} than the ME. The strongest softening is observed in GRB~080607A ($G_2=0.55\pm0.04$ vs. $G_1=1.13\pm0.05$), while GRB~050505, GRB~060906, GRB~151027A and GRB~161017A show the mildest contrasts. When combined with episode–only durations, all components fall within the long–GRB (Type~II) trace on the $T_{90}$–HR plane, establishing that the ME and EE of each event share the same Type I/Type II classification while exhibiting systematic episode–level spectral softening from $G_{1}$ to $G_{2}$. The episode–level hardness ratios are reported in Table~\ref{tab:HR}. 

\begin{figure*}
\includegraphics[width=0.5\textwidth]{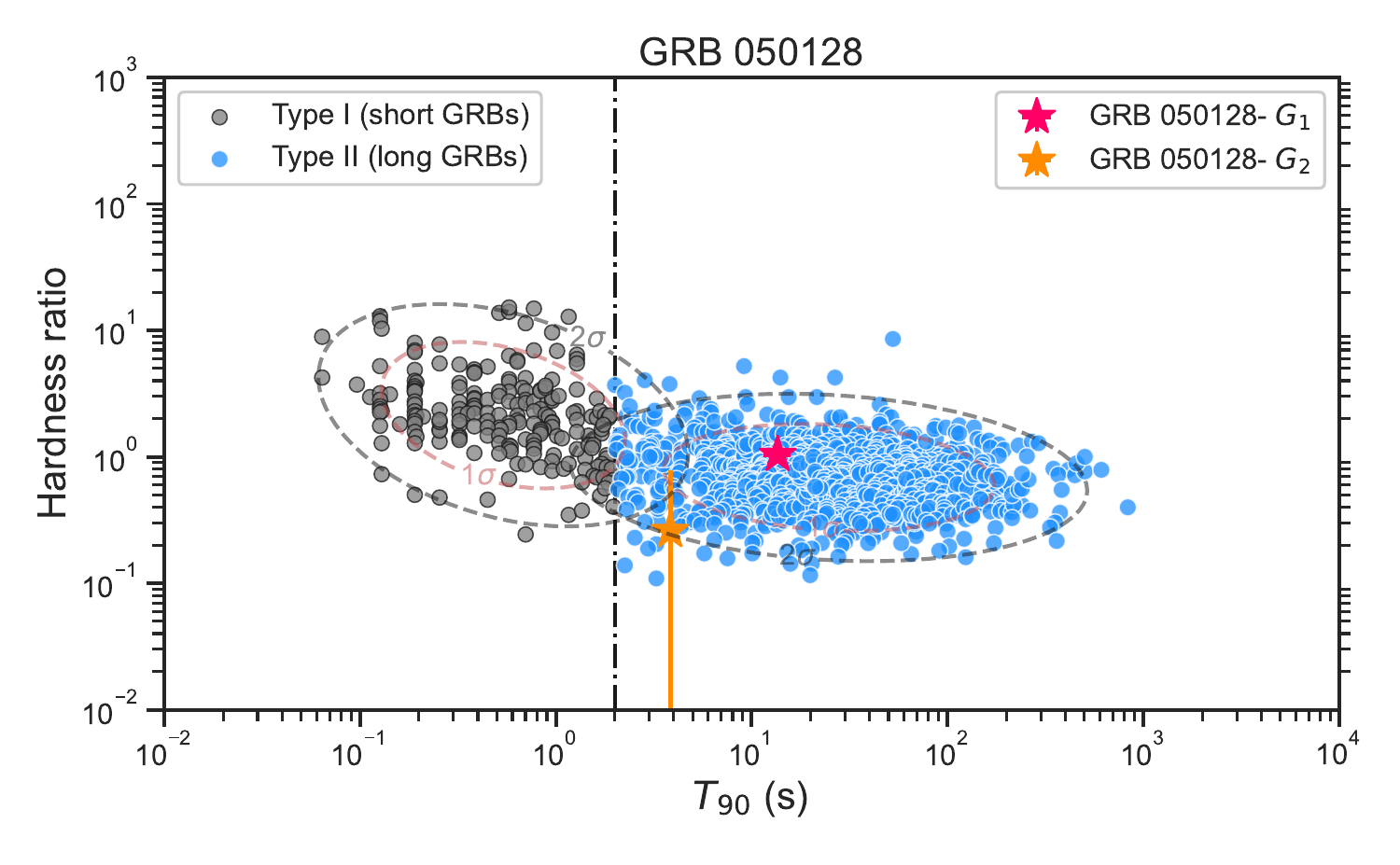}
\includegraphics[width=0.5\textwidth]{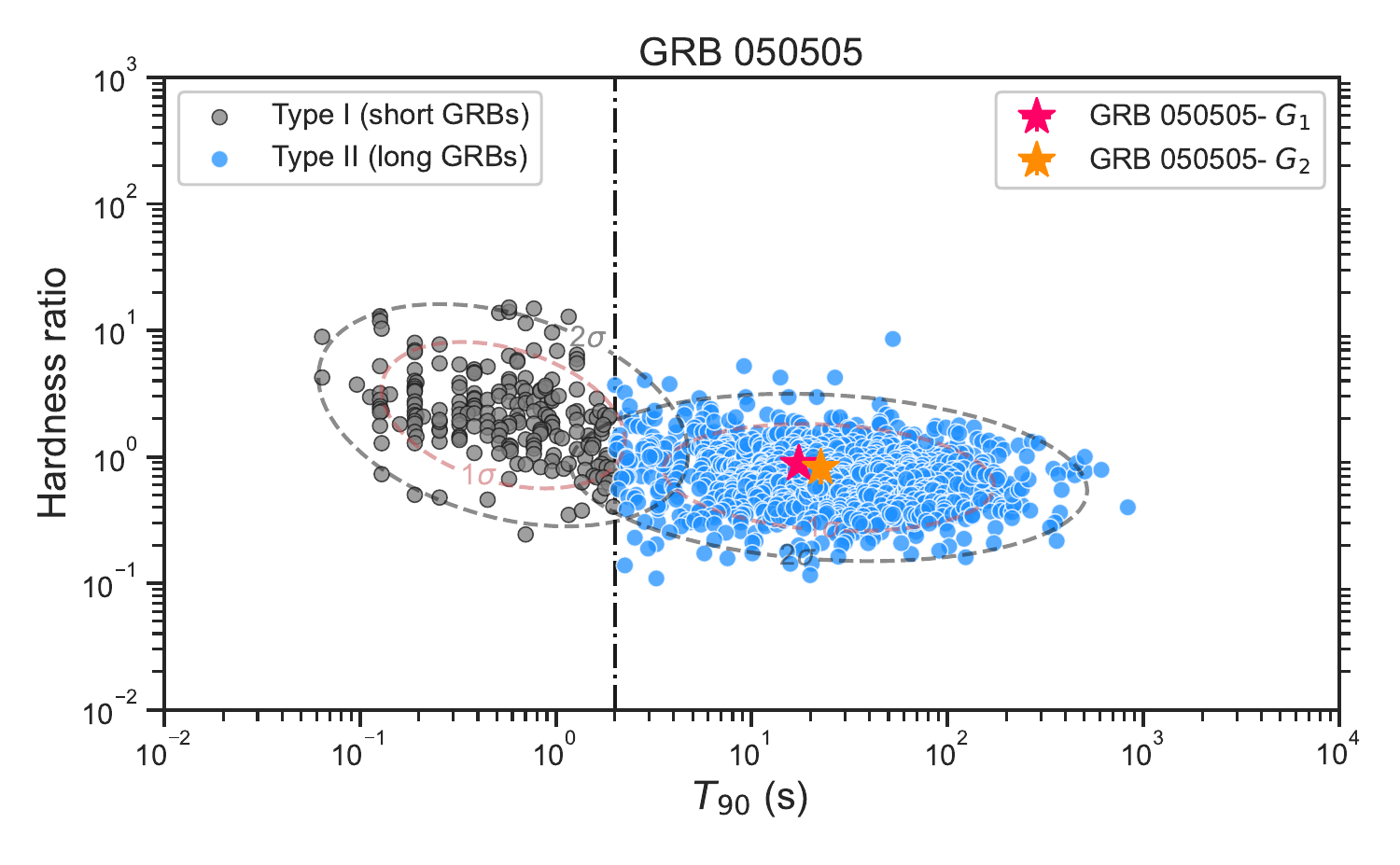}
\includegraphics[width=0.5\textwidth]{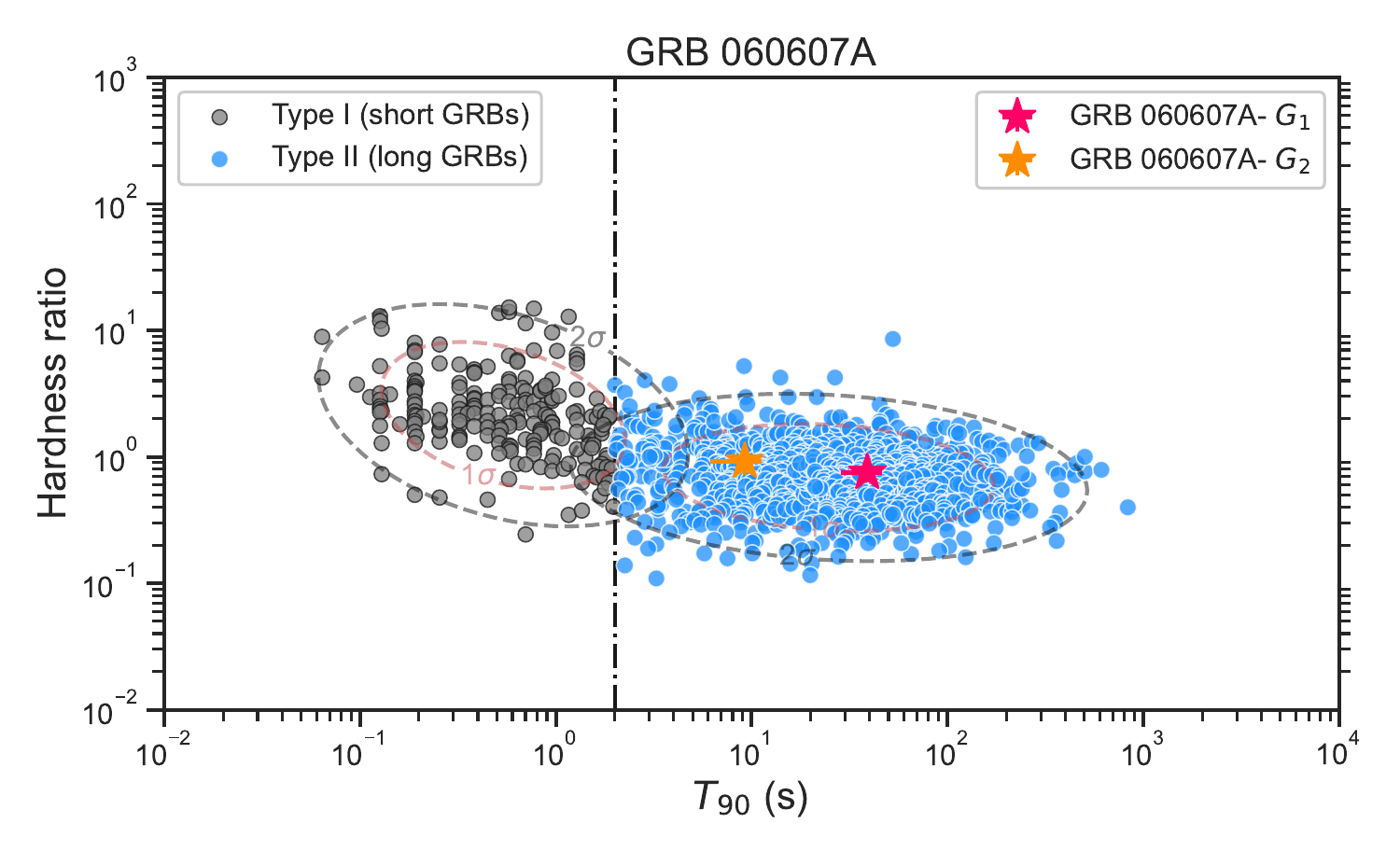}
\includegraphics[width=0.5\textwidth]{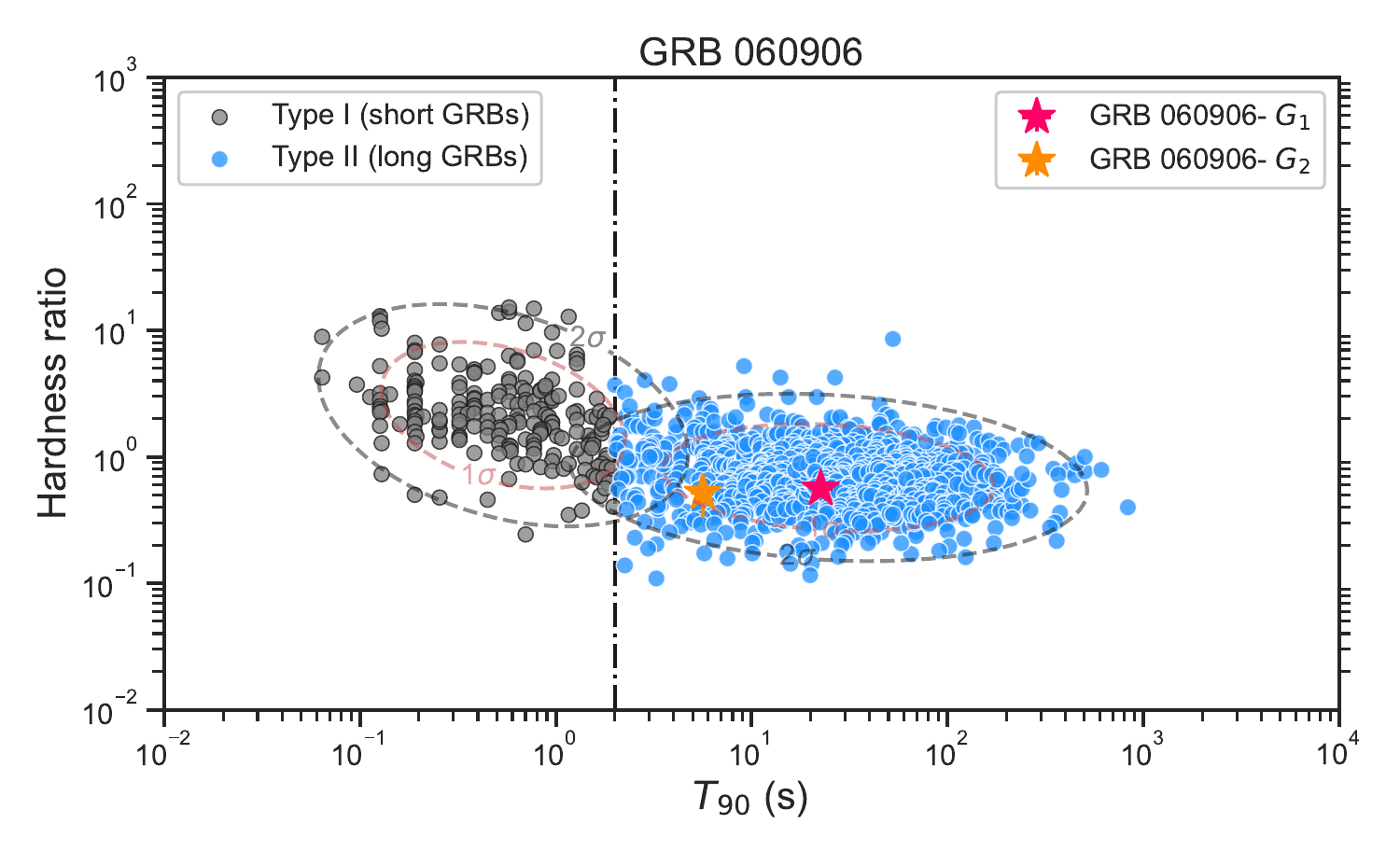}
\includegraphics[width=0.5\textwidth]{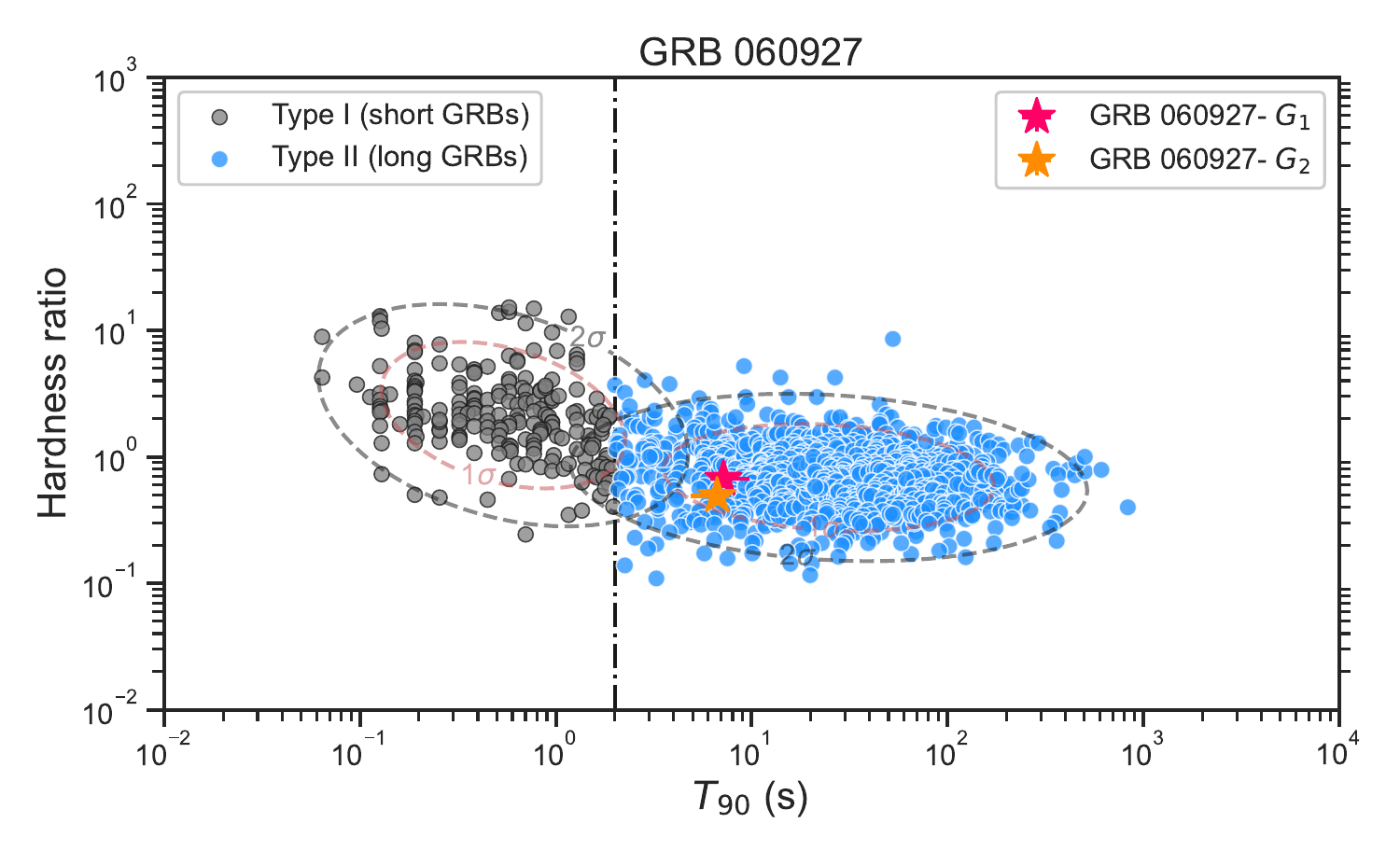}
\includegraphics[width=0.5\textwidth]{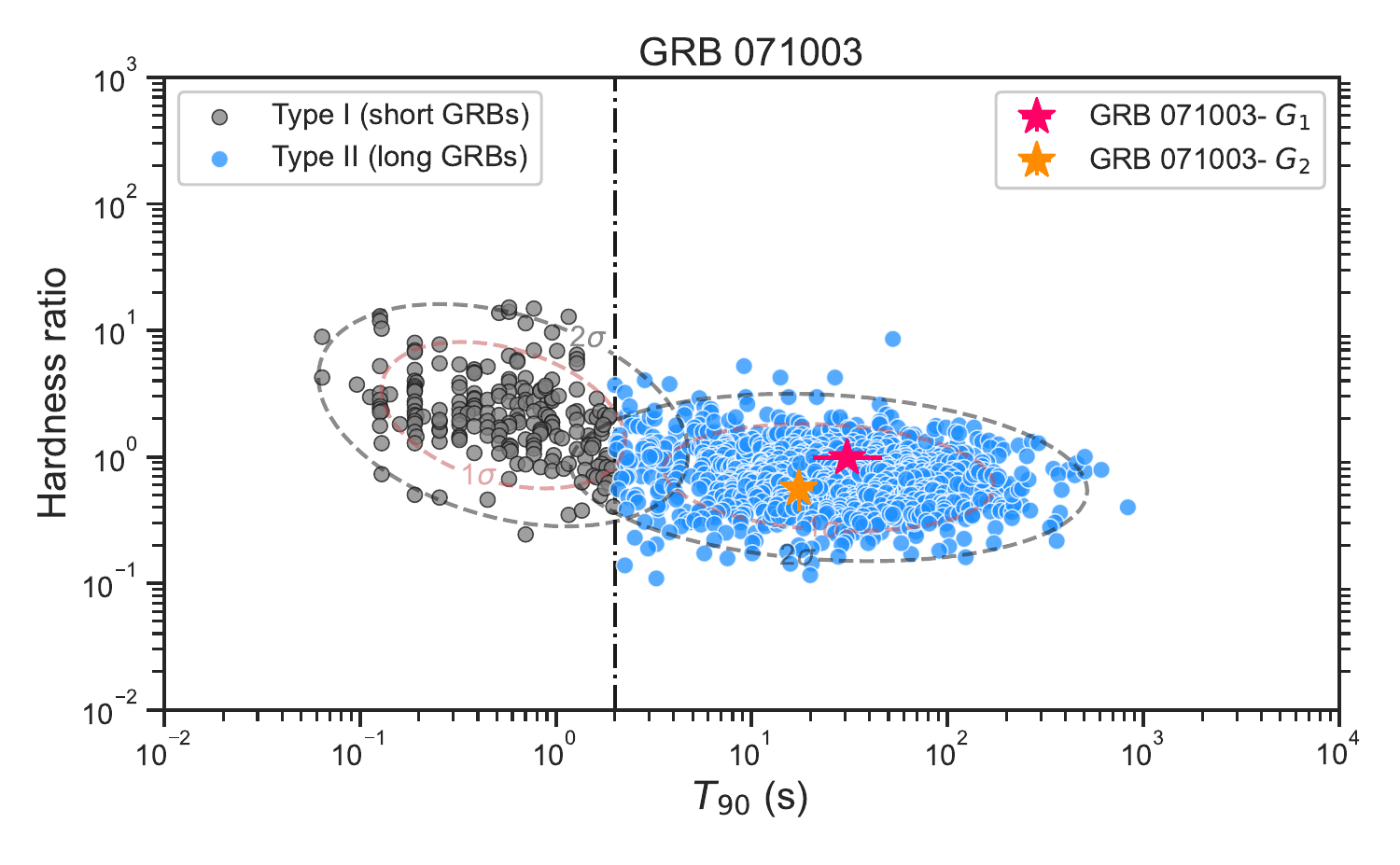}
\caption{Duration–hardness ratio diagrams for 22 representative GRBs in our sample. Each panel overlays the precursor ($G_1$, magenta star) and main emission ($G_2$, orange star) atop the broader \emph{Swift} GRB population from \citet{Horvath2010}, color-coded by Type I (short; black) and Type II (long; blue). The 1$\sigma$ ellipses correspond to the bivariate normal fits for each class. Both $G_1$ and $G_2$ fall securely within the long GRB region, while $G_2$ is systematically displaced toward lower hardness at roughly comparable episode-only $T_{90}$, visually reinforcing the sample-wide result that extended emission is spectrally softer than the main spike.}
\label{fig:HR_T90}
\end{figure*}

Figure~\ref{fig:HR_T90} present each burst’s two prompt emission episodes on the duration–hardness plane relative to the broader \emph{Swift} population. In each panel, the ME/EE-episode pair is shown as stars (magenta for $G_{1}$, orange for $G_{2}$), overlaid on the Type~I (short; black points) and Type~II (long; blue points) distributions from \citet{Horvath2010}. The dashed vertical line marks the nominal $T_{90}=2$~s boundary and the two dashed line contours ellipses trace the $1\sigma$ and $2\sigma$ of a bivariate normal fits for the two classes \citep{2026arXiv260121693L}. As we can see in the duration–hardness plane (Figure~\ref{fig:HR_T90}), both episodes ($G_1$ and $G_2$) occupy the long-GRB region, indicating that ME and EE remain consistent with a Type~II (collapsar) population globally, even though their spectra differ systematically at the episode level. 

This episode-level spectral softening of EE in long GRBs mirrors the well-known behavior of the EE component in short GRBs \citep[e.g.,][]{Norris2006,Perley2009,Kaneko2015,ZhangXL2020}, and here we demonstrate it for the first time in a systematic long-GRB sample. In short GRB+EE events, the EE component is usually much softer than the initial spike, sometimes to the point of resembling X-ray afterglow emission. Physically, lower hardness implies a lower characteristic photon energy, which in GRB prompt emission spectra typically correlates with the peak energy $E_{\rm p}$ \citep{Band1993, Preece2000}. The systematic softening of EE can be explained if the GRB engine in the EE phase is less powerful, leading to a lower radiation temperature or a transition to a less energetic emission mechanism that produces fewer high-energy photons.

\subsection{Episode (pulse)-wise MVT}\label{sec:MVT}

The MVT $\Delta t_{\min}$ serves as a proxy for the smallest physical time structure in the prompt emission, often linked to the central engine’s activity time or the size of the emitting region \citep{Golkhou2014, Golkhou2015, Ahumada2021NatAs}. A shorter MVT means the light curve has very fast, fine time structure (spikes, steep rises), whereas a longer MVT implies a smoother, more slowly varying profile.

Overall, across all twenty-two events, the median ratio $R_{\rm MVT} \sim 1.81$, with a range from $\sim0.64$ to $\sim13.57$ for the sample, indicating a major change in the internal timescales of the emission region between $G_1$ and $G_2$. In three out of twenty-two bursts, the MVT values for $G_2$ is an order of magnitude (or more) larger than $G_1$, indicating that EE is substantially smoother. Thus $R_{\rm MVT}$ typically spans a few to $\sim$an order of magnitude across the sample, establishing that the EE tend to be systematically \emph{smoother} and less erratic than the ME. The \emph{episode–wise} minimum variability timescales, $\Delta t_{\rm min}$, for each event is reported in Table~\ref{tab:MVT}.

\begin{figure*}
\includegraphics[width=0.5\textwidth]{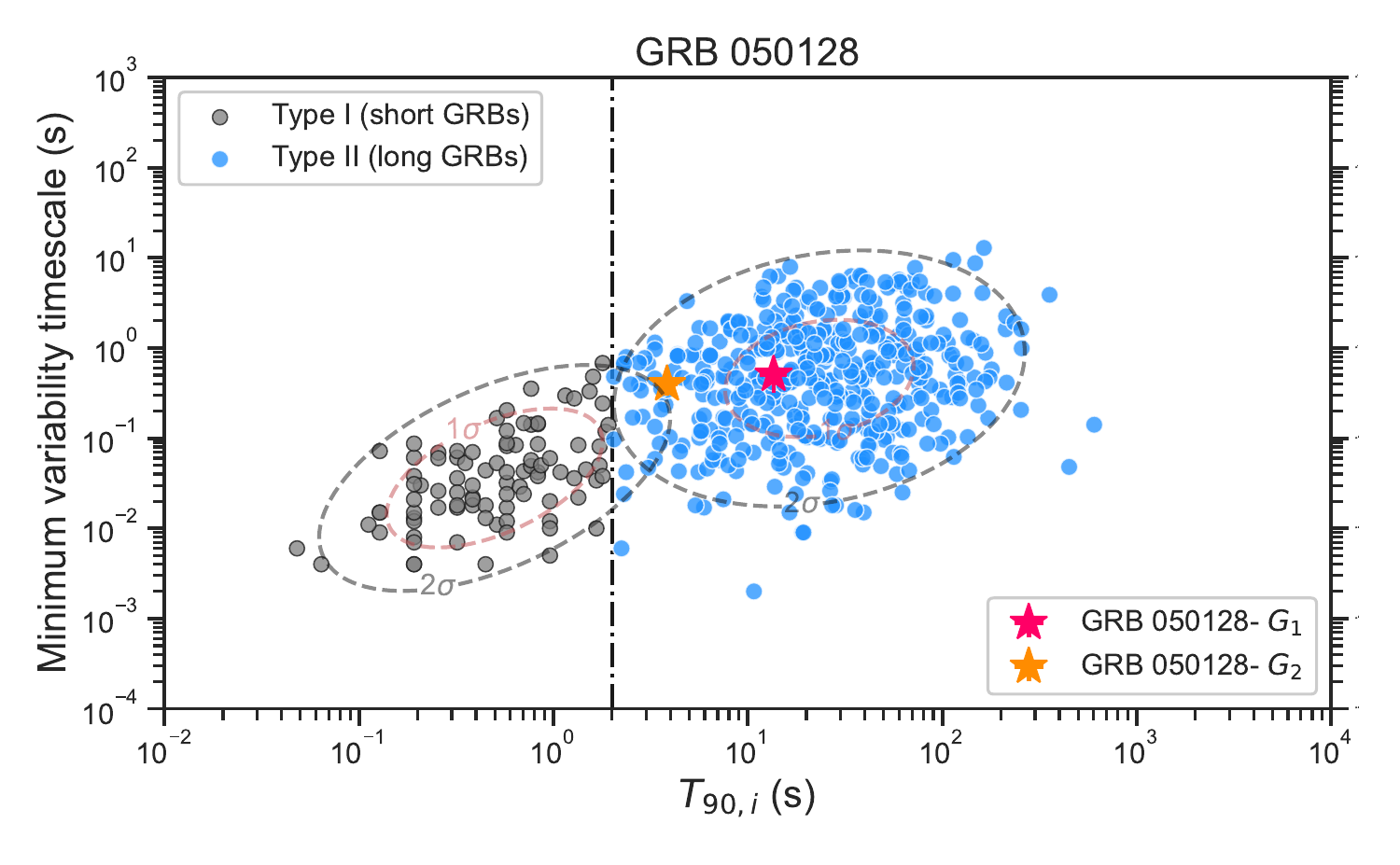}
\includegraphics[width=0.5\textwidth]{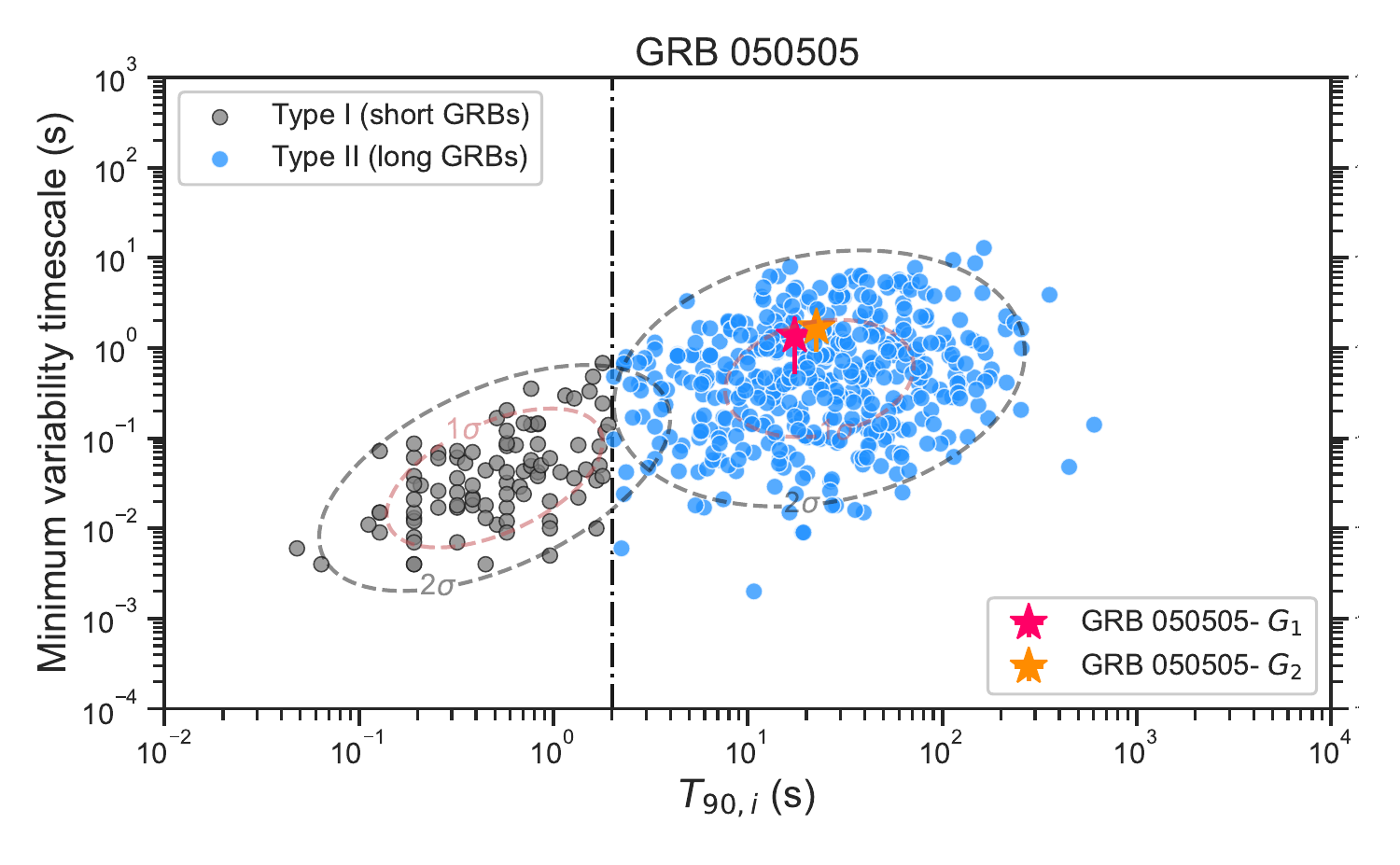}
\includegraphics[width=0.5\textwidth]{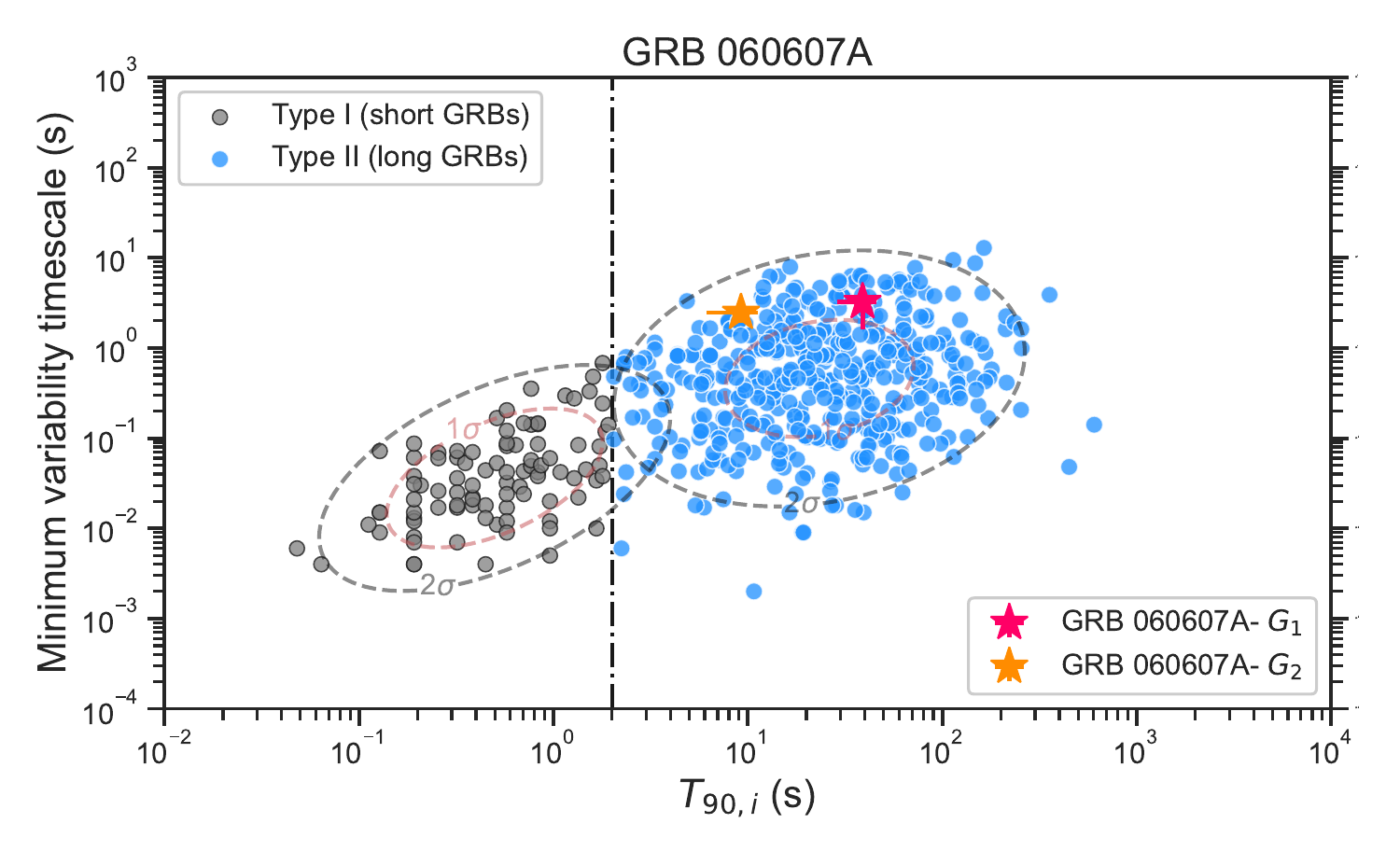}
\includegraphics[width=0.5\textwidth]{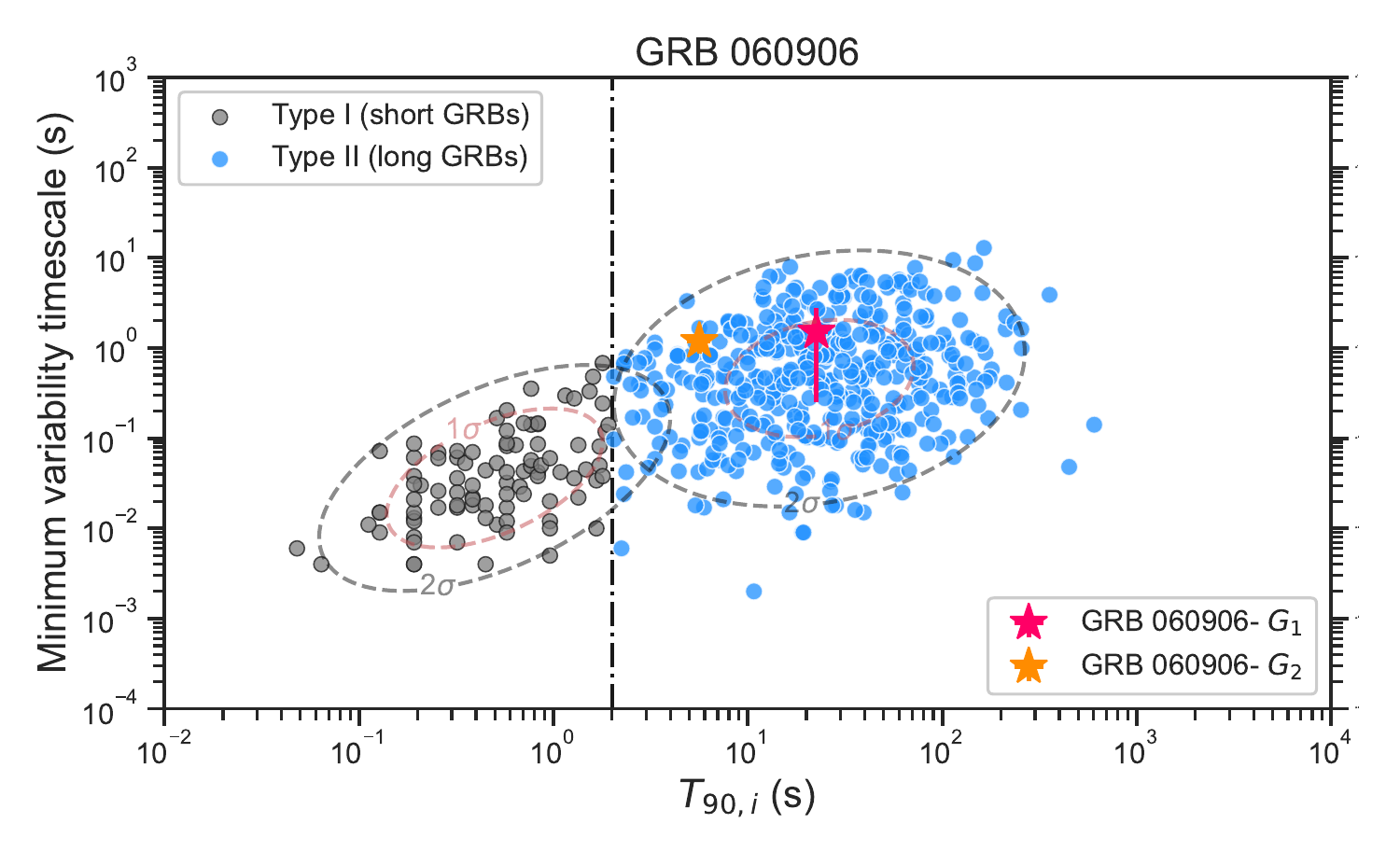}
\includegraphics[width=0.5\textwidth]{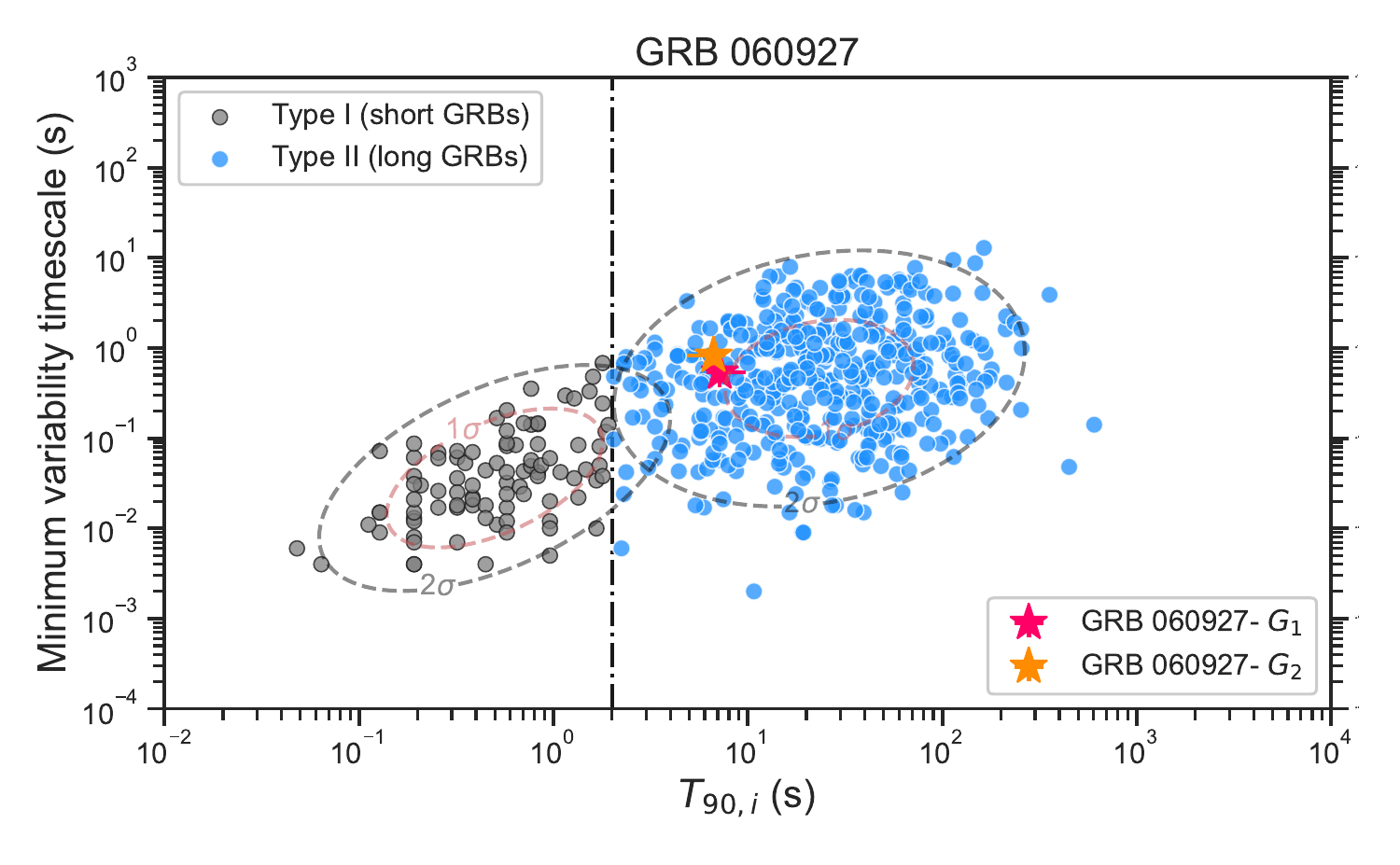}
\includegraphics[width=0.5\textwidth]{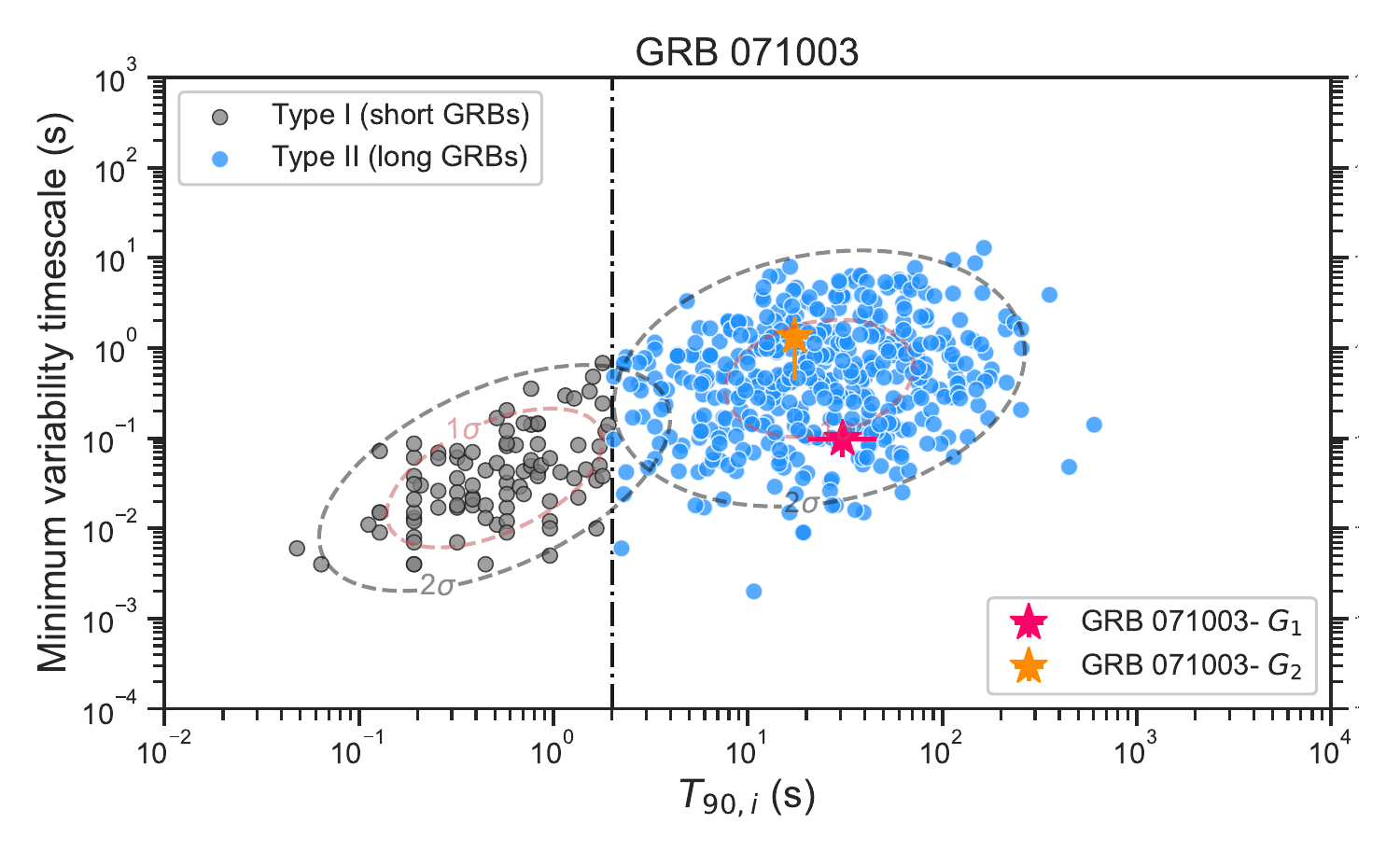}
\caption{Duration ($T_{90}$) versus minimum variability timescale (MVT) for main burst spike ($G_1$, red star) and extended emission ($G_2$, orange star) pulses in the sample. Gray and blue points show the distributions of Type I (short) and Type II (long) GRBs, respectively, from \citet{Golkhou2015}. The ellipses represent 1$\sigma$ clustering regions for each population. While both the ME and EE lies squarely within the Type II region, EE typically exhibits longer MVTs, indicating smoother temporal structure, yet remain consistent with the collapsar (Type II) population.}
\label{fig:MVT_T90}
\end{figure*}

Figure~\ref{fig:MVT_T90} places each burst’s two prompt episodes on the $T_{90}$–MVT plane, overlaid on the \emph{Swift} reference distributions from \citet{Golkhou2015} (gray: Type~I; blue: Type~II) with $1\sigma$ and $2\sigma$ clustering ellipses. In every panel, the main emission ($G_{1}$; red star) and the extended emission ($G_{2}$; orange star) both fall within the long‑GRB region, establishing a common Type~II classification. At the same time, $G_{1}$ is typically displaced to \emph{longer} MVT than $G_{2}$ at comparable episode‑wise $T_{90}$, indicating a smoother temporal structure for EE. This systematic offset, visible across all twenty-two events, provides the timing counterpart to the spectral trends reported elsewhere, and motivates the interpretation that the early episode is produced under conditions (e.g., larger dissipation scale and/or smaller bulk Lorentz factor) that suppress rapid variability relative to the main spike. Notably, the strong separation we observe in long-GRB ME+EE echoes the smoothness of EE reported in short GRB+EE studies \citep{Norris2006,Kaneko2015}.

We checked whether the lower flux of $G_2$ could artificially raise the MVT (because one might miss finer structure when the signal is weaker). While limited signal-to-noise can indeed cause one to detect only the strongest, broadest features, we note that even in cases where $G_2$ had decent brightness (e.g. the EE observed in GRB 100704A is quite long and not extremely faint), the difference is still obvious. We performed a crude test by degrading some $G_1$ light curves to similar S/N as $G_2$ and still saw that $G_1$ had more rapid structure. Thus, while sensitivity plays a role (a very weak EE might appear smoother simply because noise dominates small variations), the magnitude of the effect and its consistency point to an intrinsic smoothing in the extended emission. This aligns with the idea that the central engine (or emission region) in the EE phase is operating on longer characteristic timescales than during the main spike. In the context of known GRB populations, long GRBs generally have longer MVTs than short GRBs \citep{MacLachlan2013,Golkhou2014, Golkhou2015}. All our measurements lie in the long-GRB regime. However, what we are highlighting is a within-burst increase in MVT from $G_1$ to $G_2$. Within a single long GRB, the ME behaves comparably to a typical long GRB in terms of variability, whereas the EE becomes even smoother, often approaching the variability levels seen in some ultra-long GRBs \citep{Levan2014} or in the EE tails of short bursts \citep{Norris2006,Kaneko2015}.

\subsection{Episode (pulse)-wise Spectral lag}\label{sec:lag}

Spectral lag (the time delay between high-energy and low-energy photon arrival) is another diagnostic of pulse dynamics. Positive values indicate that the harder band peaks earlier. A negative lag means that, counter-intuitively, the lower-energy band peaked before the higher-energy band. Long GRBs usually exhibit positive lags on the order of 0.1–0.5~s, attributable to pulse broadening and hard-to-soft spectral evolution \citep{Norris2000,Ukwatta2010}, whereas short GRBs typically have negligible lags consistent with more impulsive emission \citep{Yi2006, ZhangZhibin2006, Bernardini2015}.

Episode–level spectral lags between the 25–50 and 15–25~keV bands are listed in Table~\ref{tab:lag}. We find that ME have small-to-moderate positive lags, with values of a few tens to hundred~ms (see Table~\ref{tab:lag}). As expected, the ME of these long GRBs have lags that are small-to-moderate positive values in most cases, typically of order tens to a few hundred milliseconds (hard photons leading soft ones). This is consistent with the well-known behavior of long GRB pulses, which often show softening as they progress, resulting in the softer band light curve peaking slightly later \citep{Kouveliotou1993}. By contrast, EE exhibits larger and more diverse lags. This pattern, which is ME lags modest and mostly positive versus EE lags broadly distributed, including sign changes, corroborates the episode‑level timing contrast inferred from the light curves and supports the picture that the EE phase is produced under conditions (e.g., larger dissipation radii and/or smaller bulk Lorentz factor) that yield slower spectral evolution and stronger curvature effects than in the ME phase. Both episodes in all bursts remain classified as “long GRB” in terms of their global characteristics (Table~\ref{tab:lag}). 

We do not find a statistically robust, sample-wide lag difference between ME and EE. For $N=22$ paired measurements, the median lag is $\tau_{\rm ME}=89.6$~ms (range $-249.6$ to $1536$~ms) and $\tau_{\rm EE}=161.6$~ms (range $-2355.2$ to $2976$~ms). While EE has a larger lag than ME in 15 of 22 cases, the distribution is broad and includes sign reversals and long tails. Accordingly, paired tests do not reject the null hypothesis (Wilcoxon $W=125$, $p=0.71$; sign test $p=0.21$; paired $t$-test $t=-0.11$, $p=0.92$). Therefore, unlike HR and MVT, the spectral lag does not provide a stable, sample-level discriminator between ME and EE in this dataset, suggesting that the energy-dependent pulse evolution during EE is more heterogeneous and possibly multi-component.

Overall, the EE phase lags are both larger in absolute magnitude and more scattered in sign compared to the ME phase lags. This implies a more varied and complex temporal-spectral behavior during the extended emission. Large positive lags typically correspond to strong hard-to-soft evolution in a long-lasting pulse. The fact that EE often has a much larger positive lag than ME fits the narrative that EE pulses are stretched out in time and gradually soften. The occurrence of negative lags might hint at distinct sub-components in the EE (e.g., an early soft X-ray rise or an afterglow onset that appears first in X-rays). In summary, main emission pulses behave like “classical” prompt pulses with modest positive lags, while extended emission pulses can behave quite differently, sometimes exhibiting extreme hard-to-soft delays or even soft-leading behavior. This diversity in lags further supports that the EE is not just a fainter copy of the main pulse, but potentially a different emission regime.

\section{Discussion}\label{sec:discussion}

Our episode-resolved analysis results show that the EE is systematically \emph{softer}, substantially \emph{smoother}, and \emph{more diverse in lag} than the ME, while both episodes remain within the long-GRB domain in the duration–hardness and duration–MVT planes. These results (softening, smoothing, and lag broadening in EE) provides strong evidence that the EE phase is physically distinct from the initial ME and observational boundary conditions for models invoking engine transitions (e.g., decaying power injection, slower magnetically moderated outflows, or larger emission radii) between $G_1$ and $G_2$. We now discuss the possible implications for GRB central engines and prompt emission physics, and how they relate to the broader picture of GRB phenomenology, including short GRBs with EE. In particular, we explore what the softer spectra, longer variability timescales, and unusual lags of EE suggest about the physical conditions during that phase, and how these compare to known scenarios like short-GRB extended emission.

\subsection{Origin of the Extended Emission: Engine Evolution vs. External Shock}

One immediate question is whether the EE in these long GRBs represents continued internal-shock activity (prompt emission) from the central engine, or instead reflects early external-shock signature, i.e., the onset of afterglow emission that overlaps temporally with the prompt emission phase. Early afterglow emission from the external forward shock origin typically rises and decays smoothly and can peak tens to hundreds of seconds after the burst for a dense environment or slow-moving ejecta, which usually peaks in X-ray or optical \citep{Meszaros1997, Sari1999, Granot2002}. 

We argue for an internal origin on three grounds. 
First, the EE episodes, while smoother than the ME pulses, still exhibits clear variability, distinct spectral peaks, and pronounced hard-to-soft spectral evolution accompanied by large positive spectral lags, all of which are signatures of internal dissipation driven by particle cooling \citep{Rees1994,Daigne1998}, rather than the smooth, single-peaked, and achromatic decay expected of an external forward shock. 
Second, the EE spectrum remains significantly harder than predicted for an external forward shock, with emission humps peaking near $50$-$100$ keV \citep{Granot2002, Zhang2006}, whereas external-forward-shock emission is expected to follow a broken power-law spectrum peaking in the X-ray or optical band. 
Third, analogous analyses of short GRBs with EE \citep{Norris2006,Kaneko2015} demonstrated that neither the spectral indices nor the temporal decay slopes of EE are compatible with high-latitude emission or a forward-shock afterglow, and the same reasoning applies to the long GRBs studied here. 
Further disfavoring an afterglow interpretation, the EE light curves frequently display complex substructure, including re-brightening and multiple sub-peaks as seen in GRB~210619B (see Figure \ref{fig:lc_sample}), which cannot be reconciled with the simple power-law decay (with photon index $\Gamma \sim 2$ and temporal decay $\sim t^{-2}$ or steeper) expected from the curvature effect or self-similar forward-shock deceleration. 
We therefore conclude that the EE represents genuine prolonged prompt emission, requiring the central engine to remain active well beyond the main pulse under modified physical conditions.

\subsection{Modelings for Long-lived GRB Central Engine Activity}

What powers the late, softer, smoother EE in long GRBs? Two leading scenarios can account for long-lived GRB engine activity on timescales of tens to hundreds of seconds. 

Fallback accretion onto a black hole. In collapsar models, after the initial jets are launched by the accretion of the stellar core, there can be continued infall of stellar material (fallback) onto the newly formed black hole. This fallback accretion can persist for hundreds of seconds with a decaying mass accretion rate $\dot{M} \propto t^{-5/3}$ (in a simple spherical fallback model, \citealt{Chevalier1989,Woosley1993,Kumar2008a,Kumar2008b,ZhangWeiqun2008}) with the jet luminosity following $L(t) \sim \eta \dot M c^{2}$ \citep{Chevalier1989, MacFadyen2001, Kumar2008a}. The result is a long-duration power-law decay of jet power. The main burst would correspond to the peak accretion and jet power when the core collapses, and the extended emission would correspond to the dwindling accretion rate at late times. As the engine power $L(t)$ drops, the jet’s bulk Lorentz factor may decrease (less energy per baryon), the dissipation radius may increase if the outflow becomes less tightly collimated, and the variability of the inflow smooths out as small fluctuations in $\dot{M}$ produce less visible variability when overall luminosity is low).  Hydrodynamical simulations of collapsar accretion \citep{MacFadyen1999,Lazzati2009,Lindner2010,Lopez-Camara2013} indeed show a prompt bright phase followed by a rapid decline and long-lived, smoother power injection phases \citep{Lindner2010}. All of these effects qualitatively align with our observations of softer spectra, longer variability timescales, and larger lags in the extended emission phase. Fallback accretion is thus an attractive explanation for the engine’s behavior in EE.

Millisecond magnetar spin-down. Alternatively, the central engine could be a rapidly spinning, highly magnetized neutron star (a millisecond magnetar) formed in the core collapse \citep{Usov1992, Thompson1994, Dai1998, Zhang2001, Metzger2008, Metzger2011, Bucciantini2012, Gompertz2013}. A newly born magnetar with spin period $\sim$1 ms and field $B \sim 10^{15}$ G can inject energy into a wind on timescales of order $10^2$ s (the spin-down timescale) \citep{Metzger2011, Yu2010, Lv2014, Li2018b}. The spin-down luminosity follows roughly $L(t) \approx L_0/(1 + t/t_{\rm sd})^2$, where $t_{\rm sd}$ is the spin-down timescale (which can be tens to hundreds of seconds for reasonable magnetar parameters). Early on ($t \ll t_{\rm sd}$), the magnetar outputs a roughly constant high power (which could drive the main GRB spike if the magnetar forms essentially at the time of core collapse); later ($t \gg t_{\rm sd}$), the power falls off as $t^{-2}$. This naturally leads to an extended emission tail. If the magnetar wind becomes less tightly collimated or undergoes internal dissipation at larger radii over time, one would again get a softer spectrum and smoother light curve as time goes on. Magnetar models have often been invoked to explain extended emission in short GRBs \citep{Metzger2011,Rowlinson2013,Lv2015}, but they could also apply to long GRBs if a magnetar (temporarily) survives the collapse. Some long GRBs might indeed have plateau phases or extended energy injection consistent with magnetar central engines \citep{Zhang2006,Lyons2010,Lv2014} (especially those without bright supernovae, hinting the core formed a magnetar rather than immediately collapsing to a BH).

These two scenarios are not mutually exclusive. It is conceivable that some GRBs began with a collapsar accretion jet and subsequently transition to magnetar spin-down emission (if the black hole formation is delayed or if a magnetar forms first and later collapses). The unifying theme is long-lived engine activity. Our observation that EE is still within the long-GRB phenomenology (just softer and smoother) fits the idea that the same engine (the collapsing star’s core) is responsible, just in a later phase. No observational evidence suggests the involvement of an entirely different progenitor source; rather, the EE reflects a continuation of the same central engine operating under modified physical conditions. In both cases, the predicted decline of $L$, increase of effective $R$, and (possibly) decrease of $\Gamma$ from ME to EE match our HR, MVT, and lag trends.

\subsection{Minimum variability timescale: Implications for Emission Region and Jet Properties for $G_1$ versus $G_2$}

MVT is related to the central engine’s ejection variability and/or the radial thickness of colliding shells \citep{Golkhou2014, Golkhou2015, Ahumada2021NatAs}. A longer MVT for EE could indicate that either (a) the engine is now modulating on a longer timescale (for instance, if the engine is now weaker and its activity is damped or more periodic), or (b) the emitting region for EE is larger (e.g., the shocks or magnetic dissipation happening further out in the jet, smoothing out rapid variability). 

One can attempt to relate $\Delta t_{\min}$ to the GRB outflow’s Lorentz factor $\Gamma$ and emission radius $R$ \citep{Rees1994,Sari1995,Lithwick2001}. A commonly used causal limit is 
\begin{equation}
R_{\rm em} \lesssim 2\Gamma^{2} c \frac{\Delta t_{\rm min}}{1+z},
\label{eq:Rvar}
\end{equation}
so that a larger $\Delta t_{\rm min}$ for EE implies either a larger $R$ or a smaller $\Gamma$, or both, relative to ME \citep[see, e.g.,][]{Rees1994, Kumar2015}. Quantitatively, the observed ratio $R_{\rm MVT}$ maps to
\begin{equation}
\frac{R^{\rm EE}_{\rm em}}{R^{\rm ME}_{\rm em}} \sim R_{\rm MVT} \left(\frac{\Gamma^{\rm EE}}{\Gamma^{\rm ME}}\right)^{-2}.
\label{eq:Rratio}
\end{equation}
For the typical $R_{\rm MVT} \gtrsim {\rm few}$ seen in our sample, the EE smoothness can be reproduced either by an order-unity increase in the dissipation radius at comparable $\Gamma$, or by a moderate decrease in $\Gamma$ at comparable $R$. Both possibilities are physically natural if the engine power decays and the outflow becomes progressively less variable and/or less relativistic as time proceeds.

We also considered whether the difference in MVT could be partly an artifact of lower signal-to-noise in the EE (since a weaker signal makes it harder to detect very fast variations). To test this, we examined the correlation between the flux ratio $F_{\rm p}(G_2)/F_{\rm p}(G_1)$ and the MVT ratio. There is a hint that the bursts with the weakest EE (relative to ME) do show the largest MVT ratios, suggesting some contribution from detectability limitations. However, even the relatively bright EE cases (like GRB 100704A) show a large MVT increase, and the overall separation between ME and EE MVT is so large in most events that it likely reflects an intrinsic difference, not just a sensitivity issue. For example, GRB 071003’s EE, although weaker than its ME, still had a healthy count rate yet was much smoother, it is implausible that we would miss $\sim 0.1$~s spikes in a 20~s long EE if they were present at the same relative brightness as in the ME.

\subsection{Spectral lag diversity: Implications for Spectral evolution (hard-to-soft) for $G_1$ versus $G_2$}

Spectral lag reflects energy-dependent pulse timing and width, which are shaped by intrinsic spectral evolution and curvature (high-latitude) effects. In our ME+EE events, EE shows a wide range of lags, from large positive to modestly negative. Large positive lags are expected when $E_{\rm p}(t)$ sweeps through the BAT band during a slowly decaying, softening pulse \citep{Norris2000, Kocevski2003, Ukwatta2010}. Occasional negative lags can result from (i) spectral components with different temporal ordering (e.g., a quasi-thermal soft component rising earlier than a hard nonthermal tail), (ii) overlapping multi-zone emission where the low-energy light curve is dominated by an earlier feature and the high-energy curve by a later one, or (iii) intensity-tracking behavior in which the high-energy pulse is broader and peaks later in the observer band \citep{Ryde2005, Hakkila2011}. The fact that EE exhibits larger scatter in lag than ME is therefore expected if late-time dissipation occurs at larger radii (enhanced curvature times) and/or involves mixed radiation zones.

EE exhibits softening and large positive lags indicate strong hard-to-soft evolution during each EE pulse. In prompt GRB pulses, a common pattern is that the spectral peak energy $E_{\rm p}$ decreases over time as the flux decays, causing the light curve to peak earlier in harder bands and later in softer bands. The longer the pulse lasts, the more $E_{\rm p}$ can shift during the emission, hence the larger the lag. Our results are consistent with EE pulses being more stretched out and having a more pronounced drop in $E_{\rm p}$, hence their lags are larger. The extreme case is when $E_{\rm p}$ might start below the hard band, potentially causing a scenario where the soft band peaks first (if, say, the spectrum initially peaks in X-rays and later moves through BAT’s hard band). This could cause the negative lags we saw. Another contributor to lag is the curvature effect \citep{Fenimore1996,Ryde2002, Li2021a} (photons arriving later from higher latitudes of the expanding radiating shell are softer due to relativistic beaming geometry). A longer-lasting, lower-$\Gamma$ emission episode will have a longer “tail” of high-latitude emission, effectively broadening pulses more in softer bands, again yielding larger lags. If the Lorentz factor is lower during EE, the curvature timescale $t_{\rm curv} \sim R/(2c\Gamma^2)$ (where $R$ is the emission radius) could be longer, contributing to late-arriving soft photons \citep{Kumar2000}. Thus, large positive lags in EE might signify lower Lorentz factor and/or larger emission radii compared to the main pulse. The several cases of negative lag in our sample are worth discussing. Negative lag can occur if the light curve in the softer band has a distinct earlier peak than the hard band. One possibility is that EE consists of two components: an early, relatively soft component (e.g., a soft X-ray flare or a thermal emission bump) followed by a harder flare. The soft component could dominate the 25–50~keV flux early, while the 50–100 keV flux is dominated by the later hard component, resulting in an apparent negative lag. Another possibility is that we are seeing the onset of the afterglow or a very soft extended tail that rises in lower energies first (like an X-ray flare from the afterglow emission originate from from an external shock). However, given that we are looking in 25–50 keV, it is unlikely to be the conventional afterglow (which peaks much later and at lower energies). It may simply reflect unusual internal spectral evolution in those particular bursts. Negative lags have been reported in a minority of BATSE and \emph{Swift} bursts and often come with multi-peak structures or complex spectral shapes \citep{Hakkila2011,Uhm2016}.

\textit{Why is EE softer? Coupled $E_{\rm p}$-luminosity-radius scalings.}
The hardness decrease from ME to EE is consistent with canonical prompt-emission scalings. In internal-shock synchrotron models, the characteristic spectral peak scales approximately as
\begin{equation}
E_{\rm p}\ \propto\ \Gamma\,B'\,\gamma_e^{2}\ \propto\ L^{1/2}\,R^{-1},
\label{eq:Ep_scaling}
\end{equation}
where $B'$ is the comoving magnetic field and $L$ is the instantaneous jet power, assuming weak evolution in the typical electron Lorentz factor $\gamma_e$ \citep[e.g.,][]{Zhang2002, Daigne1998, Kumar2015}. A drop in engine power ($L_{\rm EE}<L_{\rm ME}$) and/or an outward shift of the dissipation zone ($R_{\rm EE}>R_{\rm ME}$), both expected at late times, naturally pushes $E_{\rm p}$ to lower energies, thereby reducing the observed hardness ratio in the \emph{Swift}/BAT band. Analogous softening also appears in dissipative photosphere scenarios, where $E_{\rm p}$ tracks the photospheric temperature and decreases as the luminosity and baryon loading evolve \citep[e.g.,][]{Ryde2009, Beloborodov2013, Ito2019}.

Consequently, the observations can be summarized as a sequence in which (1) the engine launches a high-power, highly variable outflow (ME) that dissipates at smaller radii and/or larger $\Gamma$, producing hard spectra, short MVT and modest lags; then (2) a lower-power, longer-lived outflow (EE) dissipates at larger radii and/or smaller $\Gamma$, shifting $E_{\rm p}$ downward (softer HR), lengthening MVT through Eq.~(\ref{eq:Rvar}), and broadening the distribution of spectral lags through a combination of slower spectral evolution and enhanced curvature timescales. Within internal-shock synchrotron scalings (Eq.~\ref{eq:Ep_scaling}), the observed softening from ME to EE implies
\begin{equation}
\frac{E_{{\rm p},\,{\rm EE}}}{E_{{\rm p},\,{\rm ME}}}
\ \sim\
\left(\frac{L_{\rm EE}}{L_{\rm ME}}\right)^{1/2}
\left(\frac{R_{\rm EE}}{R_{\rm ME}}\right)^{-1}
\ \lesssim\ 1,
\end{equation}
consistent with $L_{\rm EE} < L_{\rm ME}$ and $R_{\rm EE} \gtrsim R_{\rm ME}$. Combining with Eq.~(\ref{eq:Rratio}) one can break the $R$--$\Gamma$ degeneracy if independent $\Gamma$ constraints (e.g., via opacity limits or afterglow onset) are available for future events.

\subsection{Toward a unified classification approach}

Our results contribute to the ongoing discussion of GRB classification schemes. There is a growing sense that the simple long/short split is insufficient, and one must incorporate additional information including spectra, lags, host galaxy properties, and afterglow characteristics \citep{Zhang2009, Steinhardt2023}. One approach is to classify bursts by their physical origin (Type I vs II) rather than by arbitrary duration. In this context, all episodes of a given burst share the same origin, so it makes sense that in our sample, both the main and extended emissions point to a Type II origin (they occur in known long bursts with likely massive-star progenitors). Fortunately, since they are still long and relatively soft, they would still look like long GRBs. The real confusion happens in cases like 060614 or 211211A, where the entire event acts like a merger (Type I) despite a long duration. Those are likely examples of a short-GRB-like engine but with extended activity, i.e. Type I with EE. In fact, GRB 060614’s light curve was a short spike + long soft tail \citep{Della2006,Gehrels2006}, very analogous to short spike+EE events, which is why it was hypothesized to be a merger origin, a conclusion supported by the absence of an associated supernova. GRB 060614 and GRB 211211A are most naturally interpreted as Type I systems with EE (SGRB+EE), whose long total $T_90$ arises from the EE tail rather than from an intrinsically long main burst pulse. Our studies emphasize the significance of episode-resolved analysis in classifying and understanding GRBs. Rather than relying on a single $T_{90}$, one can analyze the burst structure: if there is a separable second episode with properties like what we have quantified (softer, smoother), that should be taken as evidence of extended engine activity. Future automated analyses of GRB light curves could flag potential “EE episodes” and “precursor episodes” and classify bursts accordingly (some catalogs have started to do this for short GRBs with EE, \citealt{Sakamoto2011, vonKienlin2020}).

\section{Conclusions}\label{sec:conclusion}

We have conducted a \emph{episode-level} (pulse-resolved) comparative study of twenty-two long-duration GRBs observed with \emph{Swift}/BAT that display a two-episode prompt emission morphology: a bright \emph{main emission} (ME, $G_1$) spike followed after a background-consistent quiescent gap by a lower-intensity weaker \emph{extended emission} (EE, $G_2$) tail. 
To avoid two-main-pulse cases, we required a conservative flux-contrast cut, $F_{\rm p}(G_2)/F_{\rm p}(G_1) < 0.5$. Using homogeneous \emph{Swift}/BAT data products and computing all measurements strictly \emph{within} each episode, we compared $T_{90}$, hardness ratio, minimum variability timescale, and spectral lags for 22 bursts with clean quiescent separations and sufficient signal-to-noise in both $G_1$ and $G_2$. By analyzing the classification and characteristics of each episode independently, we obtained the following key results: 

\begin{itemize}
\item EE is significantly softer than ME with $R_{\rm HR}<1$ for the majority of the events. The hardness ratio drops by $\sim 30\%$ on average from ME to EE. This indicates a systematic shift to lower characteristic photon energies during the EE phase. Both episodes still occupy the long-GRB region in the $T_{90}$–hardness plane, indicating the overall event is a collapsar-type GRB, but EE shifts toward the softer end of that distribution.
\end{itemize}

\begin{itemize}
\item EE is substantially smoother than ME with $R_{\rm MVT}$ typically of order a few up to $\sim$ an order of magnitude with a median $\sim1.8$. In three of twenty-two cases, the MVT in EE is an order of magnitude larger (longer) than in ME for our sample. This points to a drastic change in the characteristic timescales of energy release between the two episodes, suggesting a larger emission region and/or lower Lorentz factor during EE.
\end{itemize}

\begin{itemize}
\item EE exhibits larger and more diverse spectral lags.
ME lags are modest and positive (a few hundred milliseconds) whereas EE often exhibits much larger lags (up to several seconds), both positive and negative. This diversity implies a more complex temporal-spectral evolution in the EE phase. The presence of large positive lags means EE undergo pronounced hard-to-soft evolution, whereas the negative lags hint at multi-component or soft-lead scenarios during EE.
\end{itemize}

\begin{itemize}
\item Episode-wise $T_{90}$ analysis shows that both ME and EE have $T_{90} > 2$ s therefore pointing to “long” in the traditional classification and they occupy the long-GRB region in the $T_{90}$–hardness and $T_{90}$–MVT planes, implying a common progenitor family within each sub burst. Some EEs last longer than the MEs, others shorter. The quiescent intervals $T_{\rm gap}$ between episodes are substantial (tens to hundreds of seconds) and do not scale simply with the episode-wise durations. This confirms that EE could be a distinct emission episode, not merely the tail of ME.
From these findings, we infer that EE is a physically distinct phase of the GRB prompt emission, arising from continued central-engine activity under altered conditions (lower power, perhaps different outflow magnetization or baryon loading) relative to the initial prompt spike. 
\end{itemize}

\begin{itemize}
\item The EE is not just the onset effect of ``afterglow emisison’’ or a trivial continuation, it has its own unique signature (softer, smoother, delayed).
\end{itemize}

\begin{itemize}
\item These long-GRB results exhibit a strong resemblance to the EE phenomenon known in short GRBs. In both cases, the EE phase is softer and longer than the initial ME spike, and is thought to arise from prolonged engine activity (fallback accretion or magnetar spin-down) after the main energy release. This suggests a common engine behavior across the GRB population: after an initial burst of energy, the engine can enter a lower-intensity, longer-duration state. For collapsars, this could be due to fallback of material feeding the black hole or a phase of spinning down if a magnetar survives. For mergers, it could be a remnant magnetar wind or delayed accretion onto a formed black hole. In both, the observable is an extended emission component.
\end{itemize}

EE is consistently \emph{smoother} than ME, suggesting that the late-time dissipation operates under conditions that suppress rapid variability, either because the characteristic emission region is larger, the bulk flow is less relativistic, or both. Meanwhile, the EE is \emph{softer}, consistent with a decline in engine power and a shift of the characteristic spectral peak to lower energies as the emission site moves outward and cools. The broadened, and in some cases sign, reversed-lags in the EE naturally follow if spectral evolution proceeds more slowly and curvature effects become more prominent at later times, or if multiple radiation zones contribute and reorder the energy–dependent pulse centroids. Consequently, these trends favor interpreting the EE as a second, physically distinct \emph{prompt} emission episode sustained by continued central engine activity rather than the onset of an external–shock afterglow.

Despite these pronounced episode–level differences, both ME and EE remain within the phenomenological long–GRB (Type~II) domain, implying a common progenitor family within each sub burst. In this light, the observed diversity, which is \emph{softer}, \emph{smoother}, and \emph{lag–diverse} EE relative to the ME, most plausibly reflects changes in engine state, dissipation scale, and flow kinematics within a single progenitor channel, rather than a switch of progenitor. Two practical caveats accompany this interpretation. First, the \emph{Swift}/BAT bandpass is narrow, so hardness serves as a proxy for true peak–energy evolution; broader coverage (e.g., simultaneous GBM+BAT) should directly confirm a systematic shift of the spectral peak from ME to EE. Second, limited signal–to–noise can inflate measured variability timescales for weak, late–time emission; nevertheless, the sample–wide separation in hardness, variability, and lag persists after accounting for this effect.  
Looking forward, expanding the ME+EE sample, adding rest-frame timing, and obtaining independent constraints on $\Gamma$ (e.g., from opacity limits or afterglow onset) will help break the $R$–$\Gamma$ degeneracy implied by the variability analysis. Time-resolved spectroscopy across a broader band should test the predicted $E_{\rm p}$ migration and quantify how the engine power and dissipation radius evolve from $G_1$ to $G_2$. In this way, the episode–resolved approach adopted here provides quantitative boundary conditions for central-engine models operating over $10^{1}$–$10^{3}$~s in long GRBs and points toward a unified interpretation of multi-episode prompt emission.

Our studies highlights the importance of time-resolved, episode-specific analysis for GRBs. Many GRBs have complex light curves, and analyzing them as single entities can obscure the true relationships between different phases. By separating sub-bursts, one can avoid misclassification (e.g., a long GRB that is actually a short GRB plus a long tail, or vice versa) and gain insights into how the engine evolves. This is especially relevant in the era of gravitational-wave astronomy when identifying EM counterparts of neutron star mergers, a short GRB with extended emission could easily be mistaken for a long GRB unless one scrutinizes the prompt profile. Our criteria and methodology could be applied in real-time GRB analysis to flag events with multiple episodes.

In conclusion, the EE in long GRBs carries critical clues about the late-time behavior of GRB central engines. We find that it is systematically softer in spectrum, longer in variability timescale, and exhibits a broader range of spectral lags than the initial main burst. These observational signatures support a scenario in which the GRB central engine, after the peak of its activity, continues to inject energy in a gradually fading, less relativistic outflow, producing a secondary prompt-emission episode. Future GRB missions with greater sensitivity and time resolution \citep[e.g.,][]{Wei2016,Fiore2020,Yuan2022} will be able to detect even weaker or faster extended components, potentially increasing the prevalence of known ME+EE events. Combining such observations with multi-wavelength afterglow and neutrino/gravitational-wave information will further elucidate the link between the prompt sub-burst properties and the nature of the GRB progenitor. Ultimately, the phenomenon of extended emission, in both long and short bursts, appears to be a window into the engine’s “slow burn” after the “flash,” and understanding it is key to a holistic picture of the GRB central-engine lifecycle.

\acknowledgments
This work is supported by the Natural Science Foundation of China (grant No. 11874033), the KC Wong Magna Foundation at Ningbo University, and made use of the High Energy Astrophysics Science Archive Research Center (HEASARC) Online Service at the NASA/Goddard Space Flight Center (GSFC). The computations were supported by the high performance computing center at Ningbo University.

\facilities{{\it Swift}}
\software{
{\tt 3ML} \citep{Vianello2015}, 
{\tt matplotlib} \citep{Hunter2007}, 
{\tt NumPy} \citep{Harris2020,Walt2011}, 
{\tt SciPy} \citep{Virtanen2020}, 
{\tt $lmfit$} \citep{Newville2016}, 
{\tt astropy} \citep{AstropyCollaboration2013},
{\tt pandas} \citep{Reback2022},
{\tt emcee} \citep{Foreman-Mackey2013},
{\tt seaborn} \citep{Waskom2017}}  
\bibliography{lGRBs.bib}

\clearpage
\startlongtable
\begin{deluxetable*}{lccccccc}
\tablewidth{0pt}
\tabletypesize{\small}
\tablecaption{Temporal properties of the main emission ($G_1$) and extended emission ($G_2$) sub-bursts in our long-GRB ME+EE sample. Each entry lists the time-bin resolution used, the start and end times ($t_1\sim t_2$), total pulse width ($T_{\text{pulse}}$), pulse peak time ($t_{\rm p}$), and peak count rate ($F_{\rm p}$) with $1\sigma$ uncertainties.\label{tab:pulse}}
\tablehead{
\colhead{GRB} & \colhead{Sub-burst} & \colhead{Time bin} & \colhead{$t_1 \sim t_2$} & \colhead{$T_{\text{pulse}}$} & \colhead{$t_{\rm p}$} & \colhead{$F_{\rm p}$} & \colhead{Classified as}\\
& & \colhead{(ms)} & \colhead{(s)} & \colhead{(s)} & \colhead{(s)} & & \colhead{(ME/EE)}
}
\startdata
050128 & $G_1$ & 256 & $-8.23\sim9.69$ & 17.92 & 6.10 & $1.26 \pm 0.10$ & ME \\
 & $G_2$ & 256 & $21.21\sim26.07$ & 4.86 & 21.98 & $0.24 \pm 0.05$ & EE \\
050505 & $G_1$ & 1024 & $-11.86\sim10.66$ & 22.53 & 0.42 & $0.20 \pm 0.02$ & ME \\
 & $G_2$ & 1024 & $20.90\sim47.53$ & 26.62 & 22.95 & $0.11 \pm 0.02$ & EE \\
060607A & $G_1$ & 1024 & $-18.41\sim37.91$ & 56.32 & 2.07 & $0.18 \pm 0.01$ & ME \\
 & $G_2$ & 1024 & $92.18\sim106.52$ & 14.34 & 96.28 & $0.05 \pm 0.01$ & EE \\
060906 & $G_1$ & 256 & $-47.02\sim-9.62$ & 27.39 & -36.76 & $0.27 \pm 0.05$ & ME \\
 & $G_2$ & 256 & $1.38\sim8.55$ & 7.17 & 4.71 & $0.12 \pm 0.03$ & EE \\
060927 & $G_1$ & 256 & $-1.69\sim9.06$ & 10.75 & 0.87 & $0.37 \pm 0.03$ & ME \\
 & $G_2$ & 256 & $14.95\sim24.42$ & 9.47 & 19.30 & $0.14 \pm 0.02$ & EE \\
071003 & $G_1$ & 1024 & $-12.98\sim43.34$ & 56.32 & 0.34 & $0.78 \pm 0.03$ & ME \\
 & $G_2$ & 1024 & $139.60\sim162.13$ & 22.53 & 142.67 & $0.10 \pm 0.01$ & EE \\
080607 & $G_1$ & 256 & $-7.66\sim16.91$ & 24.58 & 2.06 & $3.43 \pm 0.16$ & ME \\
 & $G_2$ & 256 & $17.17\sim88.08$ & 70.91 & 24.34 & $0.52 \pm 0.06$ & EE \\
080905B & $G_1$ & 512 & $-1.91\sim10.38$ & 12.29 & 2.70 & $0.21 \pm 0.03$ & ME \\
 & $G_2$ & 512 & $56.97\sim95.88$ & 38.91 & 60.04 & $0.08 \pm 0.01$ & EE \\
090424 & $G_1$ & 256 & $-2.22\sim11.35$ & 13.57 & 0.60 & $8.76 \pm 0.23$ & ME \\
 & $G_2$ & 256 & $43.10\sim63.58$ & 20.48 & 51.54 & $0.15 \pm 0.02$ & EE \\
090715B & $G_1$ & 256 & $-7.62\sim23.36$ & 30.98 & 7.23 & $0.50 \pm 0.03$ & ME \\
 & $G_2$ & 256 & $55.10\sim79.94$ & 24.83 & 67.65 & $0.21 \pm 0.02$ & EE \\
100704A & $G_1$ & 256 & $-8.54\sim24.49$ & 33.02 & 1.19 & $0.57 \pm 0.03$ & ME \\
 & $G_2$ & 1024 & $138.66\sim198.06$ & 59.39 & 148.90 & $0.09 \pm 0.01$ & EE \\
100906A & $G_1$ & 256 & $-1.23\sim33.84$ & 35.07 & 10.54 & $1.29 \pm 0.05$ & ME \\
 & $G_2$ & 256 & $90.16\sim128.82$ & 38.66 & 105.52 & $0.39 \pm 0.02$ & EE \\
110715A & $G_1$ & 64 & $-2.34\sim7.13$ & 9.47 & 2.52 & $7.79 \pm 0.21$ & ME \\
 & $G_2$ & 256 & $12.25\sim20.95$ & 8.70 & 14.30 & $0.35 \pm 0.04$ & EE \\
130514A & $G_1$ & 256 & $-8.19\sim57.86$ & 66.05 & 14.08 & $0.40 \pm 0.04$ & ME \\
 & $G_2$ & 256 & $104.45\sim153.34$ & 48.90 & 112.90 & $0.16 \pm 0.02$ & EE \\
130831A & $G_1$ & 256 & $-2.21\sim29.79$ & 32.00 & 2.91 & $1.56 \pm 0.07$ & ME \\
 & $G_2$ & 256 & $30.82\sim39.01$ & 8.19 & 33.89 & $0.17 \pm 0.03$ & EE \\
140430A & $G_1$ & 256 & $-1.30\sim7.14$ & 8.45 & 2.79 & $0.37 \pm 0.03$ & ME \\
 & $G_2$ & 1024 & $146.15\sim181.99$ & 35.84 & 171.75 & $0.06 \pm 0.01$ & EE \\
140703A & $G_1$ & 256 & $-10.93\sim7.76$ & 18.69 & 0.59 & $0.42 \pm 0.08$ & ME \\
 & $G_2$ & 256 & $60.50\sim70.99$ & 10.50 & 67.41 & $0.13 \pm 0.02$ & EE \\
151027A & $G_1$ & 256 & $-1.15\sim6.78$ & 7.94 & 0.90 & $0.97 \pm 0.08$ & ME \\
 & $G_2$ & 256 & $86.40\sim136.58$ & 50.18 & 109.18 & $0.48 \pm 0.03$ & EE \\
161017A & $G_1$ & 256 & $-5.21\sim38.57$ & 43.78 & 21.67 & $0.37 \pm 0.03$ & ME \\
 & $G_2$ & 1024 & $126.38\sim163.24$ & 36.86 & 136.62 & $0.08 \pm 0.01$ & EE \\
170903A & $G_1$ & 256 & $-2.79\sim13.34$ & 16.13 & 0.79 & $0.53 \pm 0.06$ & ME \\
 & $G_2$ & 256 & $18.71\sim23.58$ & 4.86 & 21.27 & $0.21 \pm 0.05$ & EE \\
210619B & $G_1$ & 256 & $-0.54\sim11.75$ & 12.29 & 1.51 & $14.81 \pm 0.24$ & ME \\
 & $G_2$ & 256 & $28.14\sim79.85$ & 51.71 & 50.66 & $5.47 \pm 0.13$ & EE \\
220521A & $G_1$ & 256 & $-2.82\sim3.32$ & 6.14 & 0.25 & $0.70 \pm 0.06$ & ME \\
 & $G_2$ & 256 & $7.67\sim12.02$ & 4.35 & 9.46 & $0.19 \pm 0.05$ & EE \\
\enddata
\end{deluxetable*}

\setlength{\tabcolsep}{0.15em}
\begin{deluxetable*}{ccccccccc}
\tablewidth{0pt}
\tabletypesize{\normalsize}
\tablecaption{Comparative analysis of main-emission ($G_1$) and extended-emission ($G_2$) properties for each GRB in our long-GRB ME+EE sample. Listed quantities include the redshift, observed-frame quiescent gap duration and its rest-frame corrected value, and the ratios of $G_2$ to $G_1$ in terms of $T_{90}$, $F_{\rm p}$, HR, MVT, and spectral lag.\label{tab:Compara}}
\tablehead{
\colhead{GRB}&\colhead{$z$}&\colhead{$T_{\mathrm{gap}}$}&\colhead{$T_{\mathrm{gap}}/(1+z)$}&\colhead{$R_{t_{90}}$}&\colhead{$R_{F_{\rm p}}$}&\colhead{$R_{\rm HR}$}&\colhead{$R_{\rm MVT}$}&\colhead{$R_{\tau}$}\\
\hline
&&(s)&(s)&&&&&\\
}
\startdata
050128 & 5.500 & $11.5 \pm 2.4$ & $1.8 \pm 0.4$ & $0.28 \pm 0.06$ & $0.19 \pm 0.04$ & $0.25 \pm 0.50$ & $0.78 \pm 0.35$ & $0.96 \pm 1.05$ \\
050505 & 4.270 & $10.2 \pm 3.2$ & $1.9 \pm 0.6$ & $1.29 \pm 0.22$ & $0.55 \pm 0.11$ & $0.93 \pm 0.25$ & $1.24 \pm 0.96$ & $0.22 \pm 1.28$ \\
060607A & 3.082 & $54.3 \pm 7.8$ & $13.3 \pm 1.9$ & $0.24 \pm 0.08$ & $0.28 \pm 0.06$ & $1.23 \pm 0.27$ & $0.76 \pm 0.42$ & $1.15 \pm 0.04$ \\
060906 & 3.690 & $11.0 \pm 3.4$ & $2.3 \pm 0.7$ & $0.25 \pm 0.04$ & $0.44 \pm 0.14$ & $0.93 \pm 0.35$ & $0.79 \pm 0.69$ & $1.10 \pm 0.13$ \\
060927 & 5.467 & $5.9 \pm 3.1$ & $0.9 \pm 0.5$ & $0.93 \pm 0.30$ & $0.38 \pm 0.06$ & $0.73 \pm 0.13$ & $1.53 \pm 0.76$ & $-5.53 \pm 0.48$ \\
071003 & 1.604 & $96.3 \pm 15.5$ & $37.0 \pm 6.0$ & $0.57 \pm 0.25$ & $0.13 \pm 0.01$ & $0.57 \pm 0.20$ & $13.57 \pm 10.42$ & $-2.89 \pm 0.42$ \\
080607 & 3.040 & $\sim 0.3$ & $0.1 \pm 1.8$ & $4.27 \pm 1.64$ & $0.15 \pm 0.02$ & $0.49 \pm 0.04$ & $3.15 \pm 2.01$ & $10.63 \pm 3.58$ \\
080905B & 2.374 & $46.6 \pm 3.2$ & $13.8 \pm 1.0$ & $3.20 \pm 0.44$ & $0.38 \pm 0.07$ & $0.85 \pm 0.20$ & $2.75 \pm 1.05$ & $-1.52 \pm 0.60$ \\
090424 & 0.544 & $31.8 \pm 4.9$ & $20.6 \pm 3.2$ & $2.18 \pm 1.03$ & $0.02 \pm 0.00$ & $0.46 \pm 0.10$ & $13.19 \pm 8.41$ & $-15.06 \pm 0.59$ \\
090715B & 3.000 & $31.7 \pm 6.5$ & $7.9 \pm 1.6$ & $1.00 \pm 0.31$ & $0.42 \pm 0.05$ & $0.77 \pm 0.11$ & $2.12 \pm 1.63$ & $-1.11 \pm 0.51$ \\
100704A & 3.600 & $114.2 \pm 9.4$ & $24.8 \pm 2.1$ & $2.14 \pm 0.65$ & $0.16 \pm 0.02$ & $0.41 \pm 0.08$ & $5.33 \pm 1.65$ & $7.62 \pm 0.66$ \\
100906A & 1.727 & $56.3 \pm 11.7$ & $20.7 \pm 4.3$ & $1.19 \pm 0.42$ & $0.30 \pm 0.02$ & $0.46 \pm 0.06$ & $1.11 \pm 0.27$ & $5.47 \pm 2.04$ \\
110715A & 0.820 & $5.1 \pm 2.9$ & $2.8 \pm 1.6$ & $1.35 \pm 0.94$ & $0.04 \pm 0.01$ & $0.63 \pm 0.20$ & $5.33 \pm 1.65$ & $0.05 \pm 0.06$ \\
130514A & 3.600 & $46.6 \pm 10.9$ & $10.1 \pm 2.4$ & $0.85 \pm 0.18$ & $0.40 \pm 0.06$ & $0.65 \pm 0.10$ & $0.64 \pm 0.46$ & $4.00 \pm 2.07$ \\
130831A & 0.479 & $1.0 \pm 7.7$ & $0.7 \pm 5.2$ & $0.29 \pm 0.08$ & $0.11 \pm 0.02$ & $0.62 \pm 0.24$ & $1.39 \pm 0.95$ & $3.33 \pm 1.18$ \\
140430A & 1.600 & $139.0 \pm 2.6$ & $53.5 \pm 1.0$ & $5.04 \pm 1.21$ & $0.16 \pm 0.03$ & $0.78 \pm 0.40$ & $11.57 \pm 3.79$ & $-5.94 \pm 1.33$ \\
140703A & 3.140 & $52.7 \pm 1.7$ & $12.7 \pm 0.4$ & $0.55 \pm 0.07$ & $0.31 \pm 0.08$ & $0.48 \pm 0.21$ & $0.85 \pm 0.35$ & $-0.12 \pm 0.01$ \\
151027A & 0.810 & $79.6 \pm 9.1$ & $44.0 \pm 5.0$ & $5.82 \pm 1.95$ & $0.49 \pm 0.05$ & $0.91 \pm 0.14$ & $1.21 \pm 0.54$ & $-2.64 \pm 0.95$ \\
161017A & 2.013 & $87.8 \pm 8.8$ & $29.1 \pm 2.9$ & $0.87 \pm 0.19$ & $0.22 \pm 0.03$ & $0.89 \pm 0.18$ & $2.08 \pm 1.74$ & $4.00 \pm 0.96$ \\
170903A & 0.886 & $5.4 \pm 2.2$ & $2.8 \pm 1.2$ & $0.24 \pm 0.08$ & $0.40 \pm 0.10$ & $1.25 \pm 0.61$ & $1.27 \pm 0.74$ & $1.60 \pm 0.12$ \\
210619B & 1.937 & $16.4 \pm 7.1$ & $5.6 \pm 2.4$ & $3.40 \pm 1.34$ & $0.37 \pm 0.01$ & $0.68 \pm 0.02$ & $2.17 \pm 1.57$ & $1.12 \pm 0.48$ \\
220521A & 5.600 & $4.3 \pm 1.5$ & $0.7 \pm 0.2$ & $0.92 \pm 0.44$ & $0.27 \pm 0.08$ & $1.12 \pm 0.68$ & $3.96 \pm 1.03$ & $-4.80 \pm 2.60$ \\
\enddata
\tablecomments{
Ratios are defined as $R_X\equiv X^{G_2}/X^{G_1}$, i.e., 
$R_{t_{90}} \equiv \frac{t_{90}^{G_2}}{t_{90}^{G_1}}$, 
$R_{F_{\rm p}} \equiv \frac{F_{\rm p}^{G_2}}{F_{\rm p}^{G_1}}$,
$R_{\rm HR} \equiv \frac{{\rm HR}^{G_2}}{{\rm HR}^{G_1}}$, 
$R_{\rm MVT} \equiv \frac{{\rm MVT}^{G_2}}{{\rm MVT}^{G_1}}$, 
and $R_{\tau} \equiv \frac{\tau^{G_2}}{\tau^{G_1}}$.
}
\end{deluxetable*}

\begin{table*}[htbp]
\centering
\footnotesize
\caption{Individual $T_{90}$ durations (15--150~keV) of the main emission ($G_1$) and extended emission ($G_2$) sub-bursts for each GRB in our long-GRB ME+EE sample. All components satisfy the long-burst ($T_{90}>2$~s) classification.\label{tab:t90}}
\centering
\begin{tabular}{c|c|c|c|c}
\hline
\multicolumn{1}{c|}{GRB} & \multicolumn{2}{c|}{Main emission} & \multicolumn{2}{c}{Extended emission} \\
& \multicolumn{2}{c|}{($G_1$)} & \multicolumn{2}{c}{($G_2$)} \\
\hline
&Value&Classification&Value&Classification\\
\hline
GRB 050128 & 13.6$^{+2.3}_{-2.0}$~[s] & Long & 3.8$^{+0.5}_{-0.5}$~[s] & Long \\
GRB 050505 & 17.4$^{+3.1}_{-2.0}$~[s] & Long & 22.5$^{+3.1}_{-1.0}$~[s] & Long \\
GRB 060607A & 38.9$^{+7.2}_{-10.2}$~[s] & Long & 9.2$^{+2.0}_{-3.1}$~[s] & Long \\
GRB 060906 & 22.5$^{+3.3}_{-1.5}$~[s] & Long & 5.6$^{+0.8}_{-0.8}$~[s] & Long \\
GRB 060927 & 7.2$^{+2.6}_{-1.0}$~[s] & Long & 6.7$^{+1.0}_{-1.8}$~[s] & Long \\
GRB 071003 & 30.7$^{+15.4}_{-10.2}$~[s] & Long & 17.4$^{+3.1}_{-2.0}$~[s] & Long \\
GRB 080607 & 14.1$^{+6.7}_{-3.8}$~[s] & Long & 60.2$^{+7.9}_{-2.8}$~[s] & Long \\
GRB 080905B & 10.2$^{+1.0}_{-1.0}$~[s] & Long & 32.8$^{+3.1}_{-3.1}$~[s] & Long \\
GRB 090424 & 7.2$^{+4.6}_{-1.8}$~[s] & Long & 15.6$^{+3.1}_{-1.8}$~[s] & Long \\
GRB 090715B & 19.7$^{+5.9}_{-5.4}$~[s] & Long & 19.7$^{+2.3}_{-2.8}$~[s] & Long \\
GRB 100704A & 21.5$^{+7.2}_{-4.3}$~[s] & Long & 46.1$^{+7.2}_{-6.1}$~[s] & Long \\
GRB 100906A & 22.5$^{+10.2}_{-2.3}$~[s] & Long & 26.9$^{+6.1}_{-5.6}$~[s] & Long \\
GRB 110715A & 4.2$^{+2.6}_{-2.7}$~[s] & Long & 5.6$^{+1.8}_{-1.3}$~[s] & Long \\
GRB 130514A & 48.4$^{+10.5}_{-7.2}$~[s] & Long & 41.0$^{+5.1}_{-2.8}$~[s] & Long \\
GRB 130831A & 21.5$^{+7.7}_{-2.8}$~[s] & Long & 6.1$^{+1.0}_{-1.0}$~[s] & Long \\
GRB 140430A & 5.9$^{+1.5}_{-1.0}$~[s] & Long & 29.7$^{+4.1}_{-2.0}$~[s] & Long \\
GRB 140703A & 15.9$^{+1.5}_{-1.3}$~[s] & Long & 8.7$^{+1.0}_{-0.8}$~[s] & Long \\
GRB 151027A & 5.6$^{+1.3}_{-1.0}$~[s] & Long & 32.8$^{+8.4}_{-9.0}$~[s] & Long \\
GRB 161017A & 33.0$^{+7.2}_{-3.6}$~[s] & Long & 28.7$^{+3.1}_{-5.1}$~[s] & Long \\
GRB 170903A & 12.5$^{+1.8}_{-1.8}$~[s] & Long & 3.1$^{+0.5}_{-1.3}$~[s] & Long \\
GRB 210619B & 9.0$^{+2.6}_{-0.8}$~[s] & Long & 30.5$^{+14.6}_{-6.7}$~[s] & Long \\
GRB 220521A & 3.3$^{+1.3}_{-1.5}$~[s] & Long & 3.1$^{+0.5}_{-0.8}$~[s] & Long \\
\hline
\end{tabular}
\end{table*}

\begin{table*}[htbp]
\centering
\footnotesize
\caption{Summary statistics (range and median) of episode-level prompt diagnostics for our long-GRB ME+EE sample ($N=22$). Episode properties are reported separately for the main emission (ME, $G_1$) and extended emission (EE, $G_2$). Quiescent time interval $T_{\rm gap}$ and ratio quantities are defined per burst and are therefore listed once (under the ME columns).\label{tab:summary}}
\setlength{\tabcolsep}{6pt}
\begin{tabular}{l|cc|cc}
\hline
\multicolumn{1}{c|}{Quantity} & \multicolumn{2}{c|}{Main emission} & \multicolumn{2}{c}{Extended emission}\\
& \multicolumn{2}{c|}{($G_1$)} & \multicolumn{2}{c}{($G_2$)}\\
\hline
& Range & Median & Range & Median\\
\hline
Episode-only $T_{90}$~[s] & [3.3, 48.4] & 15.0 & [3.1, 60.2] & 18.6 \\
HR & [0.52, 1.13] & 0.81 & [0.26, 0.92] & 0.56 \\
${\rm MVT}$~[ms] & [98, 3489] & 599 & [397, 5378] & 1363 \\
Spectral lag $\tau$~[ms] & [-250, 1536] & 80 & [-2355, 2976] & 110 \\
\hline
$t_{\rm gap}$~[s] & [0.26, 139.01] & 31.75 & \nodata & \nodata \\
$t_{\rm gap}/(1+z)$~[s] & [0.06, 53.47] & 9.03 & \nodata & \nodata \\
$R_{T_{90}}$ & [0.24, 5.82] & 0.96 & \nodata & \nodata \\
$R_{F_{\rm p}}$ & [0.02, 0.55] & 0.29 & \nodata & \nodata \\
$R_{\rm HR}$ & [0.25, 1.26] & 0.70 & \nodata & \nodata \\
$R_{\rm MVT}$ & [0.64, 13.57] & 1.81 & \nodata & \nodata \\
$R_{\tau}$ & [-15.1, 10.6] & 0.59& \nodata & \nodata \\
\hline
\end{tabular}
\end{table*}

\begin{table*}[htbp]
\centering
\footnotesize
\caption{Hardness ratios [HR $= S_{50\text{--}100\,\mathrm{keV}}/S_{25\text{--}50\,\mathrm{keV}}$] for each main emission ($G_1$) and extended emission ($G_2$) pulse in our long-GRB ME+EE sample, computed from background-subtracted fluences in the specified BAT bands.\label{tab:HR}}
\setlength{\tabcolsep}{4pt}
\begin{tabular}{c|c|c|c|c}
\hline
\multicolumn{1}{c|}{GRB} & \multicolumn{2}{c|}{Main emission} & \multicolumn{2}{c}{Extended emission}\\
& \multicolumn{2}{c|}{($G_1$)} & \multicolumn{2}{c}{($G_2$)}\\
\hline
&Value&Classification&Value&Classification\\
\hline
050128 & 1.04$\pm$0.06 & Long & 0.26$\pm$0.52 & Long \\
050505 & 0.88$\pm$0.08 & Long & 0.82$\pm$0.21 & Long \\
060607A & 0.75$\pm$0.03 & Long & 0.92$\pm$0.20 & Long \\
060906 & 0.56$\pm$0.05 & Long & 0.52$\pm$0.19 & Long \\
060927 & 0.67$\pm$0.04 & Long & 0.49$\pm$0.08 & Long \\
071003 & 0.98$\pm$0.05 & Long & 0.56$\pm$0.19 & Long \\
080607 & 1.13$\pm$0.05 & Long & 0.55$\pm$0.04 & Long \\
080905B & 0.85$\pm$0.16 & Long & 0.72$\pm$0.10 & Long \\
090424 & 0.84$\pm$0.03 & Long & 0.39$\pm$0.08 & Long \\
090715B & 0.82$\pm$0.05 & Long & 0.63$\pm$0.08 & Long \\
100704A & 0.87$\pm$0.05 & Long & 0.36$\pm$0.07 & Long \\
100906A & 0.80$\pm$0.03 & Long & 0.37$\pm$0.05 & Long \\
110715A & 0.76$\pm$0.02 & Long & 0.48$\pm$0.15 & Long \\
130514A & 0.77$\pm$0.04 & Long & 0.50$\pm$0.07 & Long \\
130831A & 0.60$\pm$0.03 & Long & 0.37$\pm$0.14 & Long \\
140430A & 0.72$\pm$0.09 & Long & 0.56$\pm$0.28 & Long \\
140703A & 0.87$\pm$0.18 & Long & 0.42$\pm$0.16 & Long \\
151027A & 0.85$\pm$0.12 & Long & 0.77$\pm$0.04 & Long \\
161017A & 0.80$\pm$0.05 & Long & 0.71$\pm$0.14 & Long \\
170903A & 0.55$\pm$0.12 & Long & 0.69$\pm$0.30 & Long \\
210619B & 1.08$\pm$0.02 & Long & 0.73$\pm$0.01 & Long \\
220521A & 0.52$\pm$0.13 & Long & 0.58$\pm$0.32 & Long \\
\hline
\end{tabular}
\end{table*}

\begin{table*}[htbp]
\centering
\footnotesize
\caption{Minimum variability timescales ($\Delta t_{\min}$) of the main emission ($G_1$) and extended emission ($G_2$) sub-bursts for GRBs in our long-GRB ME+EE sample, measured with the Haar wavelet method from \emph{Swift}/BAT light curves.\label{tab:MVT}}
\setlength{\tabcolsep}{4pt}
\begin{tabular}{c|c|c|c|c}
\hline
\multicolumn{1}{c|}{GRB} & \multicolumn{2}{c|}{Main emission} & \multicolumn{2}{c}{Extended emission}\\
& \multicolumn{2}{c|}{($G_1$)} & \multicolumn{2}{c}{($G_2$)}\\
\hline
&Value&Classification&Value&Classification\\
\hline
050128 & (507$\pm$188)~[ms] & Long & (397$\pm$103)~[ms] & Long \\
050505 & (1365$\pm$849)~[ms] & Long & (1691$\pm$776)~[ms] & Long \\
060607A & (3246$\pm$1645)~[ms] & Long & (2467$\pm$557)~[ms] & Long \\
060906 & (1519$\pm$1270)~[ms] & Long & (1193$\pm$319)~[ms] & Long \\
060927 & (537$\pm$168)~[ms] & Long & (824$\pm$318)~[ms] & Long \\
071003 & (98$\pm$36)~[ms] & Long & (1333$\pm$897)~[ms] & Long \\
080607 & (420$\pm$129)~[ms] & Long & (1323$\pm$741)~[ms] & Long \\
080905B & (413$\pm$142)~[ms] & Long & (1137$\pm$184)~[ms] & Long \\
090424 & (184$\pm$62)~[ms] & Long & (2429$\pm$1311)~[ms] & Long \\
090715B & (890$\pm$208)~[ms] & Long & (1882$\pm$1380)~[ms] & Long \\
100704A & (599$\pm$146)~[ms] & Long & (3195$\pm$609)~[ms] & Long \\
100906A & (1015$\pm$31)~[ms] & Long & (1130$\pm$277)~[ms] & Long \\
110715A & (599$\pm$146)~[ms] & Long & (3195$\pm$609)~[ms] & Long \\
130514A & (1824$\pm$1193)~[ms] & Long & (1166$\pm$357)~[ms] & Long \\
130831A & (1015$\pm$31)~[ms] & Long & (1408$\pm$964)~[ms] & Long \\
140430A & (465$\pm$132)~[ms] & Long & (5378$\pm$876)~[ms] & Long \\
140703A & (3489$\pm$1090)~[ms] & Long & (2980$\pm$814)~[ms] & Long \\
151027A & (341$\pm$99)~[ms] & Long & (413$\pm$141)~[ms] & Long \\
161017A & (664$\pm$159)~[ms] & Long & (1380$\pm$1107)~[ms] & Long \\
170903A & (1331$\pm$760)~[ms] & Long & (1693$\pm$159)~[ms] & Long \\
210619B & (260$\pm$101)~[ms] & Long & (565$\pm$345)~[ms] & Long \\
220521A & (340$\pm$43)~[ms] & Long & (1345$\pm$306)~[ms] & Long \\
\hline
\end{tabular}
\end{table*}

\begin{table*}[htbp]
\centering
\footnotesize
\caption{Spectral lags between the 25--50~keV and 15--25~keV bands for the main emission ($G_1$) and extended emission ($G_2$) sub-bursts for GRBs in our long-GRB ME+EE sample. Uncertainties are $1\sigma$.\label{tab:lag}}
\setlength{\tabcolsep}{4pt}
\begin{tabular}{c|c|c|c|c}
\hline
\multicolumn{1}{c|}{GRB} & \multicolumn{2}{c|}{Main emission} & \multicolumn{2}{c}{Extended emission}\\
& \multicolumn{2}{c|}{($G_1$)} & \multicolumn{2}{c}{($G_2$)}\\
\hline
&Value&Classification&Value&Classification\\
\hline
050128 & (-76.8$\pm$50.9)~[ms] & Long & (-73.6$\pm$64.0)~[ms] & Long \\
050505 & (-230.4$\pm$209.9)~[ms] & Long & (-51.2$\pm$291.8)~[ms] & Long \\
060607A & (1094.4$\pm$32.0)~[ms] & Long & (1257.6$\pm$16.6)~[ms] & Long \\
060906 & (-249.6$\pm$19.8)~[ms] & Long & (-275.2$\pm$24.3)~[ms] & Long \\
060927 & (163.2$\pm$14.1)~[ms] & Long & (-902.4$\pm$6.1)~[ms] & Long \\
071003 & (-179.2$\pm$25.0)~[ms] & Long & (518.4$\pm$21.1)~[ms] & Long \\
080607 & (15.2$\pm$5.0)~[ms] & Long & (161.6$\pm$9.3)~[ms] & Long \\
080905B & (-134.4$\pm$9.6)~[ms] & Long & (204.8$\pm$79.4)~[ms] & Long \\
090424 & (-34.0$\pm$0.4)~[ms] & Long & (512.0$\pm$19.2)~[ms] & Long \\
090715B & (-44.8$\pm$19.5)~[ms] & Long & (49.6$\pm$6.9)~[ms] & Long \\
100704A & (390.4$\pm$33.9)~[ms] & Long & (2976.0$\pm$19.8)~[ms] & Long \\
100906A & (48.0$\pm$17.6)~[ms] & Long & (262.4$\pm$17.9)~[ms] & Long \\
110715A & (390.4$\pm$33.9)~[ms] & Long & (19.2$\pm$23.7)~[ms] & Long \\
130514A & (70.4$\pm$33.3)~[ms] & Long & (281.6$\pm$58.9)~[ms] & Long \\
130831A & (115.2$\pm$29.4)~[ms] & Long & (384.0$\pm$94.7)~[ms] & Long \\
140430A & (396.8$\pm$37.1)~[ms] & Long & (-2355.2$\pm$481.3)~[ms] & Long \\
140703A & (1536.0$\pm$128.0)~[ms] & Long & (-185.6$\pm$12.2)~[ms] & Long \\
151027A & (89.6$\pm$28.2)~[ms] & Long & (-236.8$\pm$41.6)~[ms] & Long \\
161017A & (313.6$\pm$60.8)~[ms] & Long & (1254.4$\pm$176.6)~[ms] & Long \\
170903A & (448.0$\pm$24.3)~[ms] & Long & (716.8$\pm$39.7)~[ms] & Long \\
210619B & (51.2$\pm$15.4)~[ms] & Long & (57.6$\pm$17.3)~[ms] & Long \\
220521A & (96.0$\pm$51.8)~[ms] & Long & (-460.8$\pm$25.6)~[ms] & Long \\
\hline
\end{tabular}
\end{table*}

\begin{figure*}
\includegraphics[angle=0,scale=0.32]{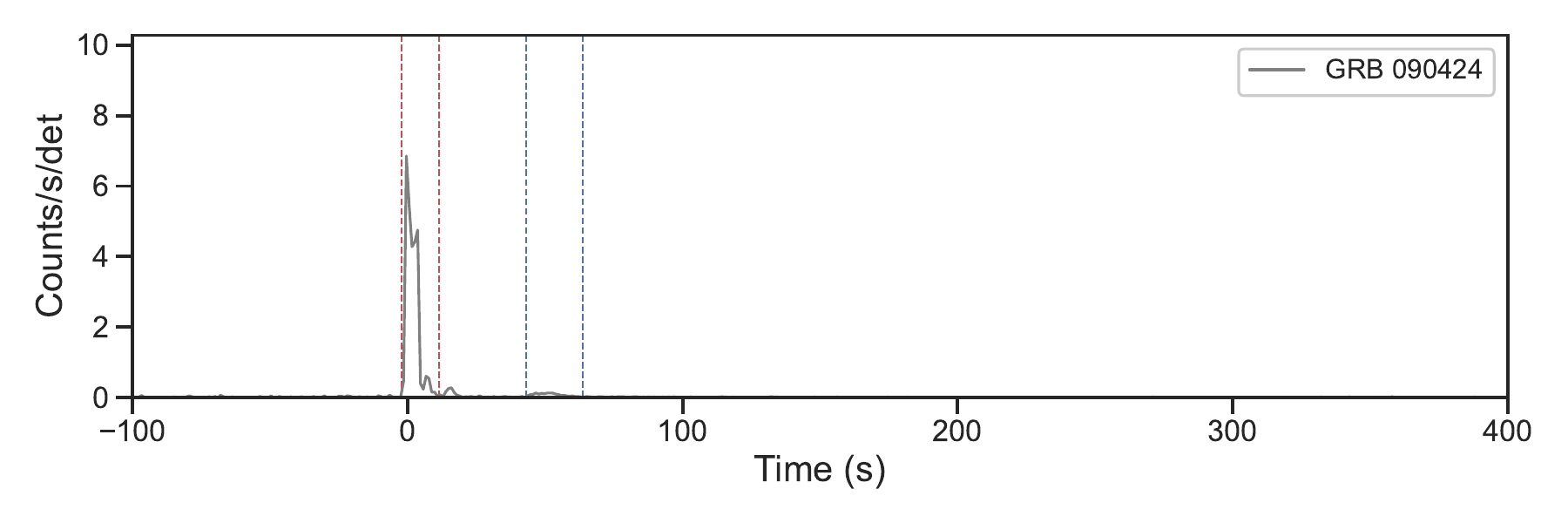}
\includegraphics[angle=0,scale=0.32]{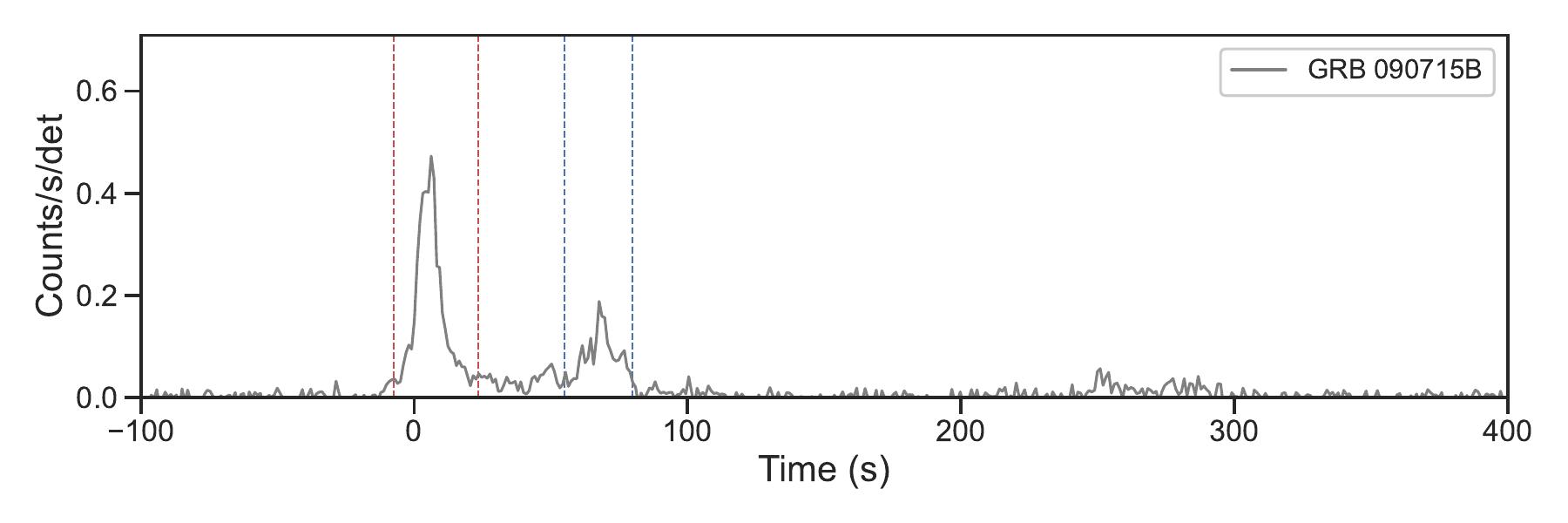}
\includegraphics[angle=0,scale=0.32]{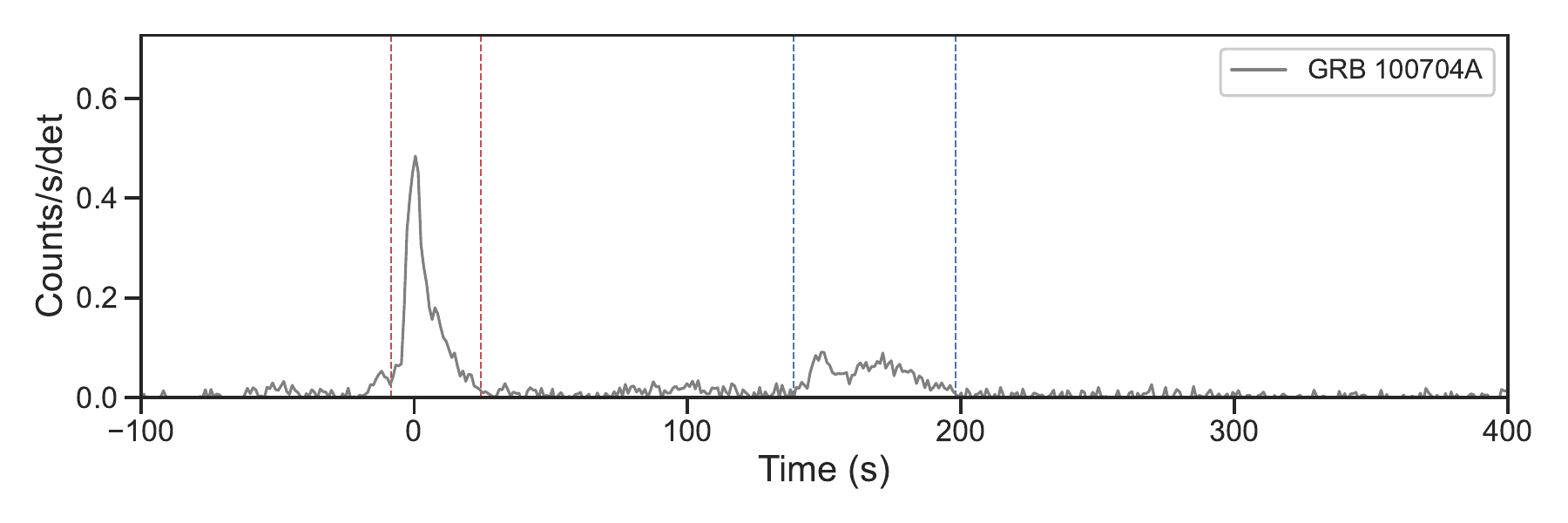}
\includegraphics[angle=0,scale=0.32]{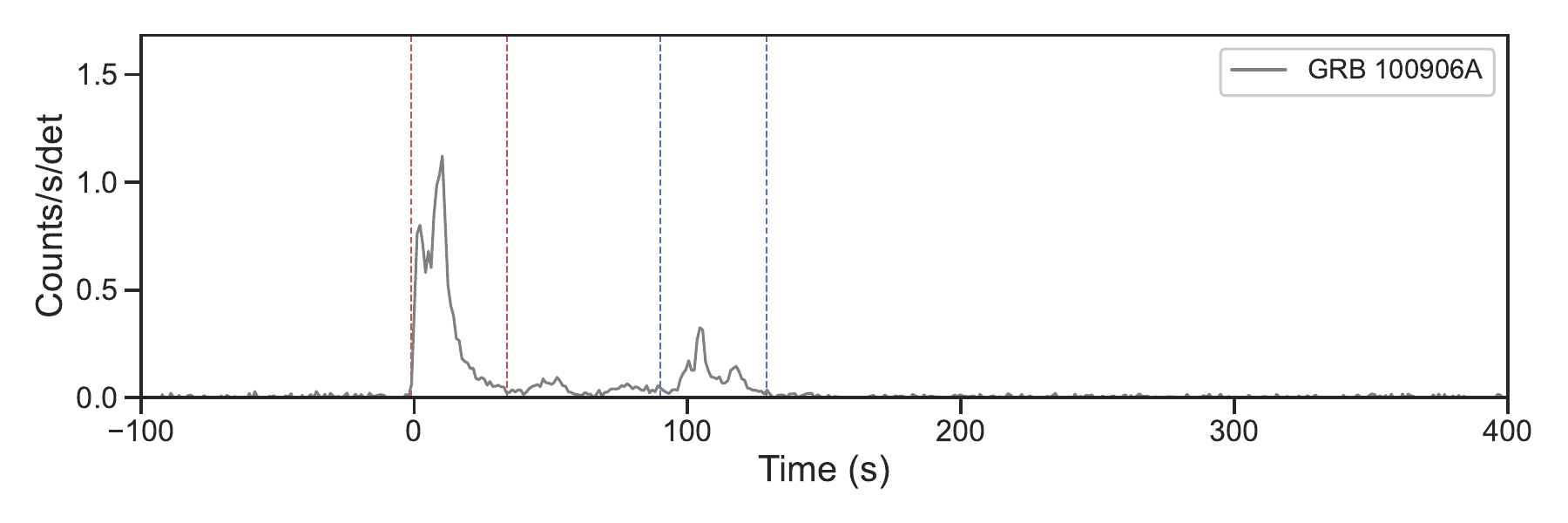}
\includegraphics[angle=0,scale=0.32]{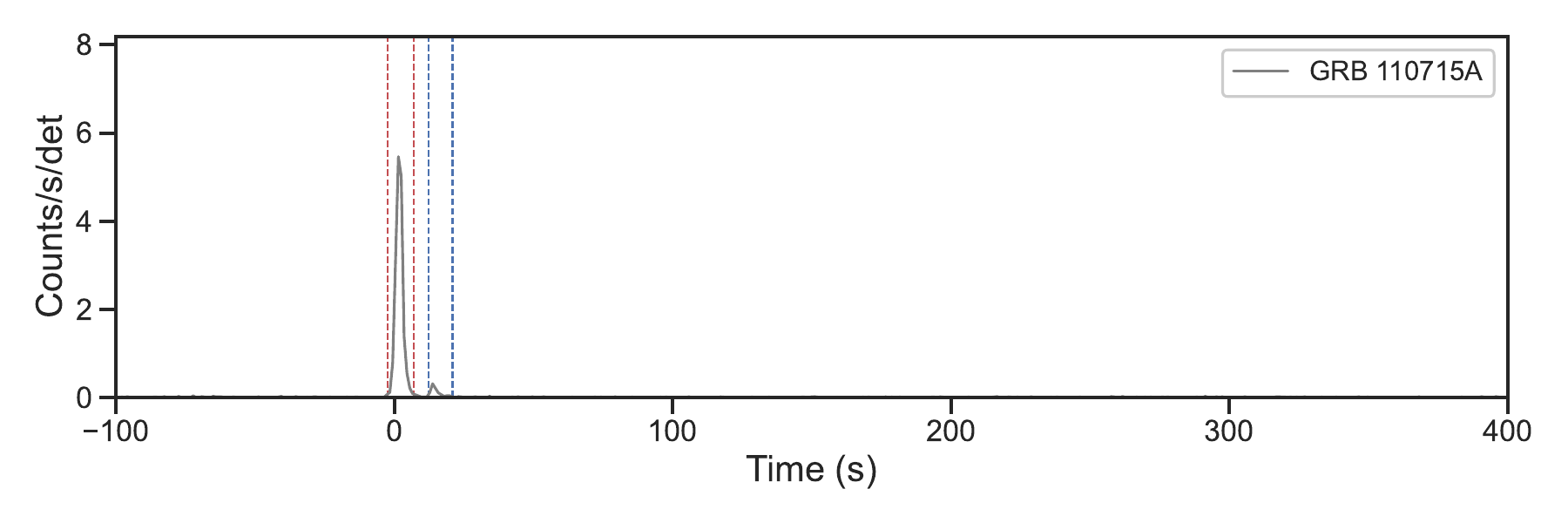}
\includegraphics[angle=0,scale=0.32]{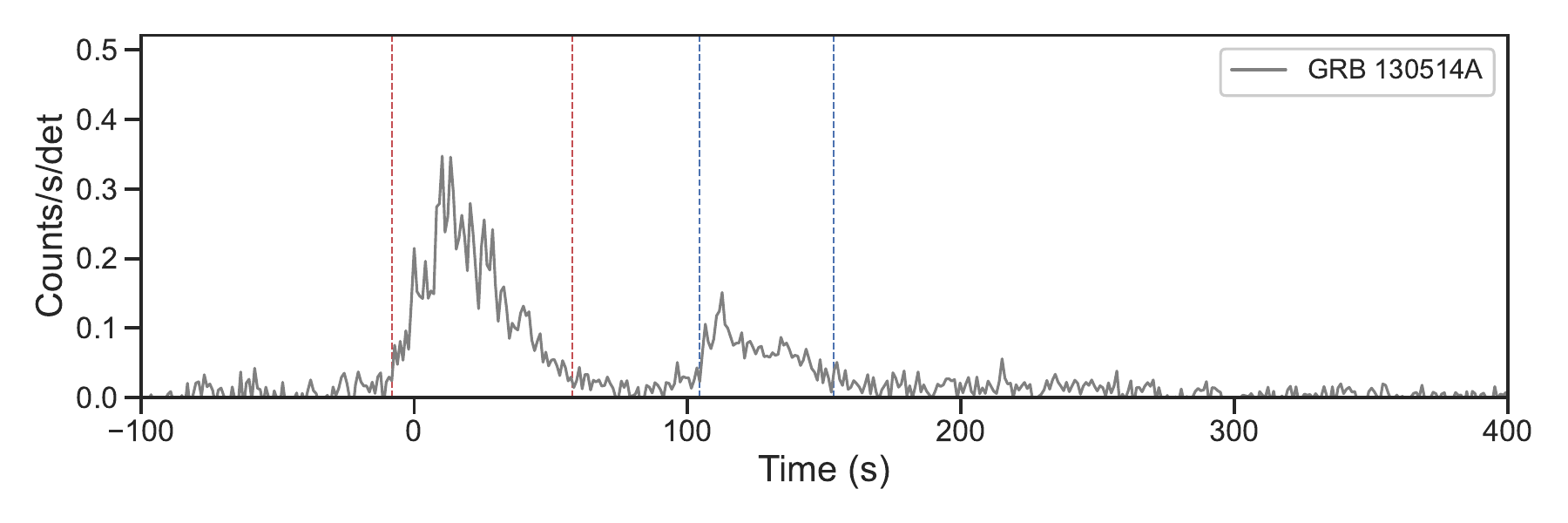}
\includegraphics[angle=0,scale=0.32]{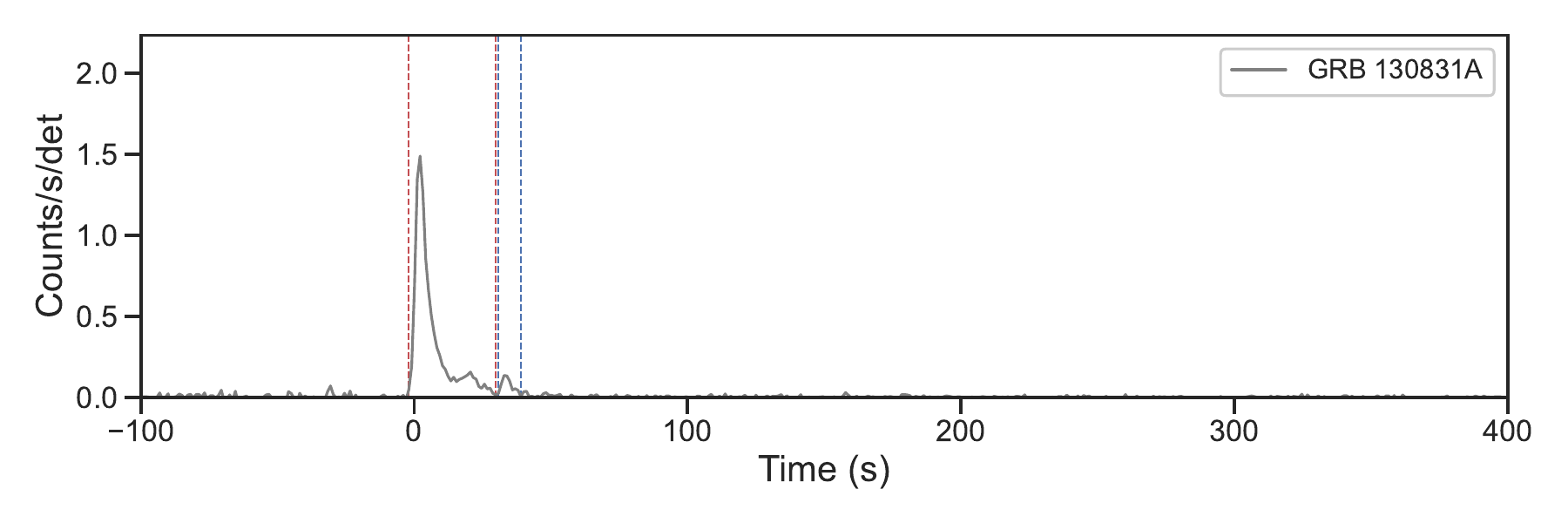}
\includegraphics[angle=0,scale=0.32]{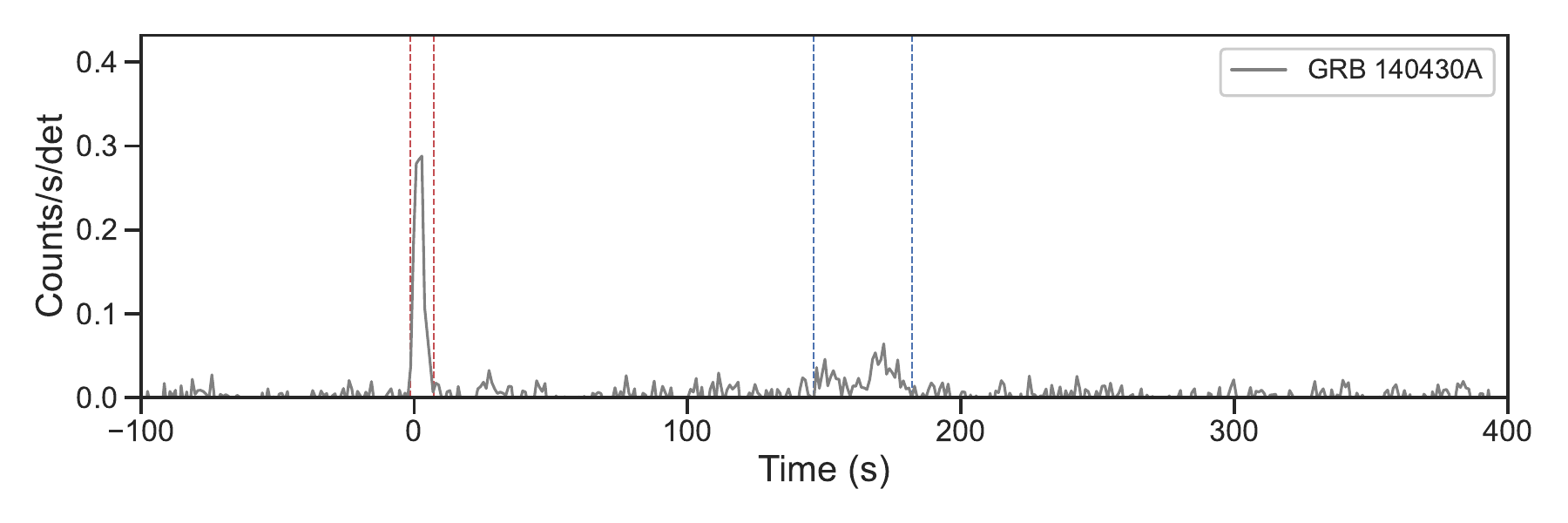}
\center{Figure \ref{fig:lc_sample}--- Continued}
\end{figure*}
\begin{figure*}
\includegraphics[angle=0,scale=0.32]{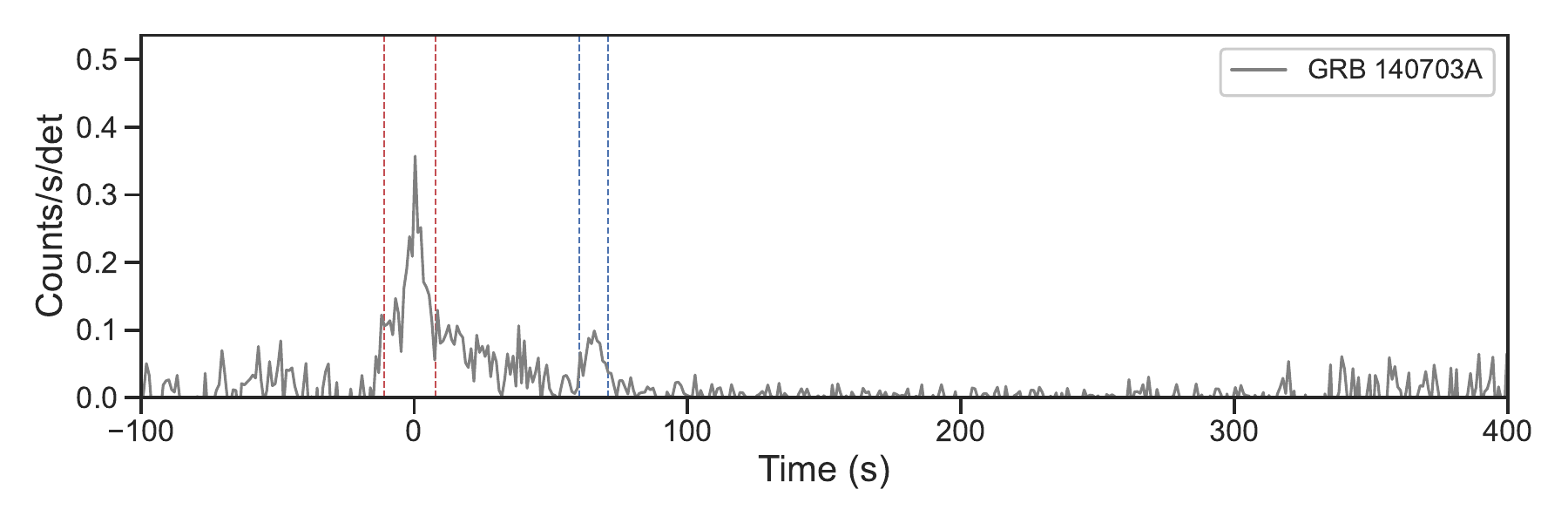}
\includegraphics[angle=0,scale=0.32]{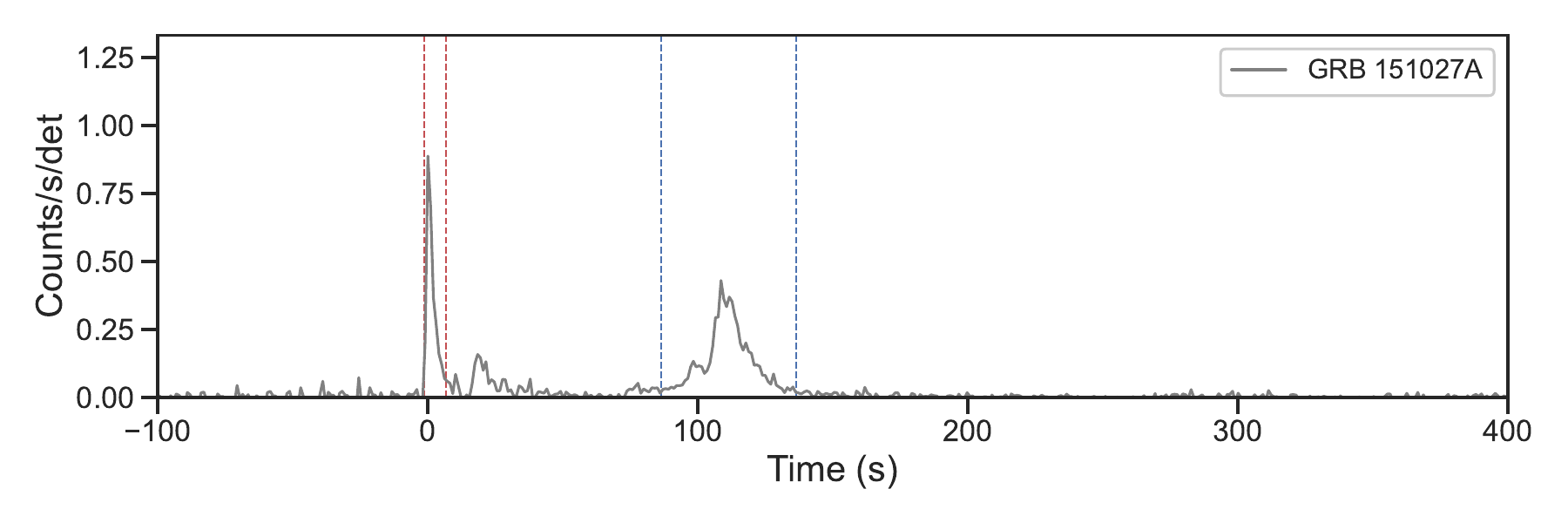}
\includegraphics[angle=0,scale=0.32]{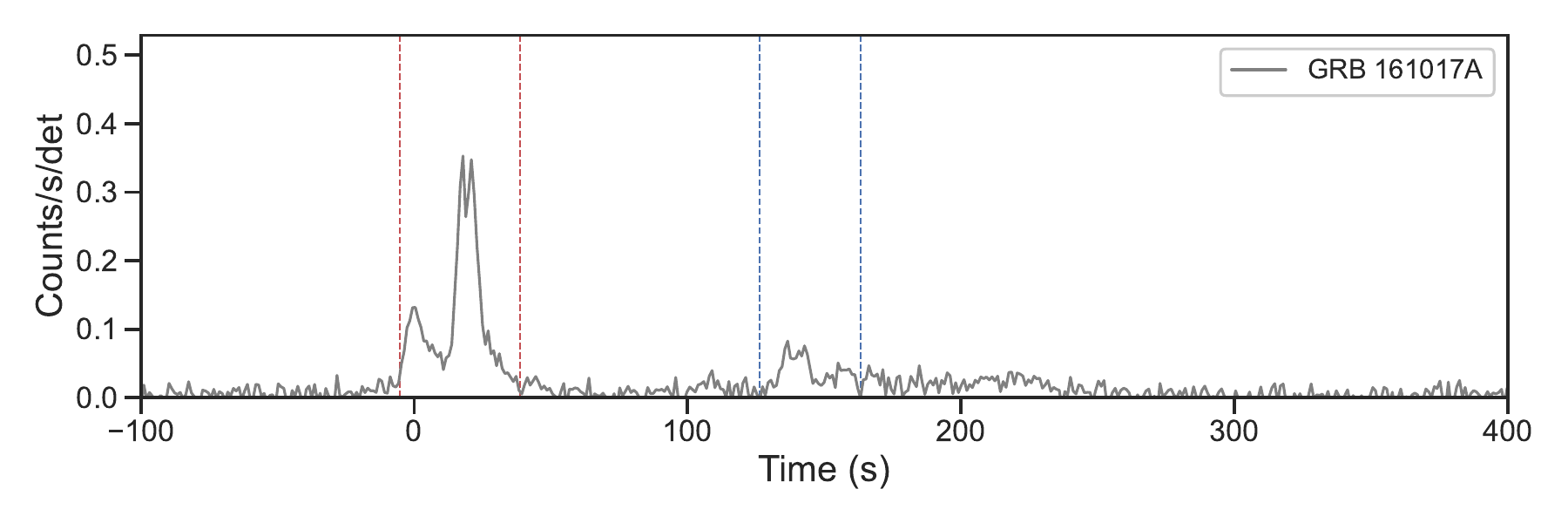}
\includegraphics[angle=0,scale=0.32]{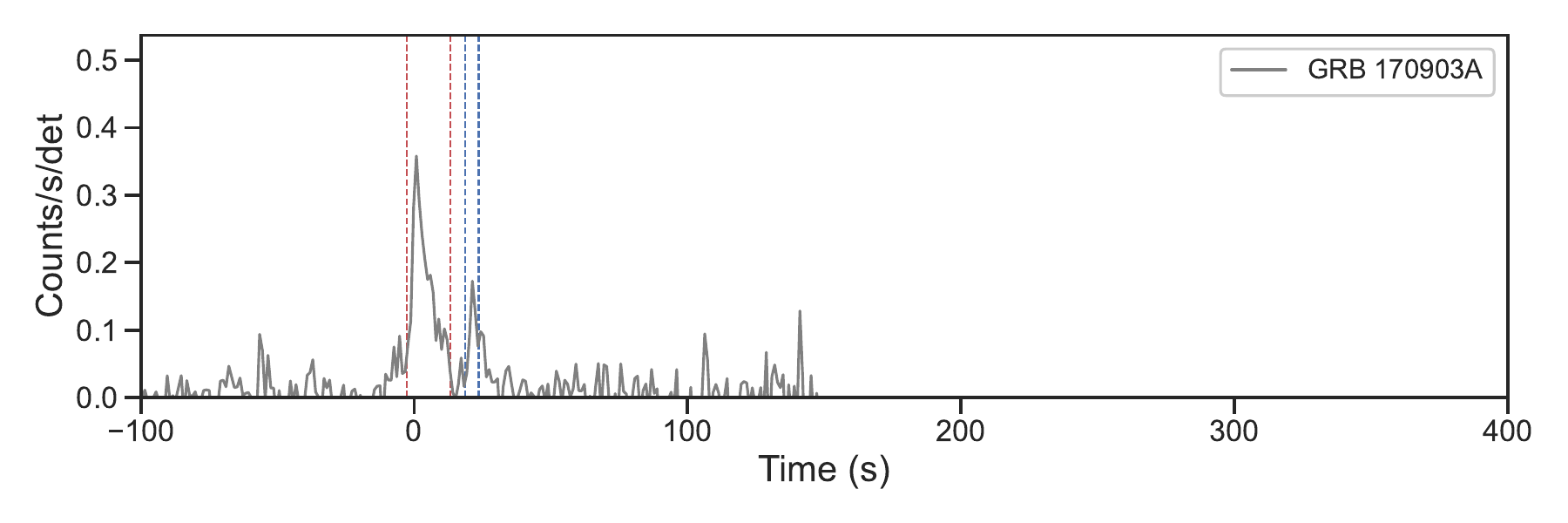}
\includegraphics[angle=0,scale=0.32]{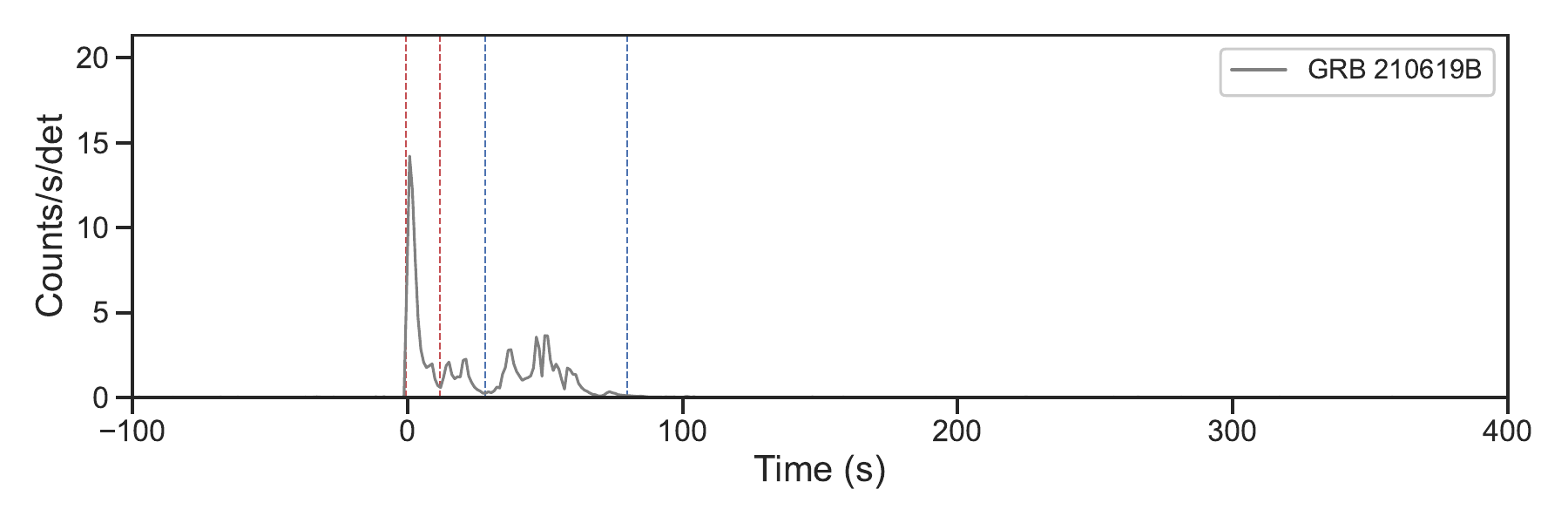}
\includegraphics[angle=0,scale=0.32]{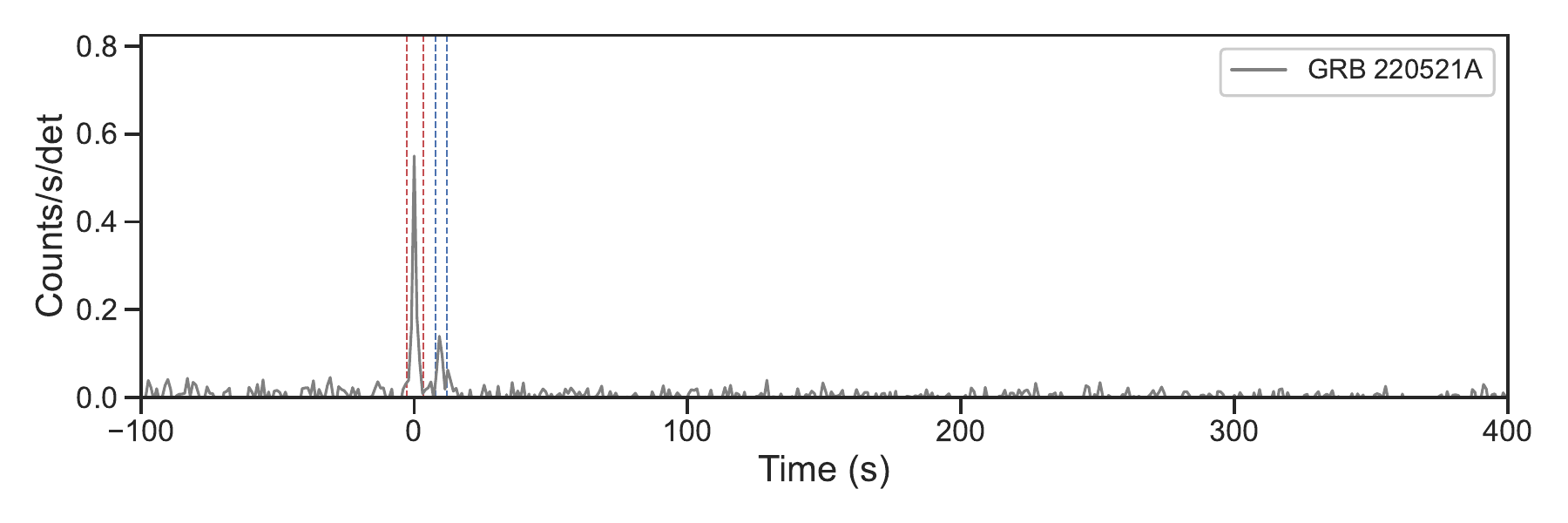}
\center{Figure \ref{fig:lc_sample}--- Continued}
\end{figure*}

\begin{figure*}
\includegraphics[width=0.5\textwidth]{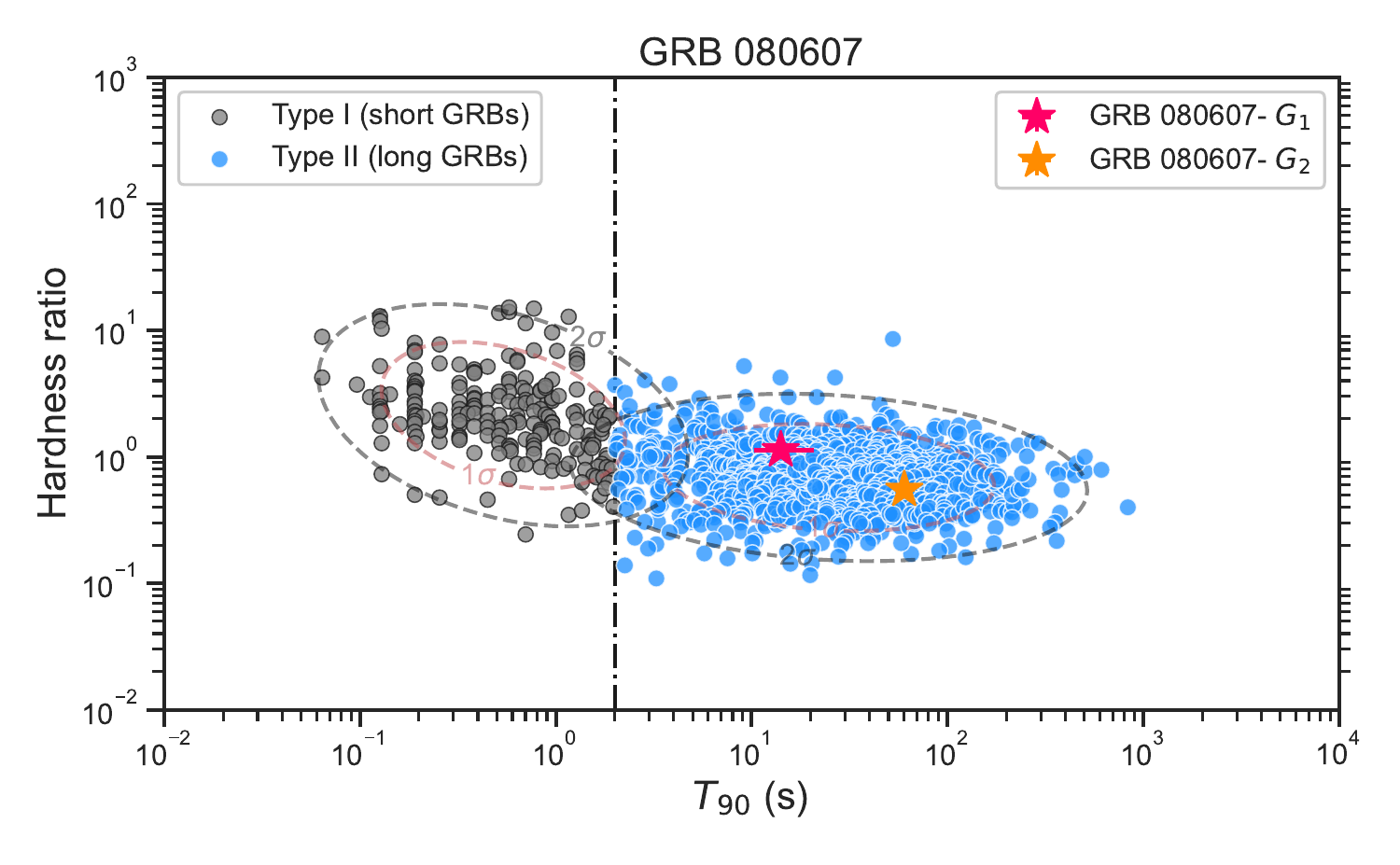}
\includegraphics[width=0.5\textwidth]{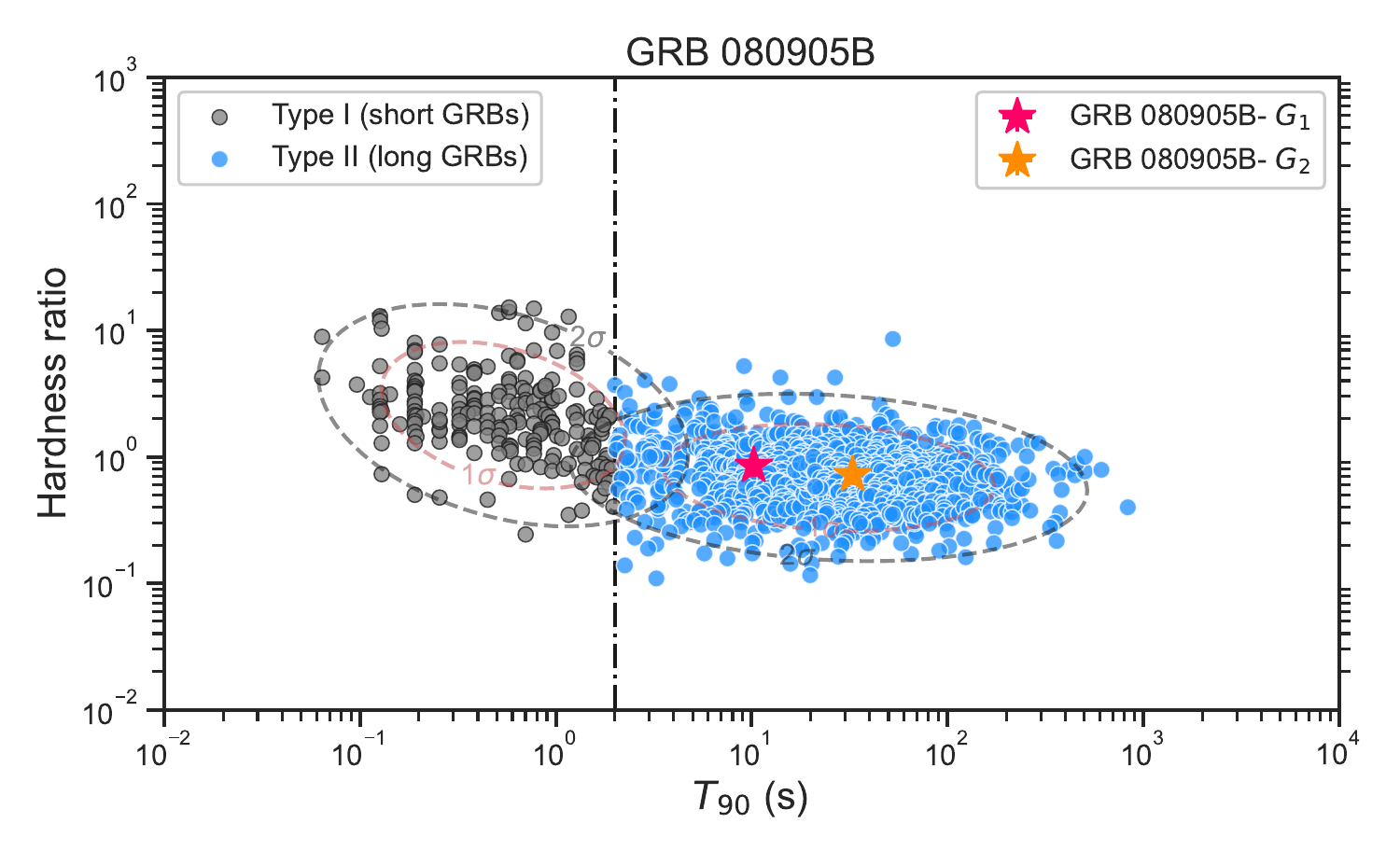}
\includegraphics[width=0.5\textwidth]{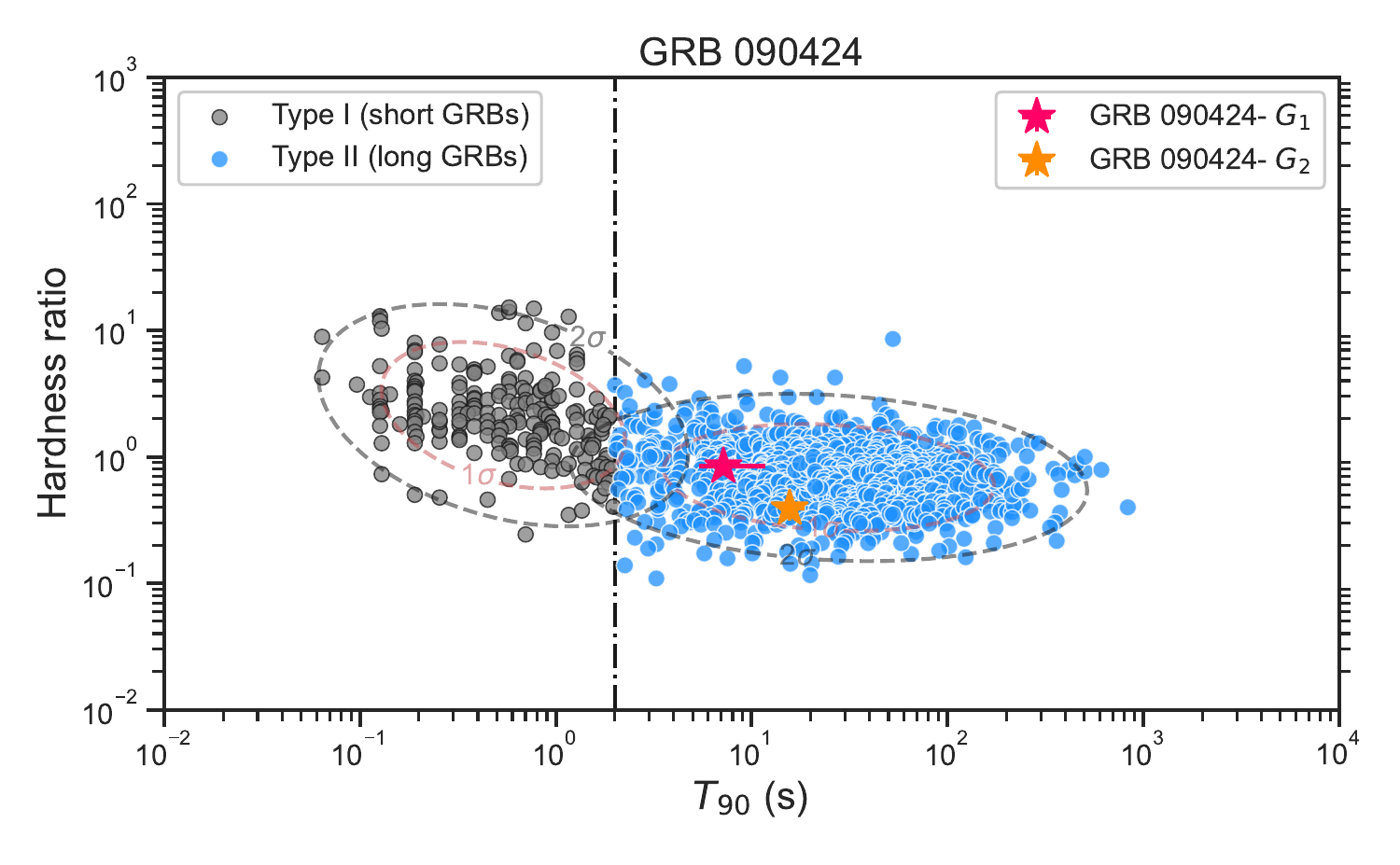}
\includegraphics[width=0.5\textwidth]{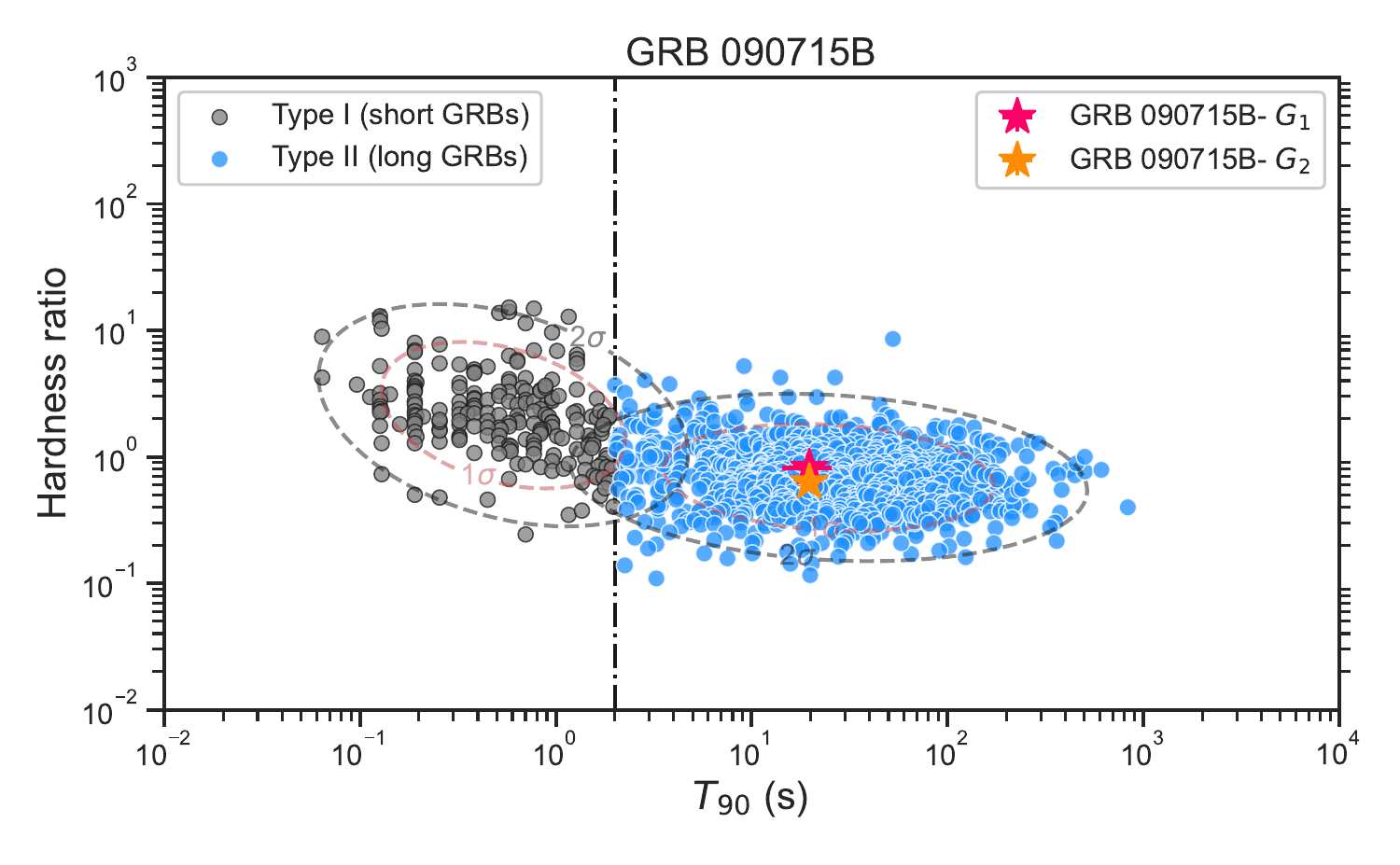}
\includegraphics[width=0.5\textwidth]{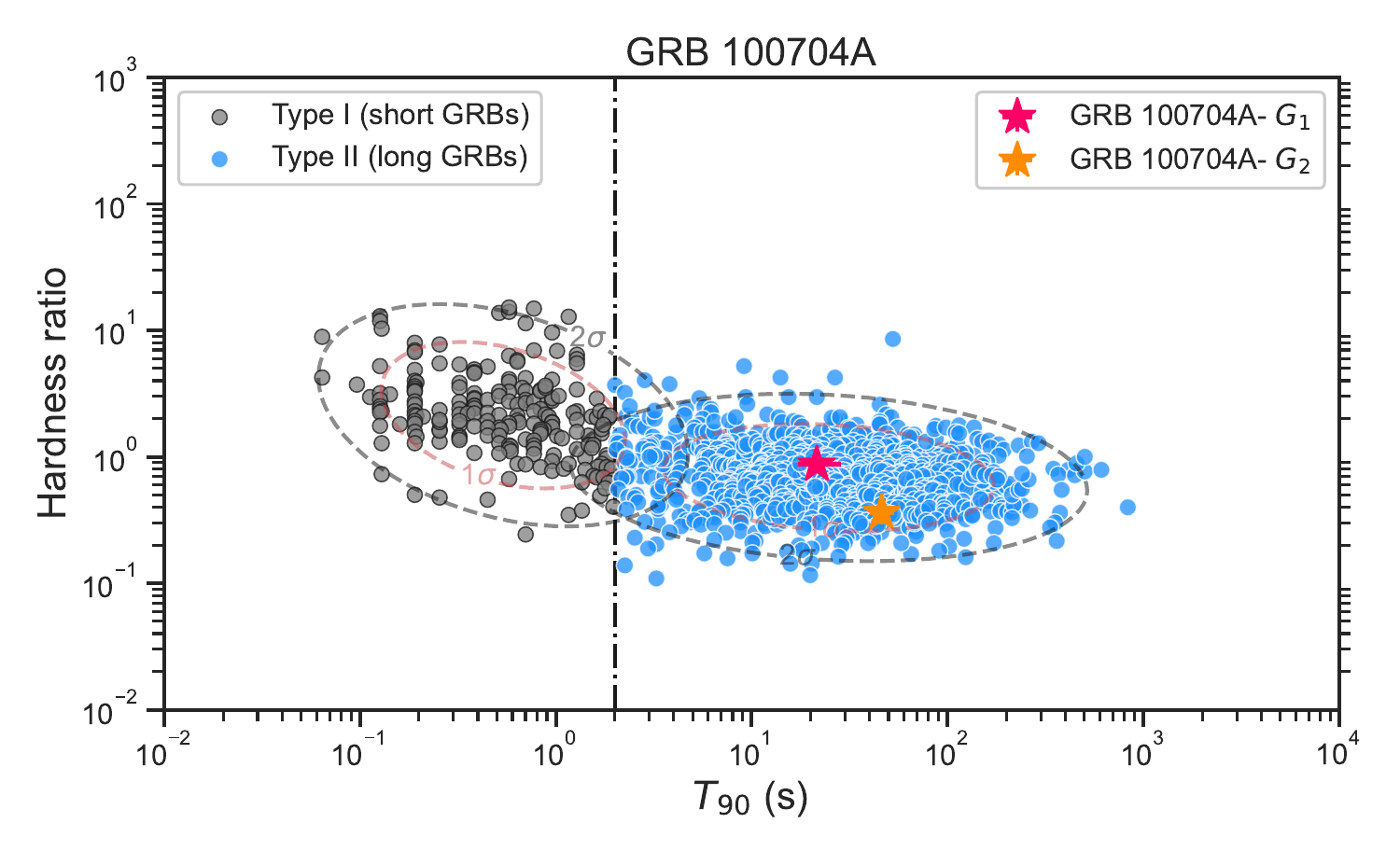}
\includegraphics[width=0.5\textwidth]{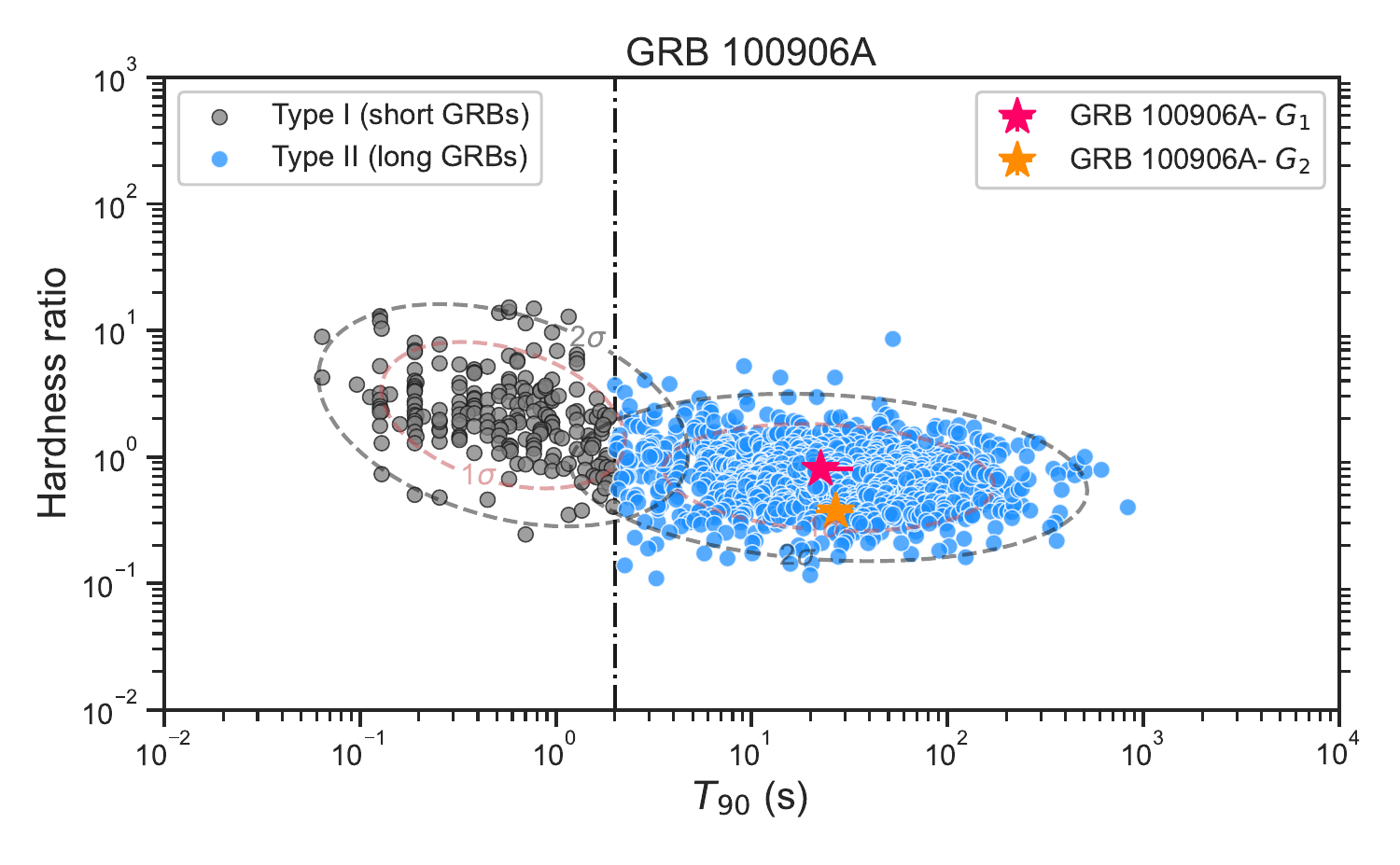}
\center{Figure \ref{fig:HR_T90}--- Continued}
\end{figure*}
\begin{figure*}
\includegraphics[width=0.5\textwidth]{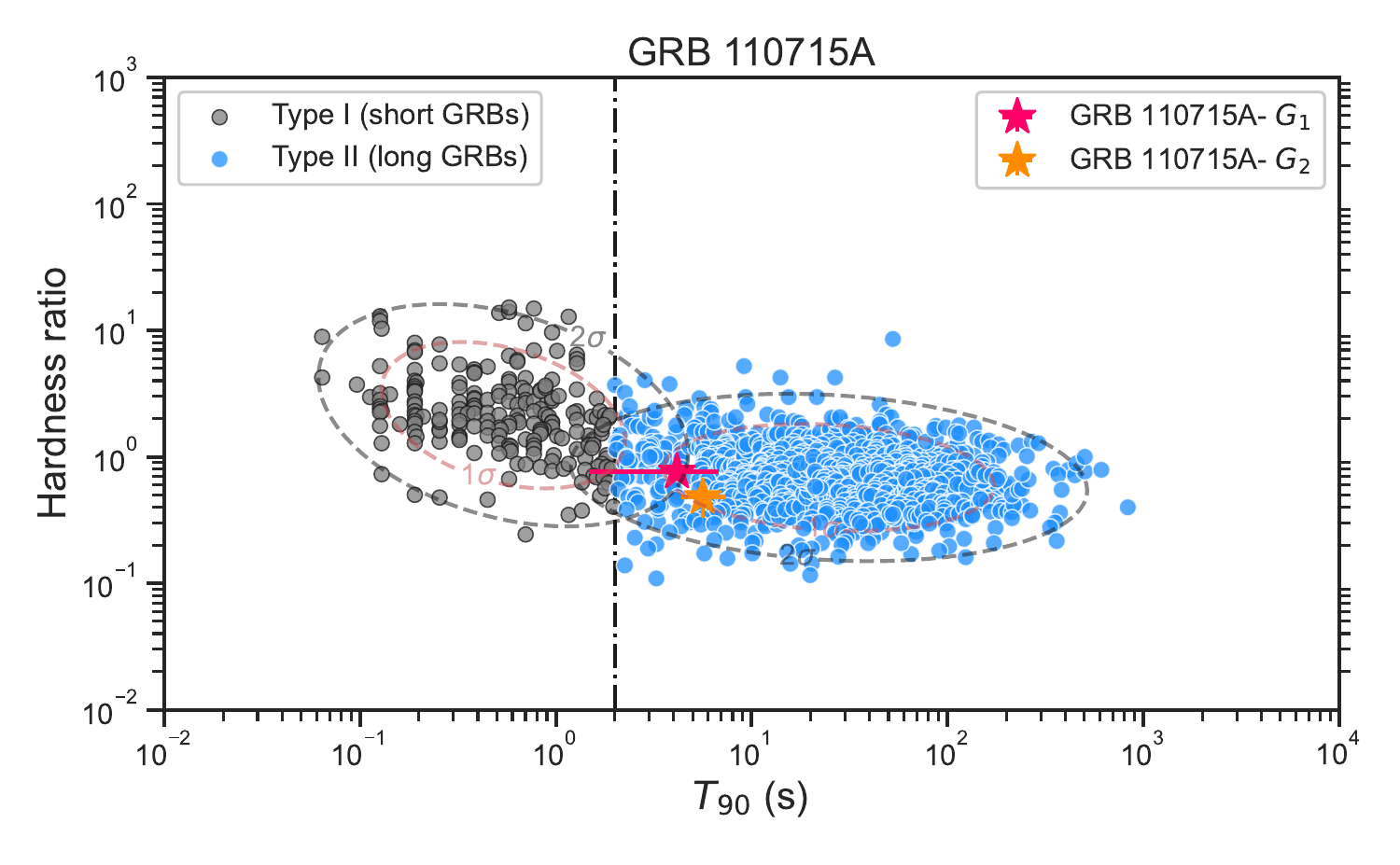}
\includegraphics[width=0.5\textwidth]{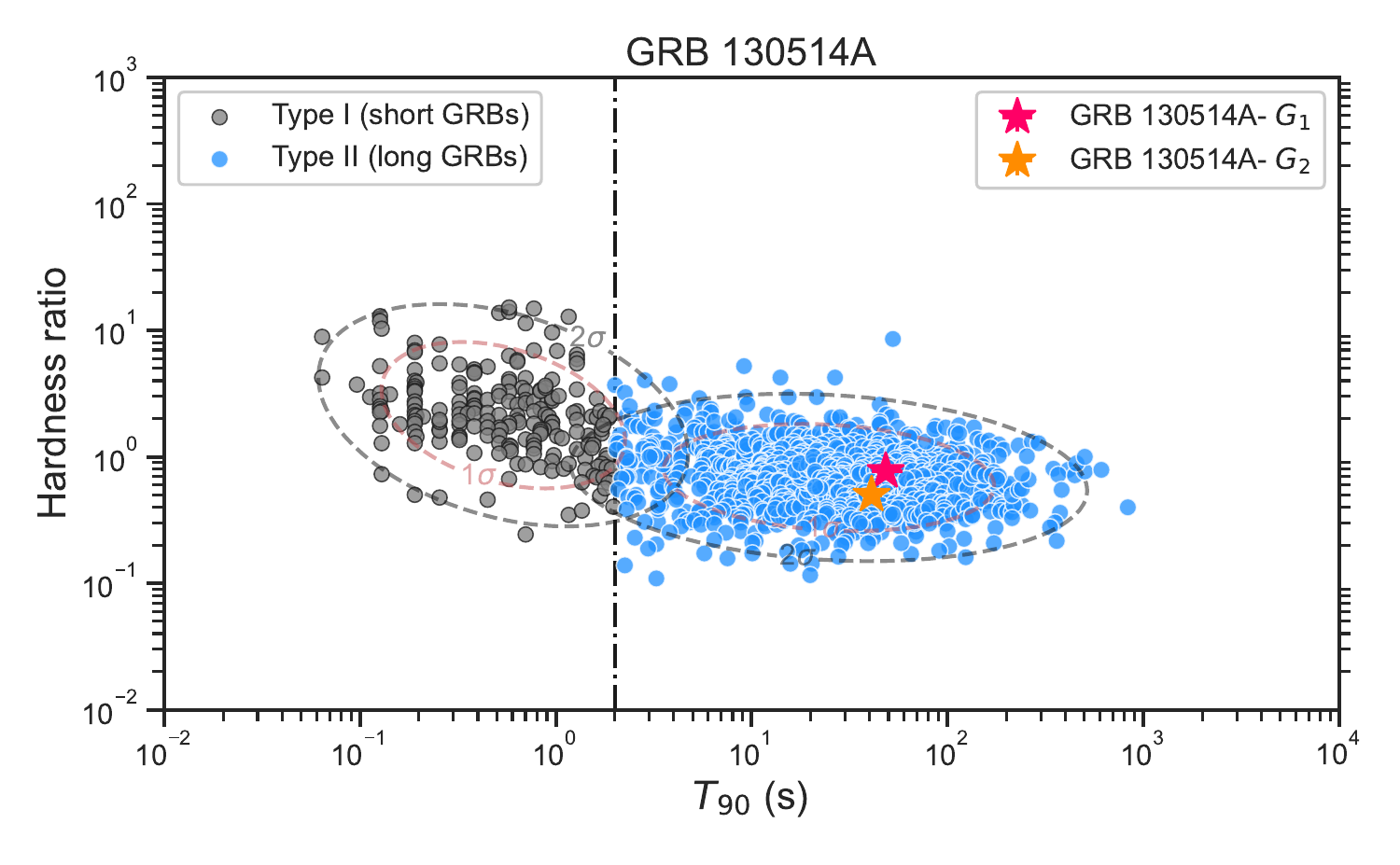}
\includegraphics[width=0.5\textwidth]{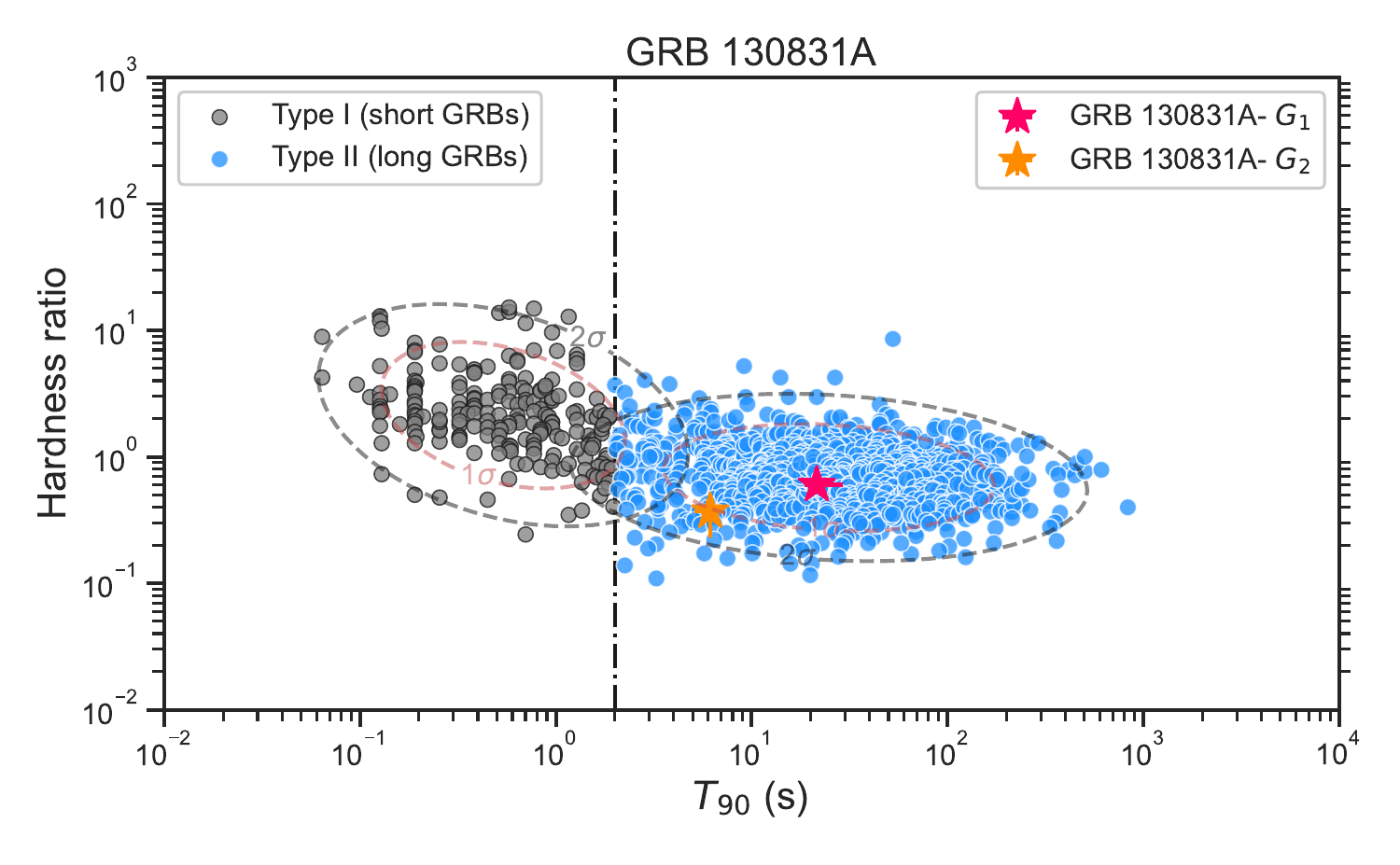}
\includegraphics[width=0.5\textwidth]{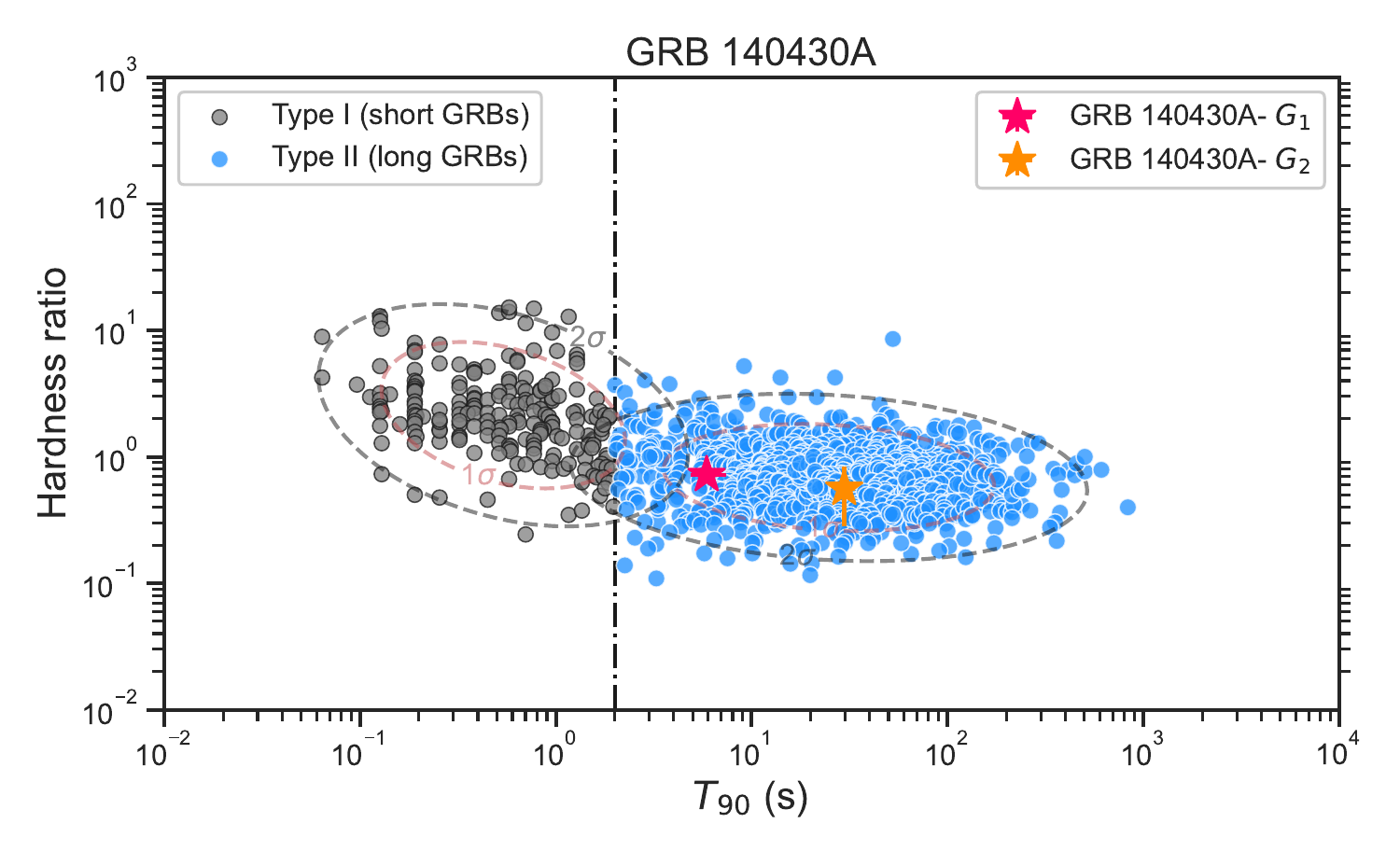}
\includegraphics[width=0.5\textwidth]{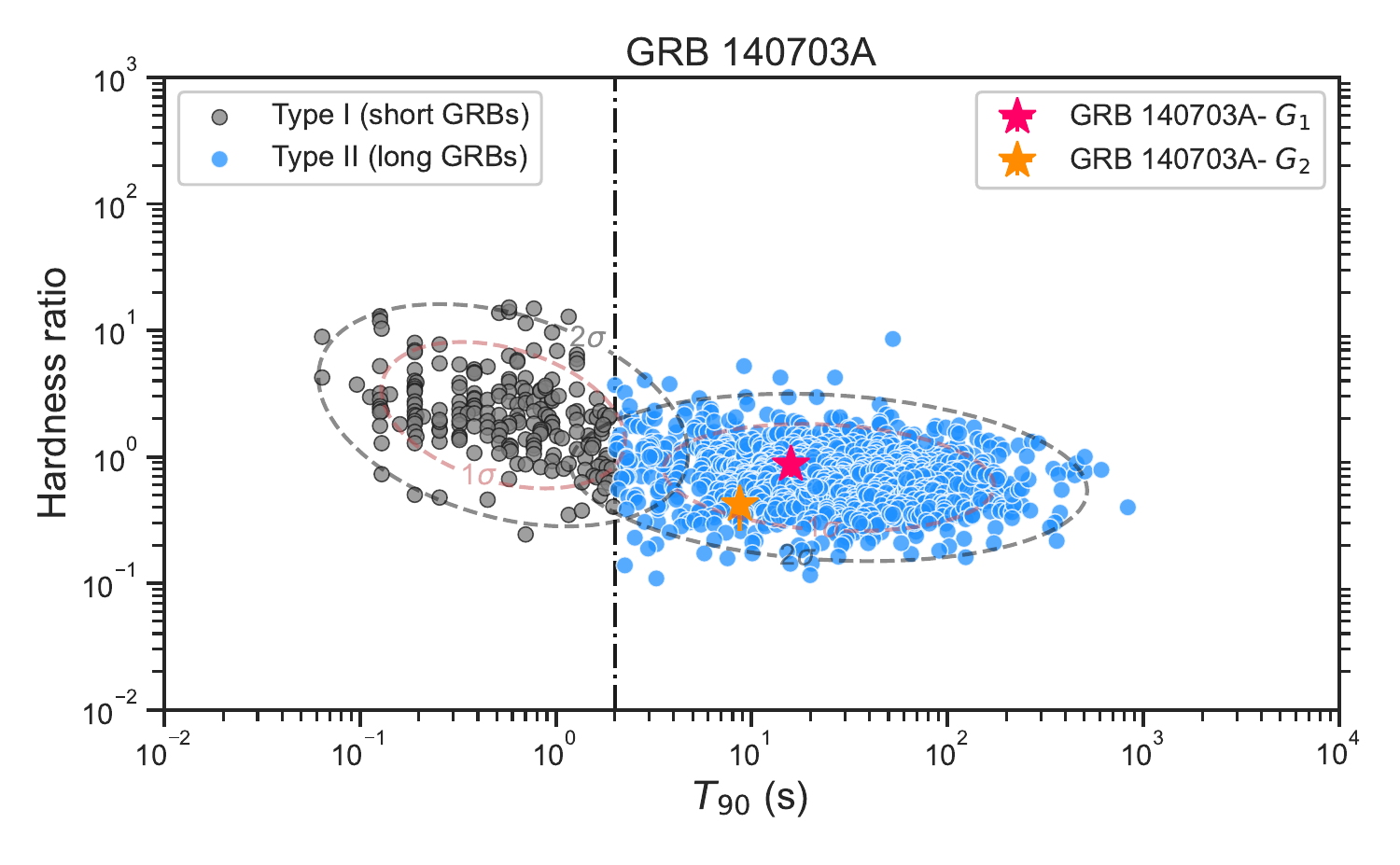}
\includegraphics[width=0.5\textwidth]{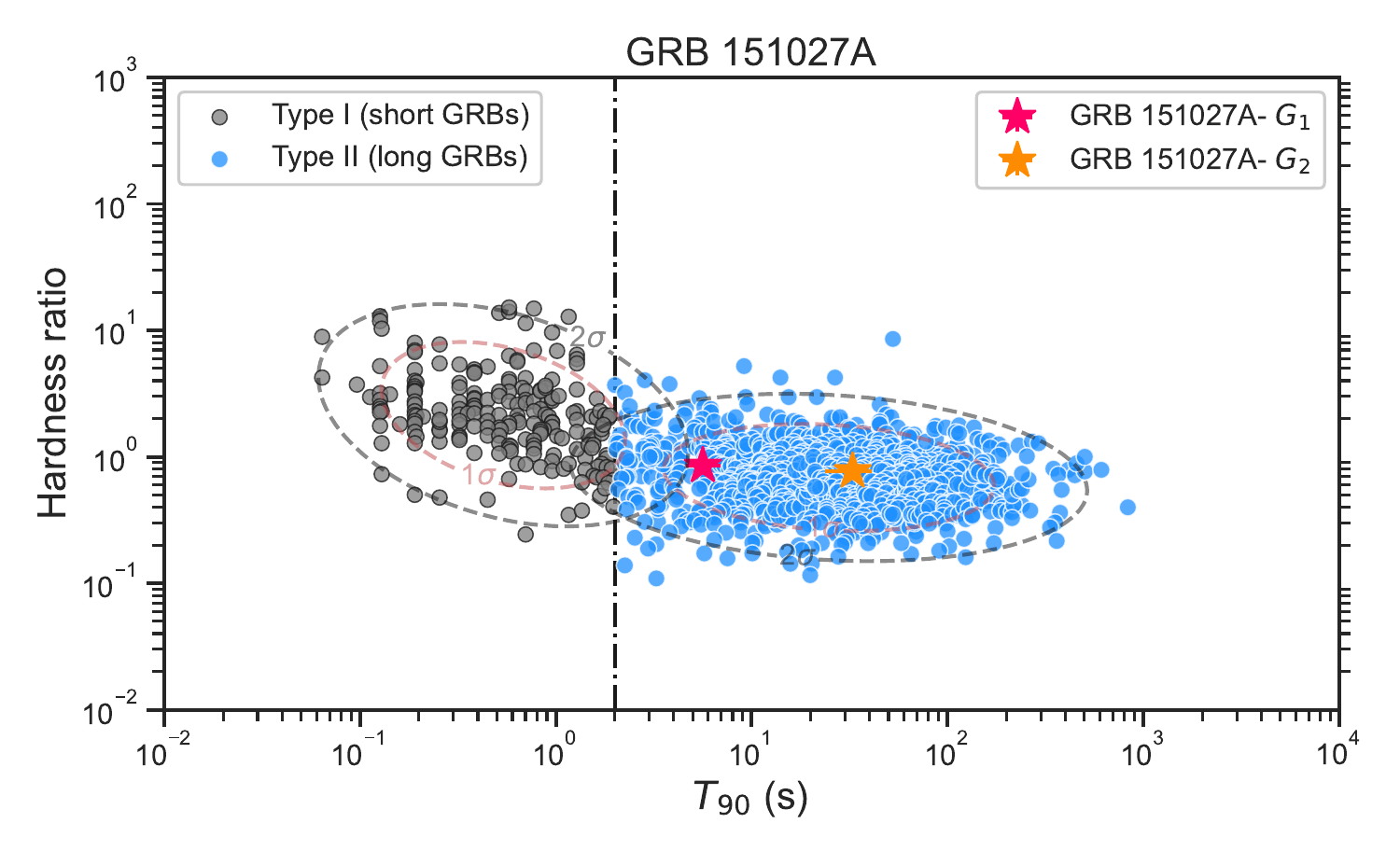}
\center{Figure \ref{fig:HR_T90}--- Continued}
\end{figure*}
\begin{figure*}
\includegraphics[width=0.5\textwidth]{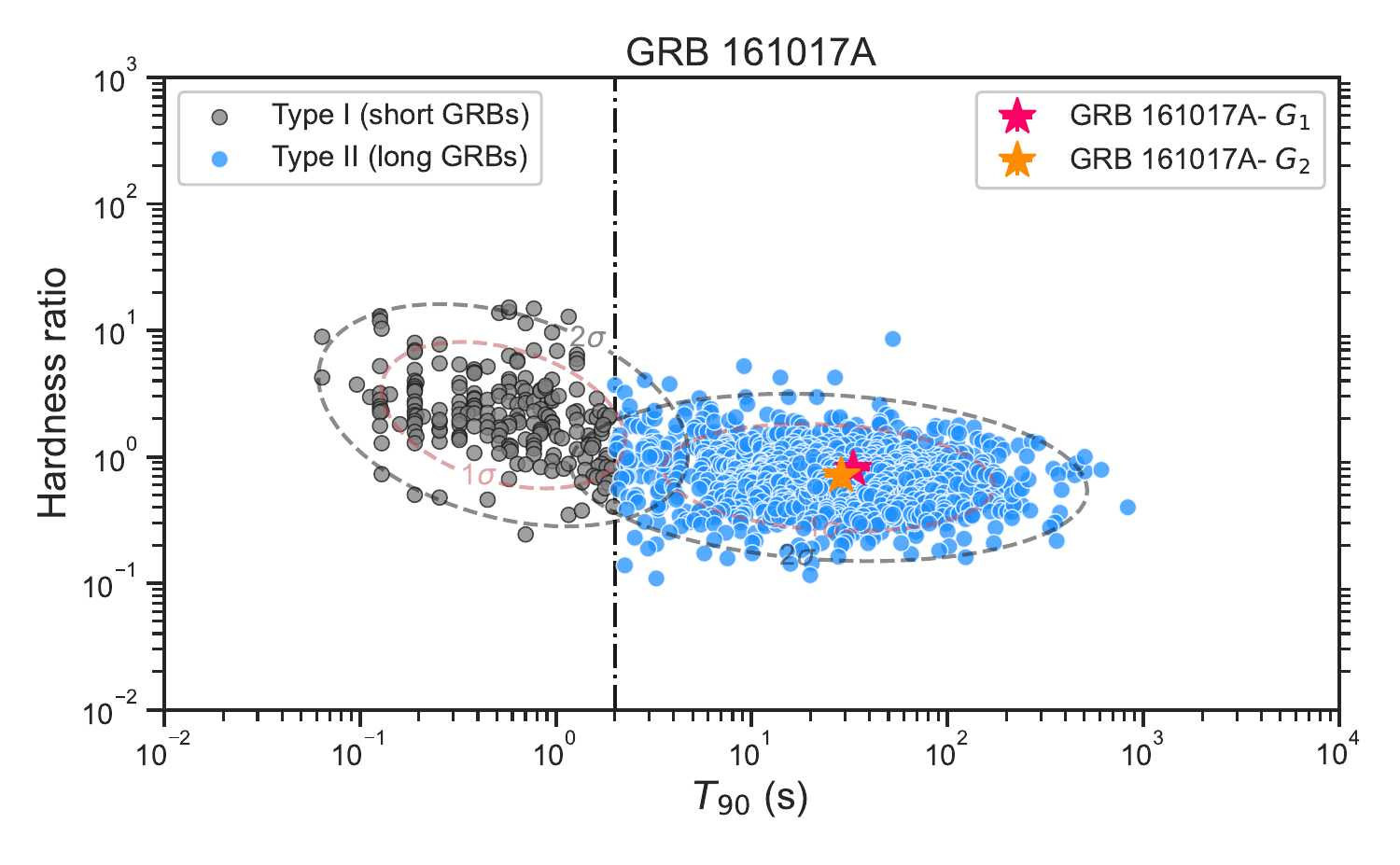}
\includegraphics[width=0.5\textwidth]{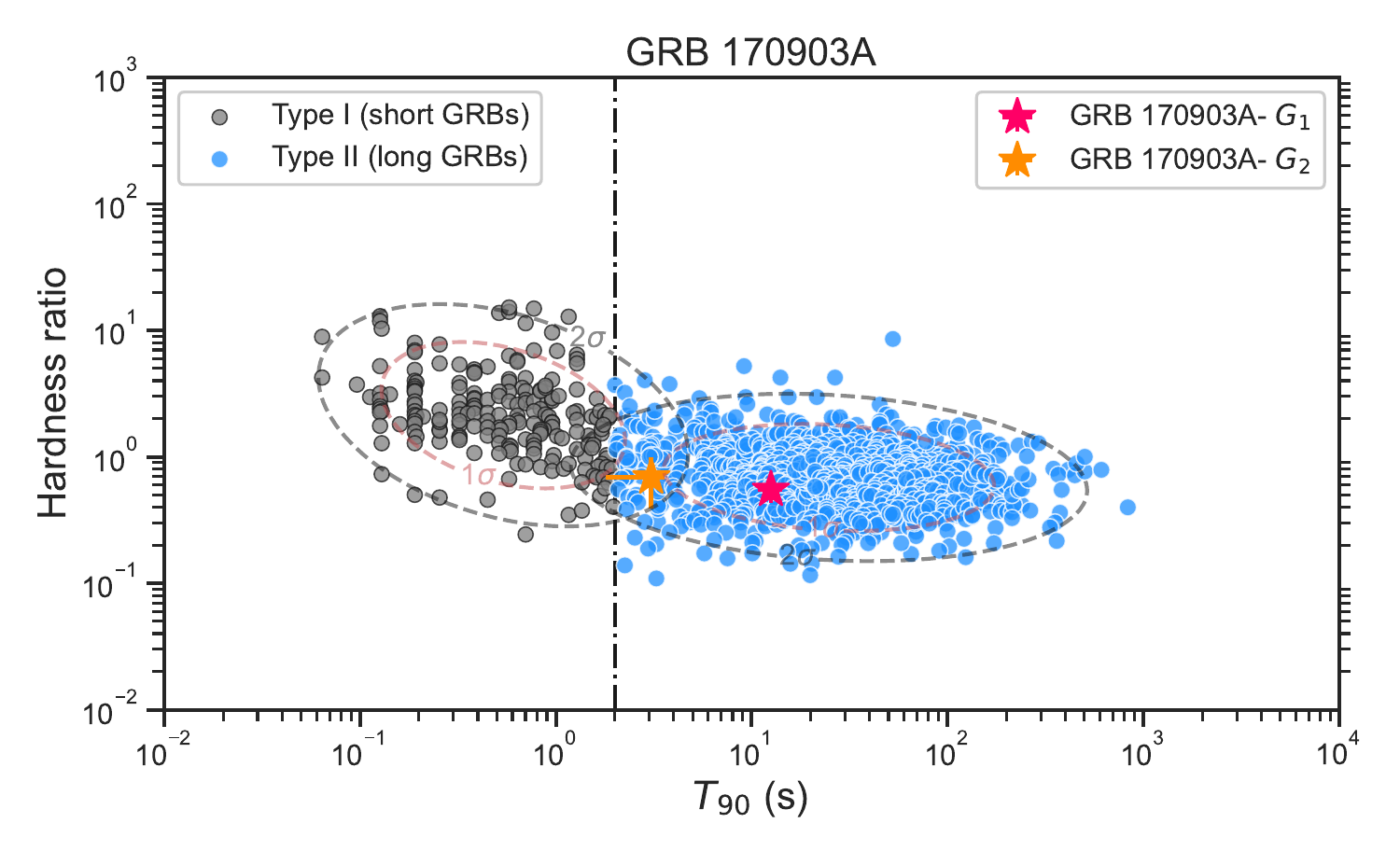}
\includegraphics[width=0.5\textwidth]{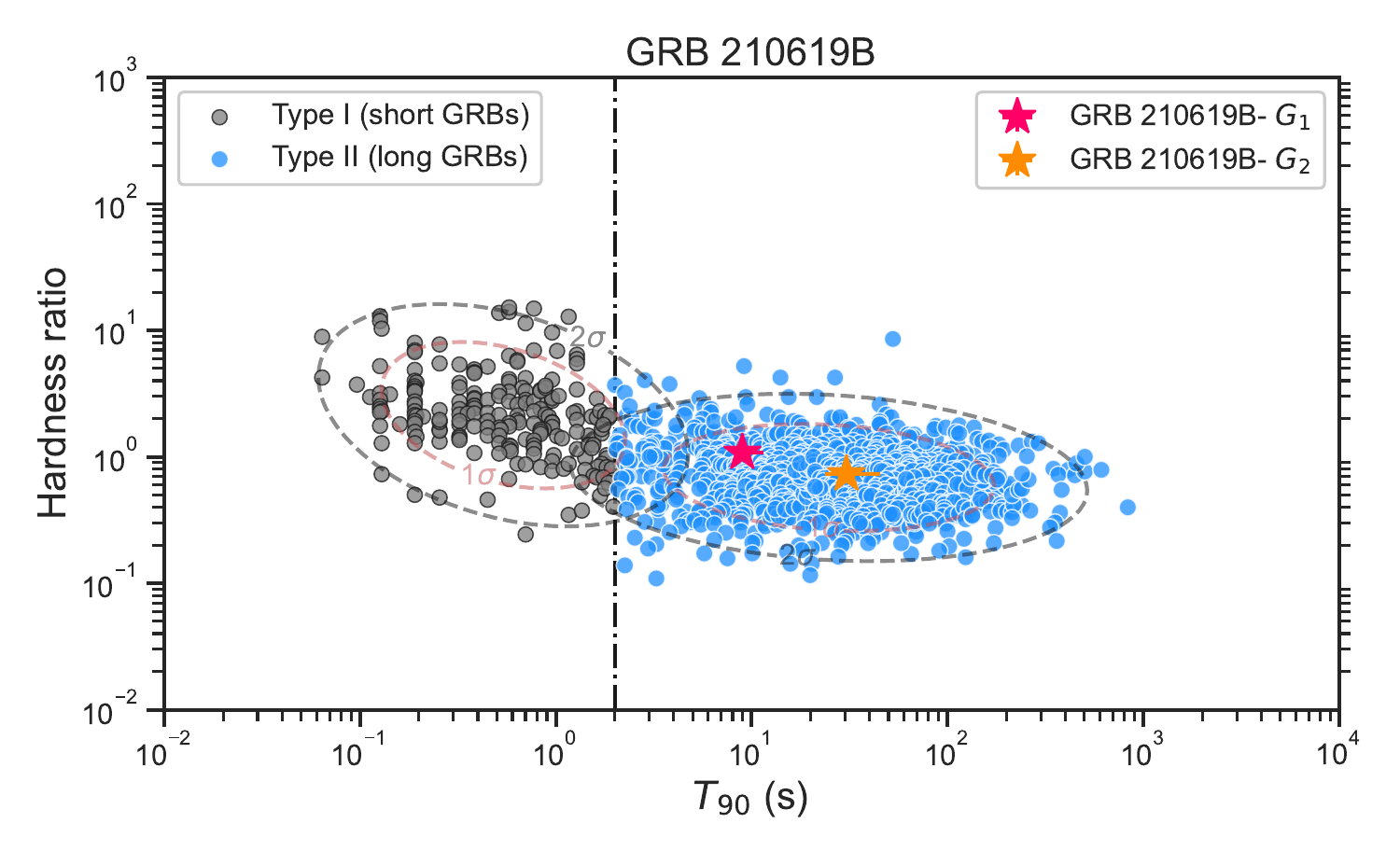}
\includegraphics[width=0.5\textwidth]{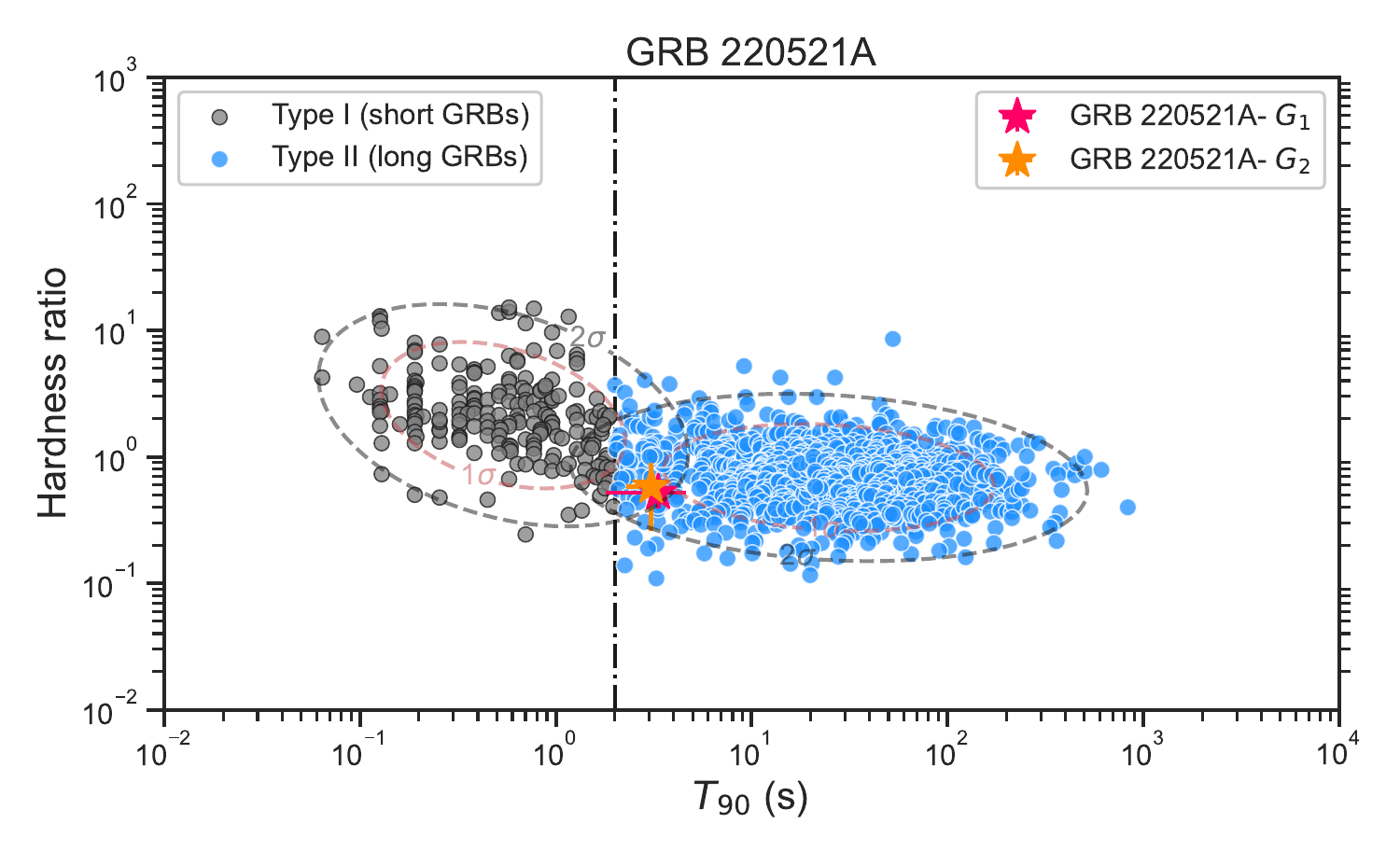}
\center{Figure \ref{fig:HR_T90}--- Continued}
\end{figure*}

\begin{figure*}
\includegraphics[width=0.5\textwidth]{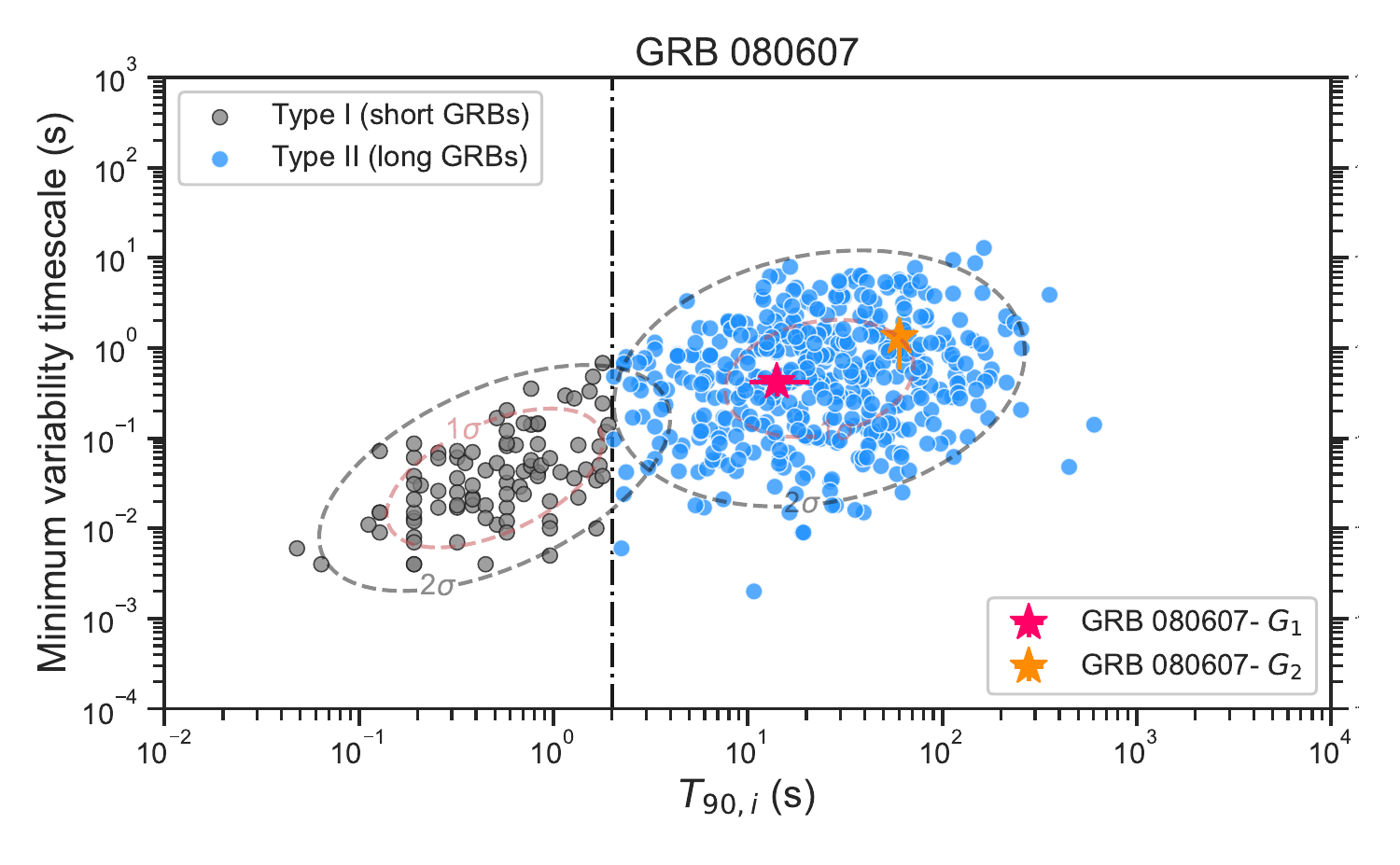}
\includegraphics[width=0.5\textwidth]{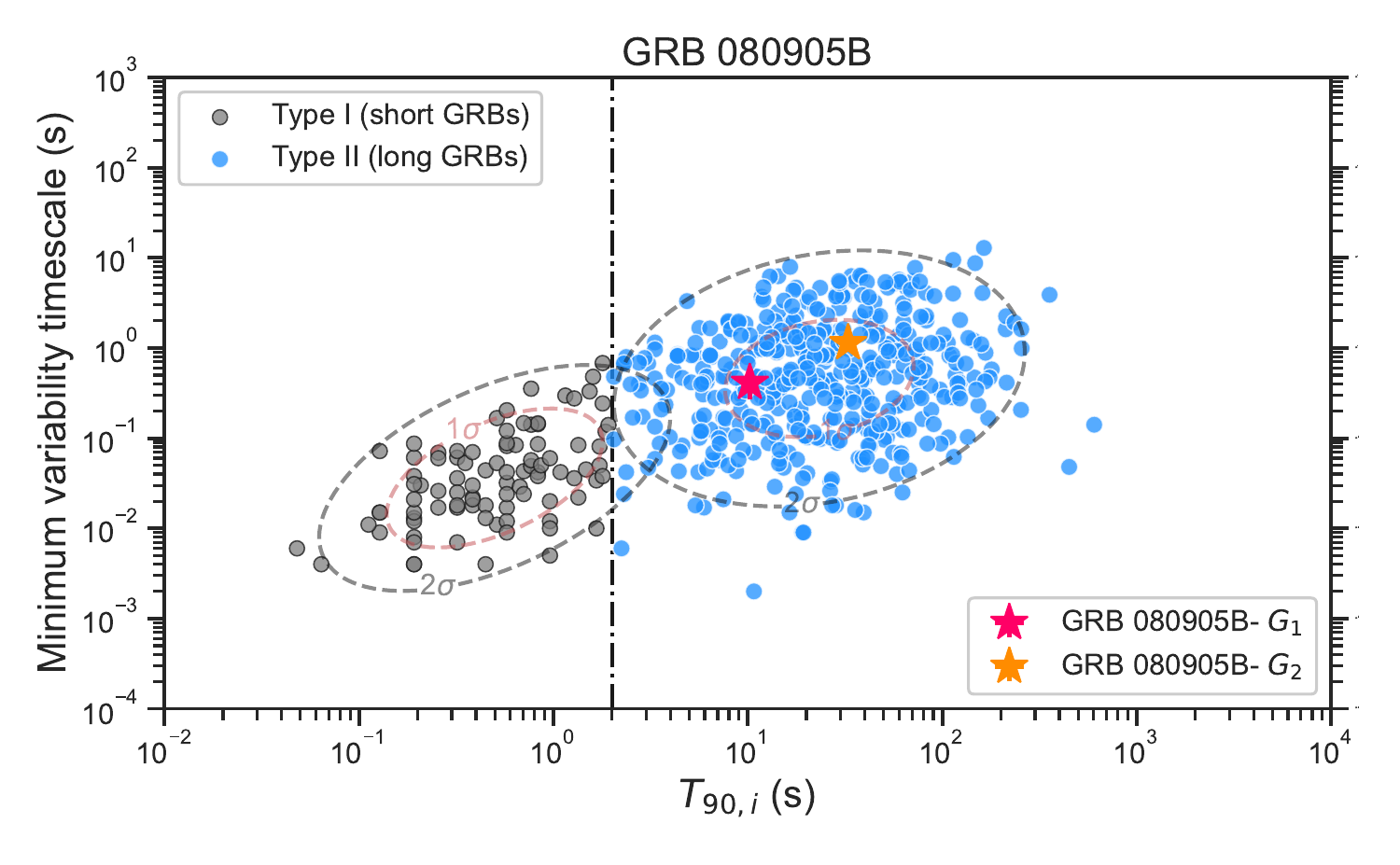}
\includegraphics[width=0.5\textwidth]{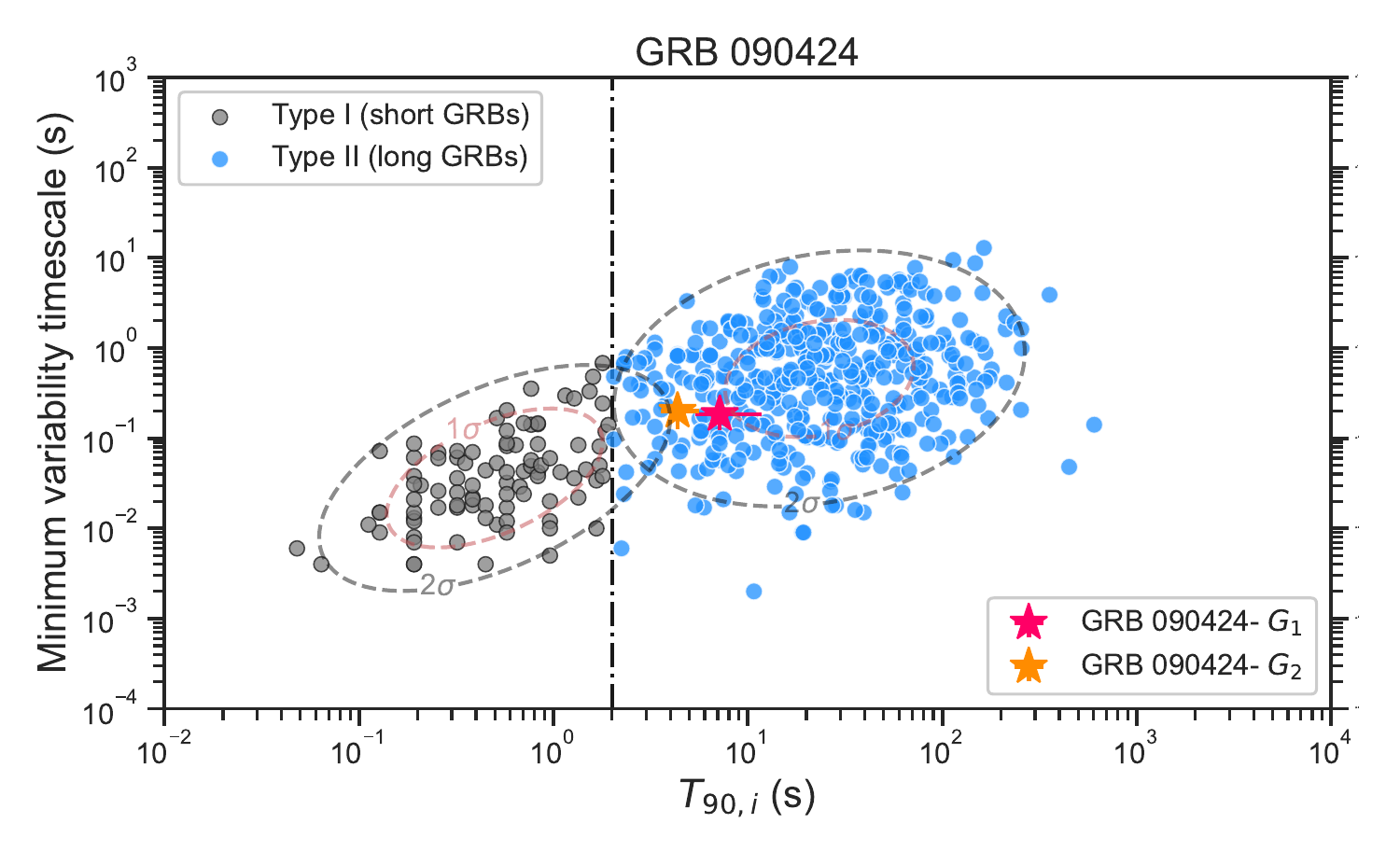}
\includegraphics[width=0.5\textwidth]{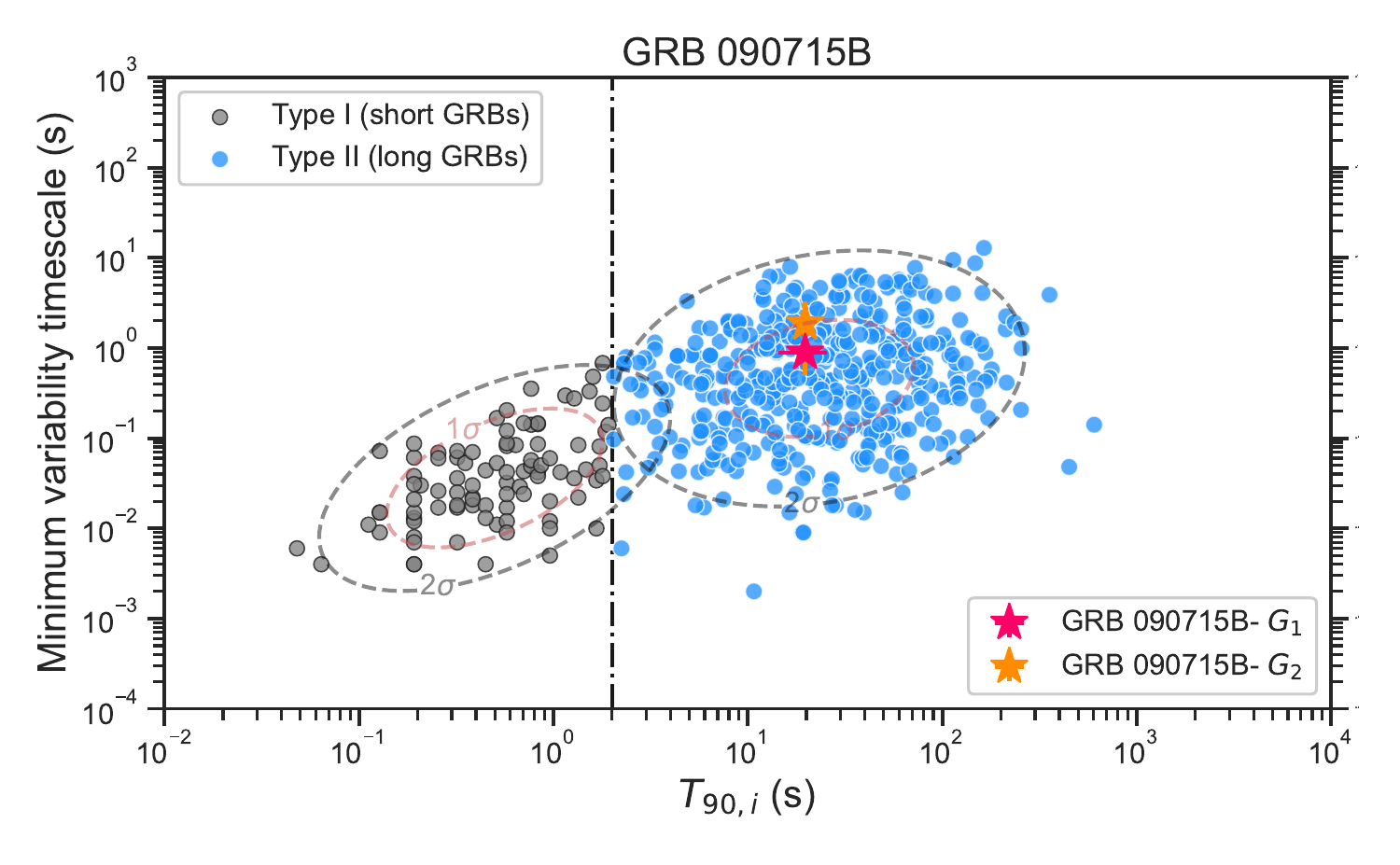}
\includegraphics[width=0.5\textwidth]{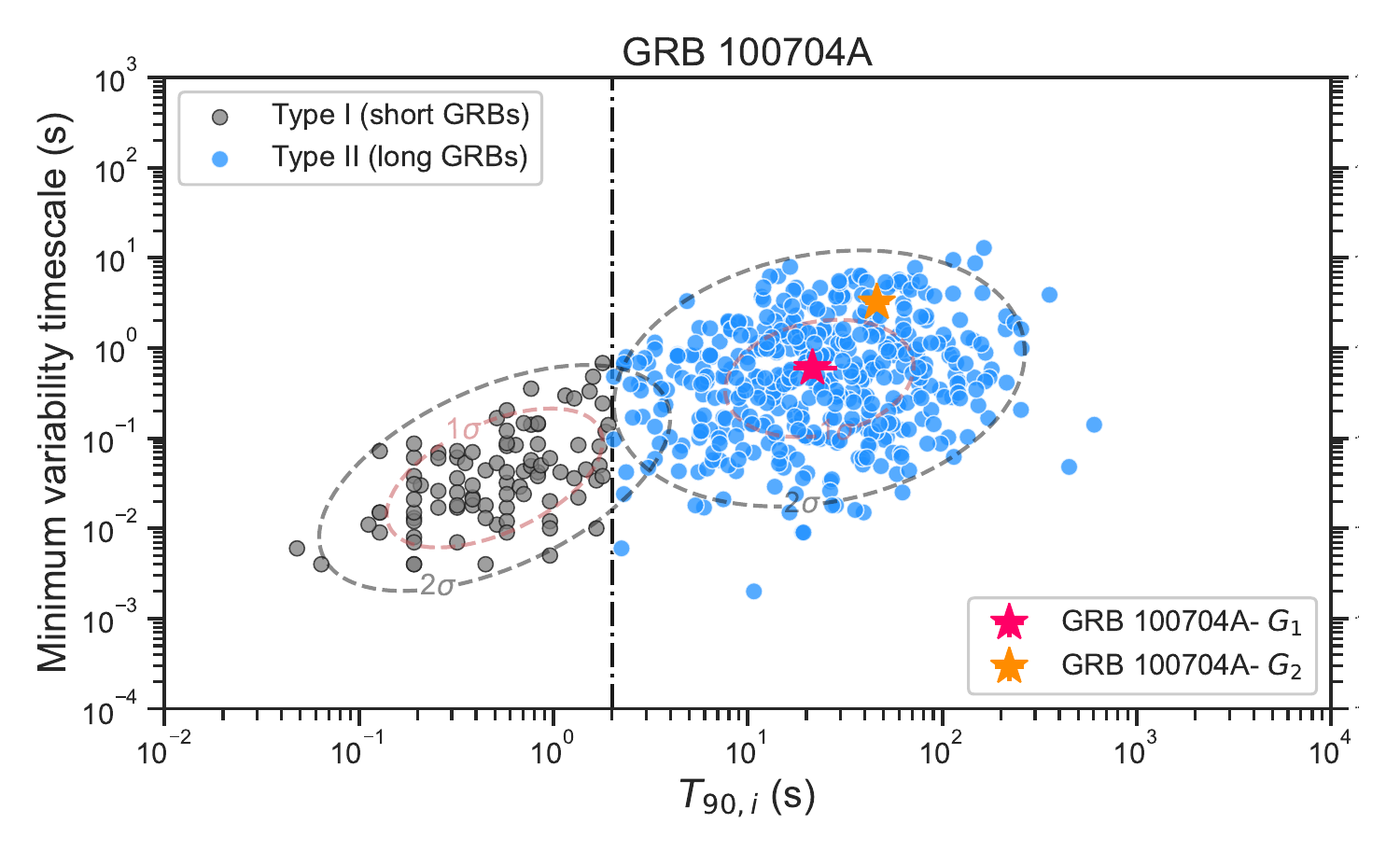}
\includegraphics[width=0.5\textwidth]{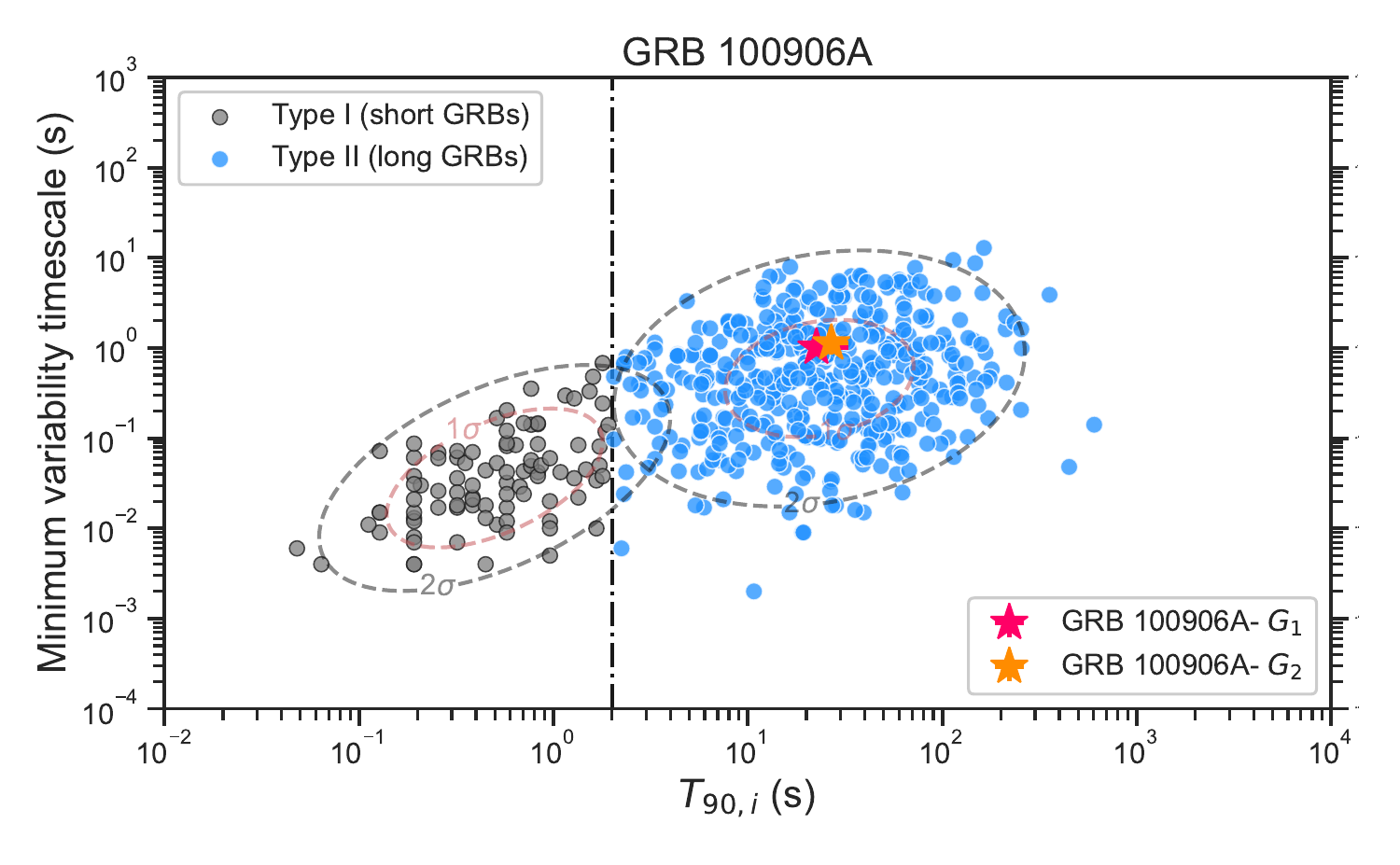}
\center{Figure \ref{fig:MVT_T90}--- Continued}
\end{figure*}
\begin{figure*}
\includegraphics[width=0.5\textwidth]{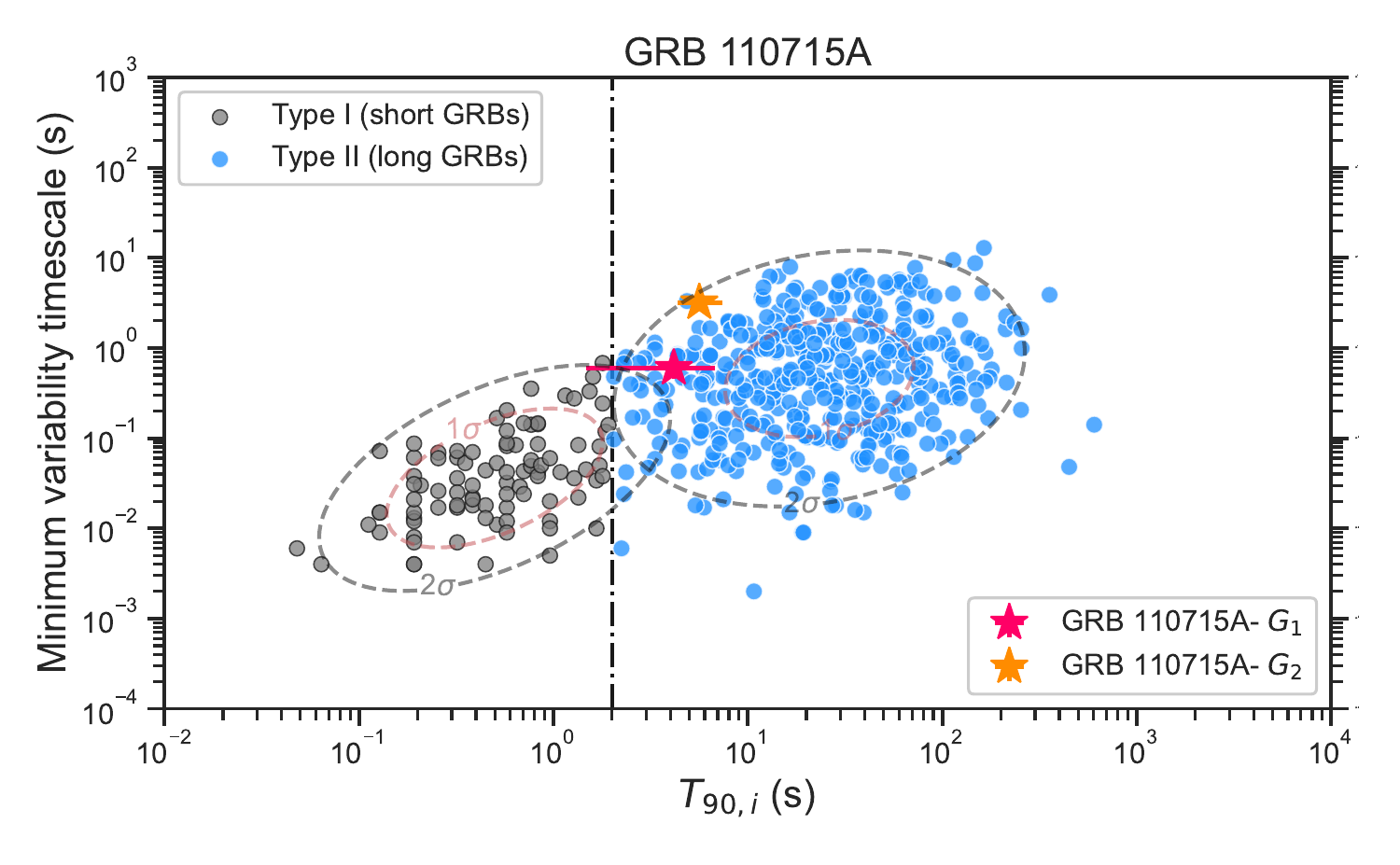}
\includegraphics[width=0.5\textwidth]{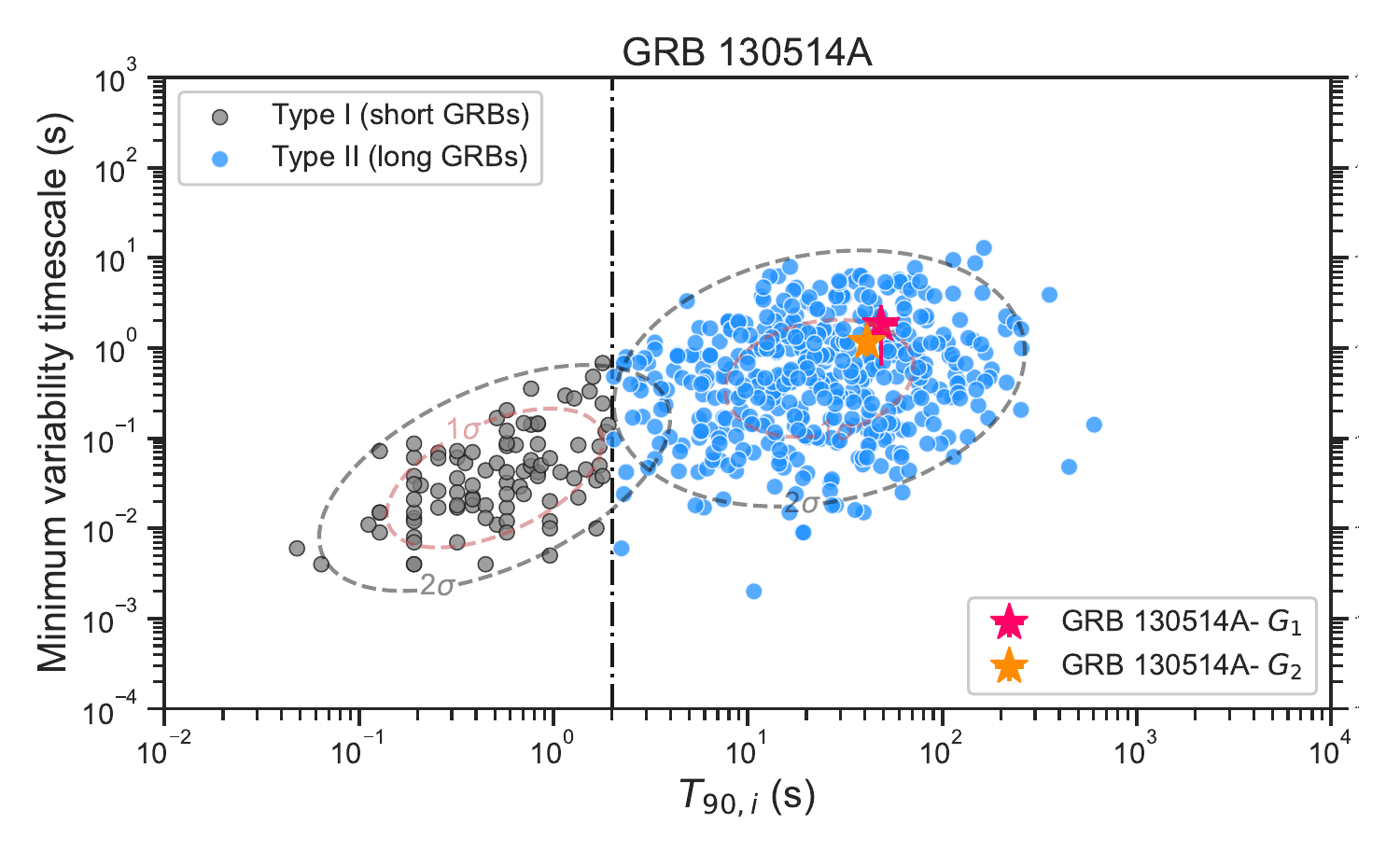}
\includegraphics[width=0.5\textwidth]{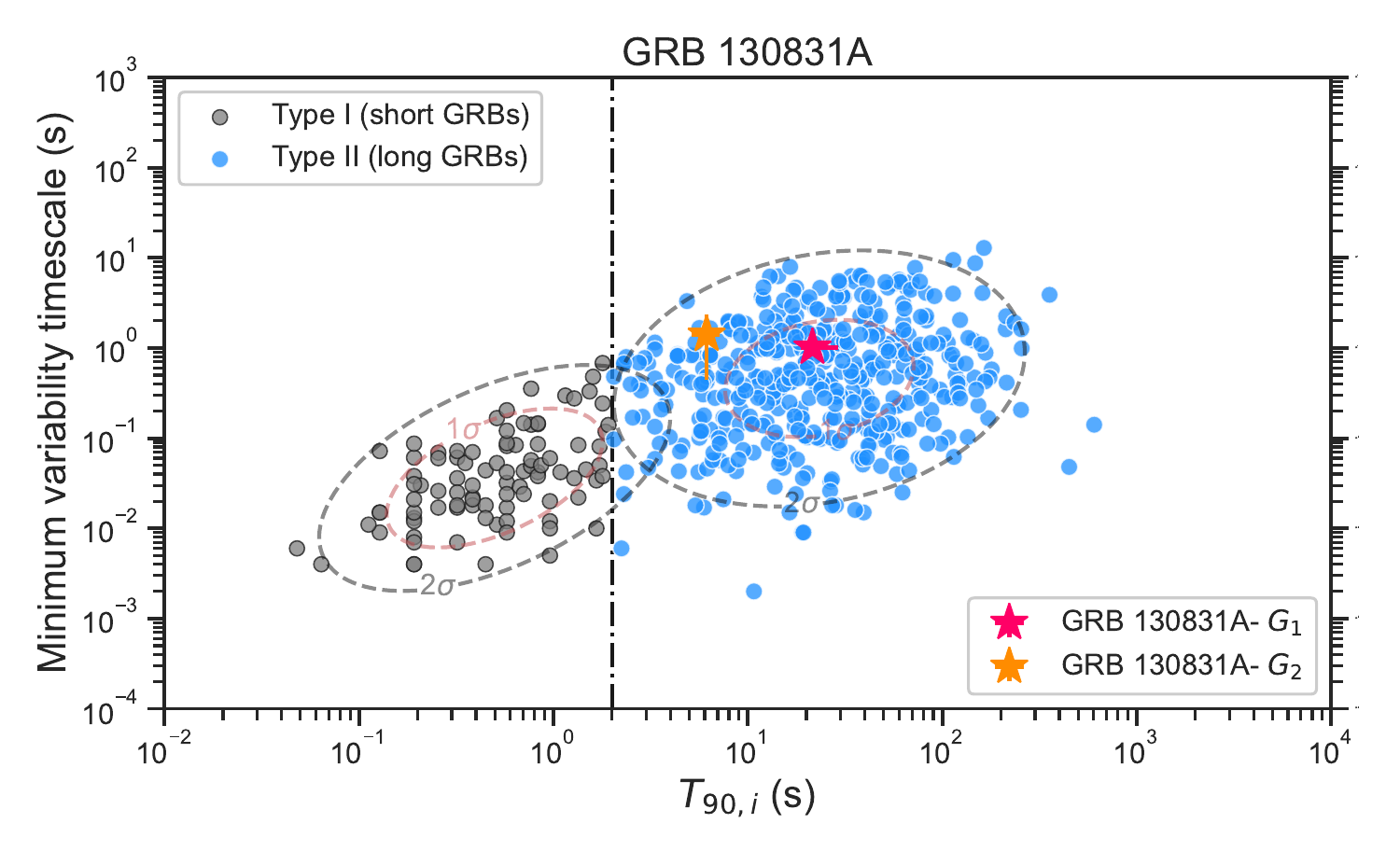}
\includegraphics[width=0.5\textwidth]{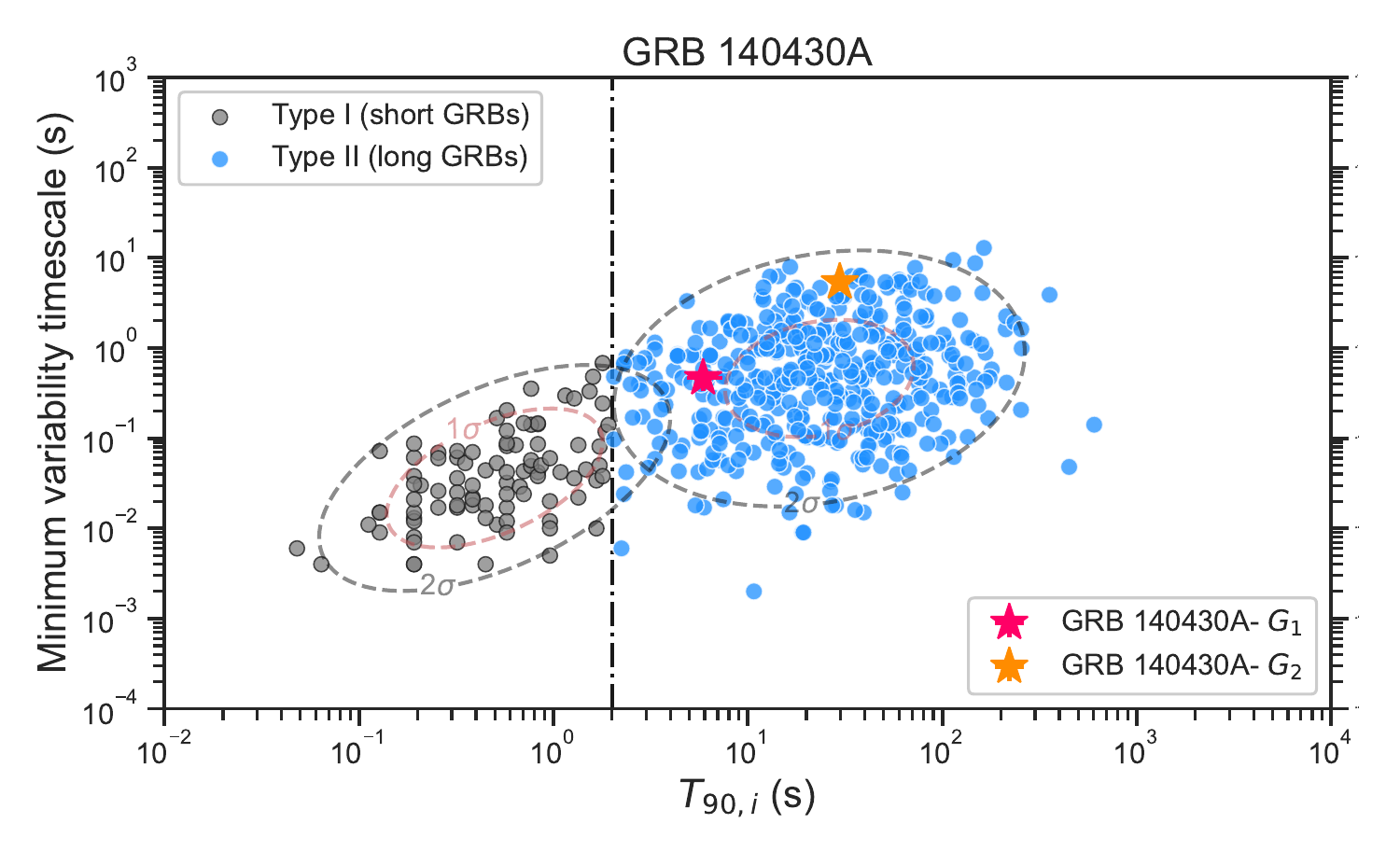}
\includegraphics[width=0.5\textwidth]{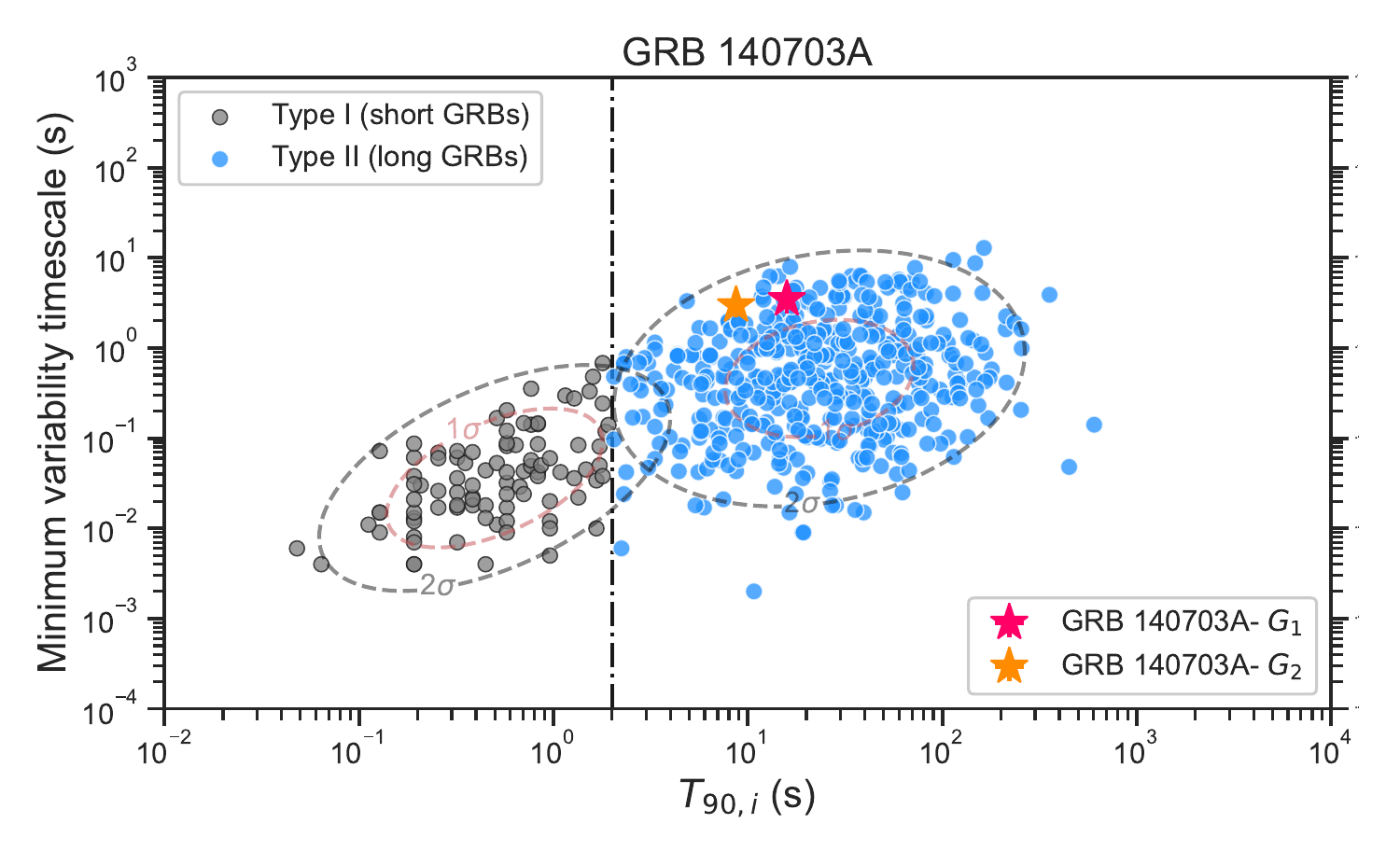}
\includegraphics[width=0.5\textwidth]{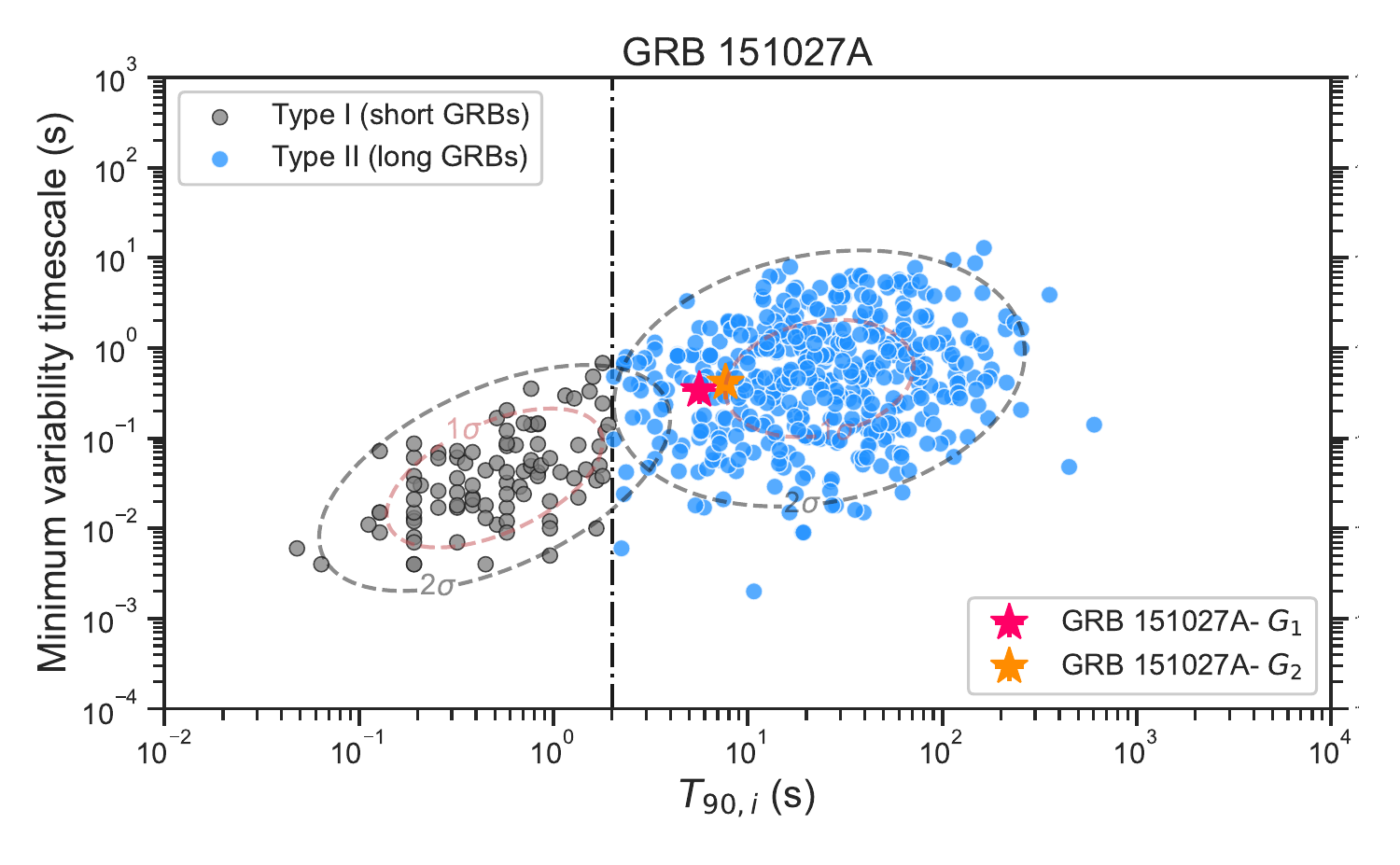}
\center{Figure \ref{fig:MVT_T90}--- Continued}
\end{figure*}
\begin{figure*}
\includegraphics[width=0.5\textwidth]{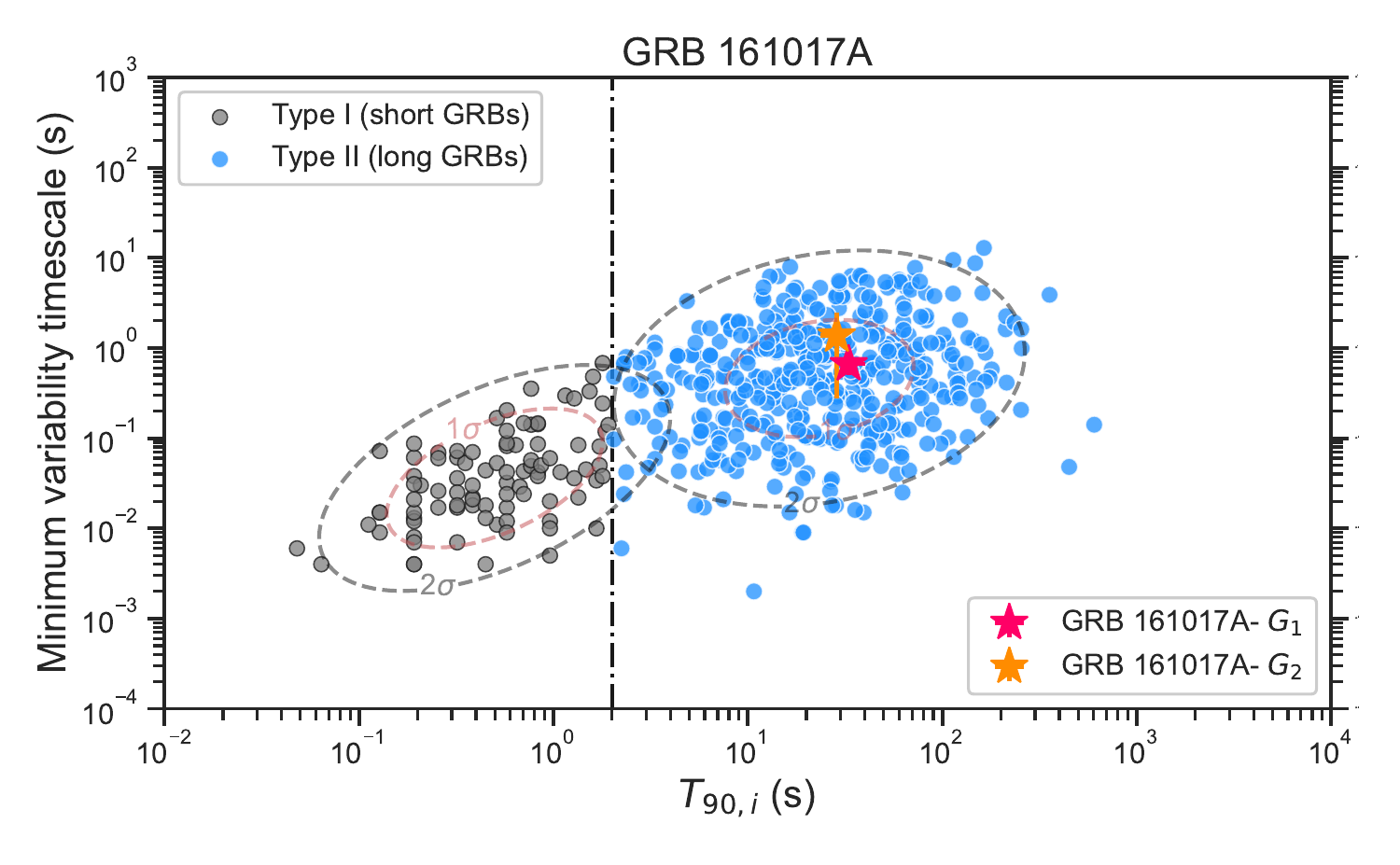}
\includegraphics[width=0.5\textwidth]{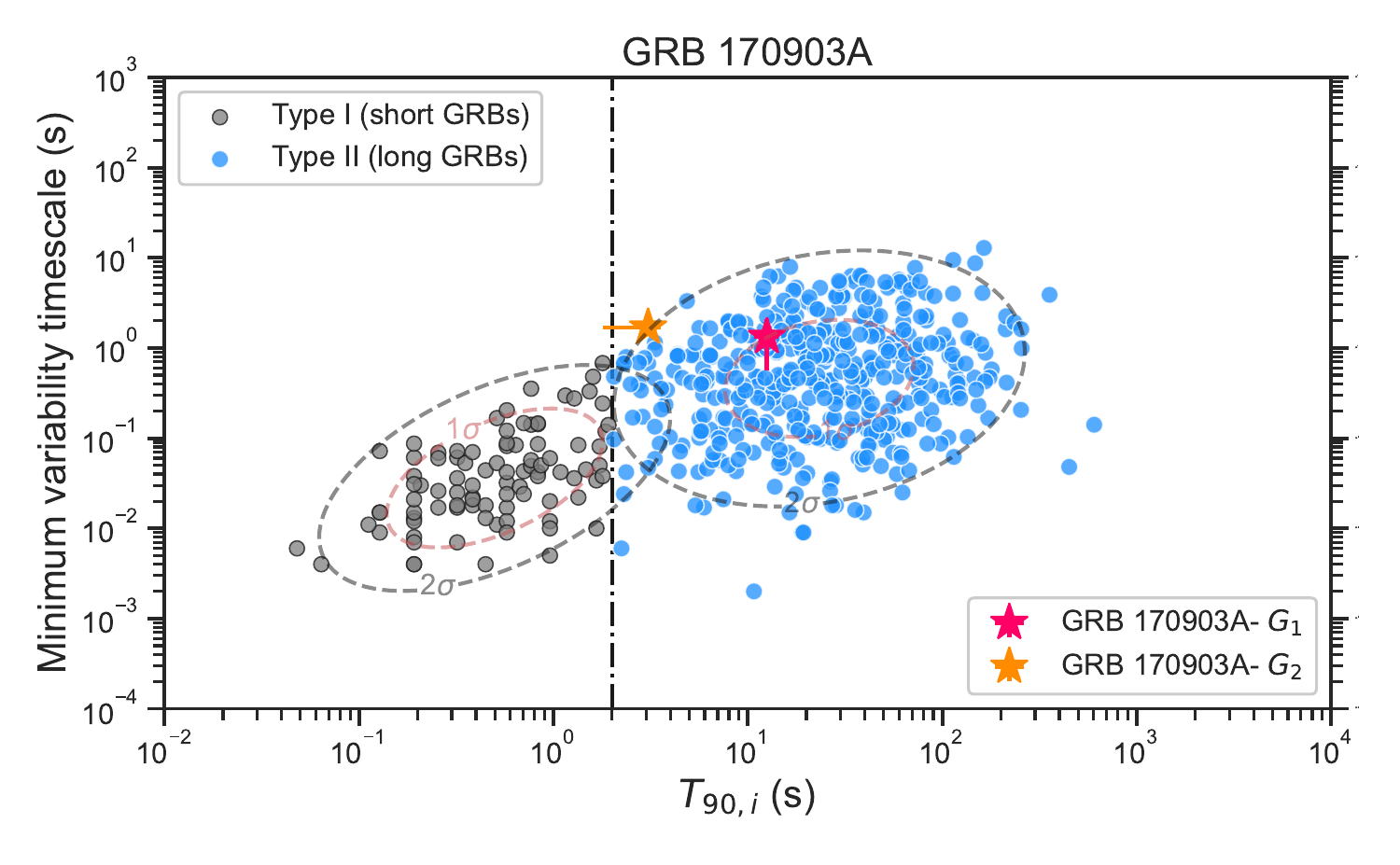}
\includegraphics[width=0.5\textwidth]{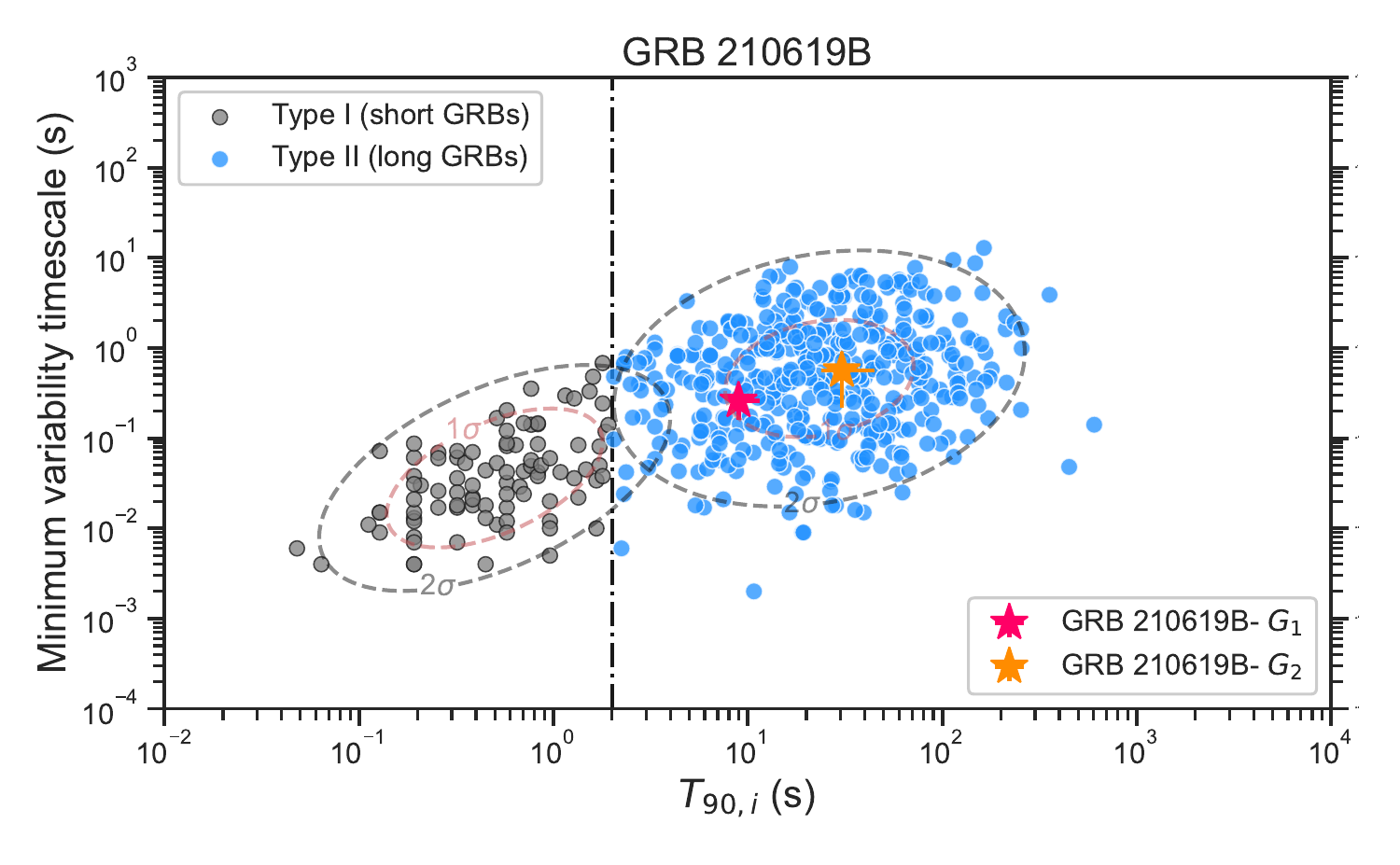}
\includegraphics[width=0.5\textwidth]{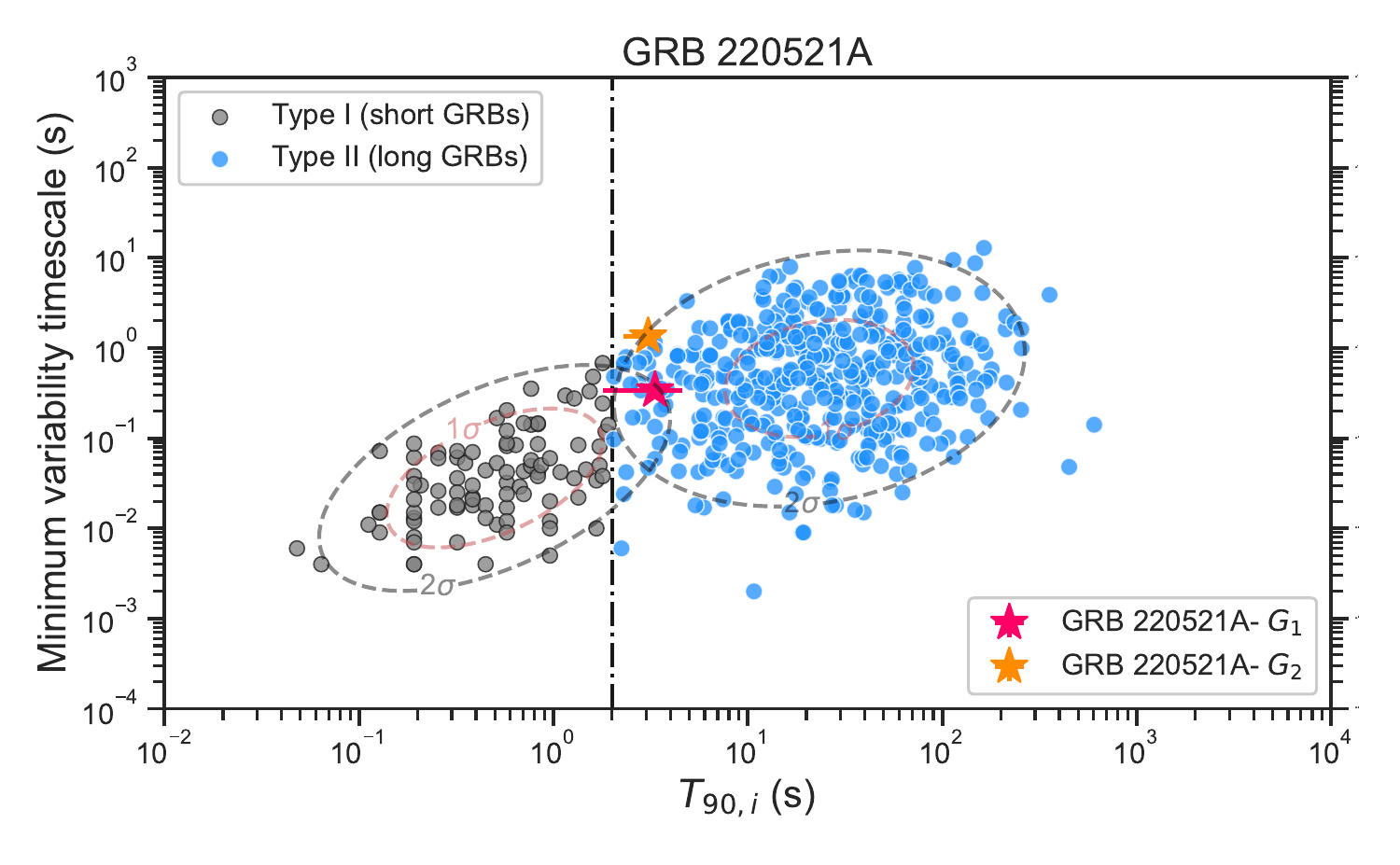}
\center{Figure \ref{fig:MVT_T90}--- Continued}
\end{figure*}

\end{document}